\documentclass[oneside]{zyx/mythesis}

\input{zyx/eth/mylogos.sty}

\begin{document}


\pagenumbering{Alph}
\hypersetup{pageanchor=false}

\begin{titlepage}
\centering

\includegraphics[width=.5\linewidth]{\logose}

\vspace{1cm}

{\Large \myThesisType}

\vfill

\begin{spacing}{1.1}
\Huge \bfseries \myTitle
\end{spacing}

\vspace{1.5em}

{\Large \myAuthor}

\vfill

\includegraphics[width=2.423cm]{\logotum} \hspace{1cm}
\includegraphics[width=2.8cm]{\logolmu} \hspace{1cm}
\includegraphics[width=2.43cm]{\logouna}

\vspace{1.2cm}

\textsc{\LARGE in cooperation with}

\includegraphics[width=.4\linewidth]{\logoeth}


\cleardoublepage
\end{titlepage}

\pagenumbering{arabic}

\frontmatter{}
\cleardoublepage

\thispagestyle{empty}
\begin{center}

\includegraphics[width=.5\linewidth]{\logose}

\vspace{1cm}

{\Large \myThesisType}

\vfill

\begin{spacing}{1.1}
\Huge \bfseries \myTitle
\end{spacing}

\vfill

\begin{tabular}{R{0.45\linewidth} p{0.45\linewidth}}
\ifthenelse{\equal{\mylanguage}{de}}{Autor}{Author}: & \myAuthor \\
\ifthenelse{\equal{\mylanguage}{de}}{Abgabedatum}{Submitted}: & \myEndDate \\
\ifthenelse{\equal{\mylanguage}{de}}{Aufgabensteller}{Supervised by}: & \myReviewFirst \\
\end{tabular}

\vfill

\includegraphics[width=2.423cm]{\logotum} \hspace{1cm}
\includegraphics[width=2.8cm]{\logolmu} \hspace{1cm}
\includegraphics[width=2.43cm]{\logouna}

\vspace{.8cm}

\textsc{\LARGE in cooperation with}

\includegraphics[width=.4\linewidth]{\logoeth}

\vspace{5mm}
{\Large
{\bfseries Institute for Software \& Systems Engineering}\\
Universitätsstraße 6a \hspace{0.25cm} D-86135 Augsburg\\
}

\end{center}

\hypersetup{pageanchor=true}
\cleardoublepage

\IfFileExists{mini/acknowledgments.tex}{
\currentpdfbookmark{\myAcknowTitle}{acknowledgments}

\thispagestyle{empty}

\vspace*{2cm}

\begin{center}
{\usekomafont{section} \myAcknowTitle}
\end{center}

\vspace{1cm}
\noindent

I thoroughly enjoyed my time at ETH Zurich and could not have imagined a more exciting (and timely) topic to research for my thesis. 

Therefore, I would like to thank Prof.~Alfons Kemper and Jan Böttcher from TUM, Prof.~Bernhard Bauer from UniA, and Prof.~Timothy Roscoe from ETH, not only for supervising me, but also for providing me with the opportunity to join ETH's Systems Group.

I am beyond grateful to my day-to-day supervisor Dr.~Vasiliki Kalavri for her great feedback, invaluable insights, and unfailing support throughout the thesis. I cannot imagine a better thesis supervision.

Further, I wish to express my sincere gratitude to Dr.~John Liagouris and Andrea Lattuada. I could not have hoped for better advice on thesis writing, SnailTrail, and the Timely Dataflow ecosystem. I would also like to thank the other members of the Systems Group, in particular Dr.~Frank McSherry and Dr.~Moritz Hoffmann, for their feedback and counsel.

Lastly, I am of course deeply thankful to my family and friends for their constant love, help, and support.

\cleardoublepage
}{}

\currentpdfbookmark{\myAbstractTitle}{abstract}

\chapter*{\myAbstractTitle}

We present \emph{ST2}, an end-to-end solution to analyze distributed dataflows in an online setting. It is powered by \emph{Timely Dataflow}, a low-latency, distributed data-parallel dataflow computational framework, and expands on its predecessor \emph{SnailTrail~1}, a system to run online critical path analysis on program activity graphs derived from dataflow execution traces. ST2 connects to a running Timely computation, creates the program activity graph representation, and runs multiple analyses on top of it. Analyses include aggregate metrics, progress and temporal invariant checking, and graph pattern matching. Through a command-line interface and a real-time dashboard, users are able to interact with and visualize ST2's analysis results.

For ST2's implementation, we discuss \emph{Differential Dataflow}, a framework that uses differential computation to incrementalize even complex relational dataflow operators, as an alternative to Timely Dataflow, but ultimately settle on using Timely. In our performance evaluations, we are able to show that ST2 is able to comfortably keep up with common streaming computations in offline and online settings, even exceeding SnailTrail~1's performance. We also showcase and evaluate ST2 from a functional standpoint in a case study. Using the dashboard to profile a faulty source computation, we manage to successfully detect the issues' root cause. We argue that ST2 is an extendable system that paves the way for users to debug, monitor, and optimize online distributed dataflows.

\microtypesetup{protrusion=false}
\cleardoublepage
\ifthenelse{\equal{\mylanguage}{de}}{
\currentpdfbookmark{Inhaltsverzeichnis}{contents}
}{
\currentpdfbookmark{Contents}{contents}
}
\setlength\cftbeforetoctitleskip{0pt}%
\setlength\cftaftertoctitleskip{1em}%
\setlength\cftbeforechapskip{0.9em}%
\tableofcontents
\microtypesetup{protrusion=true}
\cleardoublepage
\mainmatter{}


\chapter{Introduction}\label{introduction}

With today's companies' need to process ever-increasing amounts of data in a real-time fashion --- for example, in 2010 Facebook already processed more than 80 terabytes of data \emph{daily} \cite{chen2016, schroepfer2010} ---, stream processing and the dataflow programming model have become central computational paradigms in modern IT architectures. However, parallelized and distributed execution of sophisticated streaming jobs comes at a cost: the added complexity complicates tasks such as debugging, monitoring, performance analysis, and predicting future system behavior. Even if the right analyses can be formulated, carrying them out efficiently remains challenging. In many cases, offline execution or sampling do not suffice; instead, analyses have to be conducted online and under very low latency and high throughput conditions to be able to keep up with the profiled streaming computation.

To analyze distributed dataflows while they are running, \citeauthor{hoffmann2018} \cite{hoffmann2018} introduced \emph{SnailTrail~1}. It is powered by \emph{Timely Dataflow}, a low-latency, distributed data-parallel dataflow computational framework. Timely matches or surpasses the performance of many current stream processors, while still providing support for complex computations. SnailTrail~1 is able to run critical path algorithms to detect cross-worker bottlenecks in dataflow executions on top of online program activity graphs derived from stream processor execution traces.

Expanding on SnailTrail~1's ideas, the contribution of this thesis is \emph{ST2}, an end-to-end solution to analyze distributed dataflows in an online setting. This includes the following contributions:

\begin{enumerate}
  \item We compare different kinds of scaling models, time, and window semantics in the context of online dataflow analysis.
  \item We discuss \emph{Differential Dataflow}, an \enquote{extension} to Timely Dataflow that adds generalized incremental operators using differential computation, and evaluate the two systems' ST2-related differences.
  \item We present a custom adapter for timely and differential dataflows that efficiently transfers and consumes log events to create a graph representation from them.
  \item We implement new analyses on top of this graph, including aggregate metrics collection, temporal and progress invariant checking, and graph pattern matching.
  \item We provide a command-line interface and an interactive real-time dashboard that combines and visualizes these analyses to enable end users to monitor, debug, and optimize their streaming jobs. 
\end{enumerate}

Our results show that ST2 is able to provide complex analyses using rich window semantics through an easy-to-access real-time dashboard with which dataflow computations can be effectively debugged and optimized. Its implementation is decoupled from the source system, making it suitable to analyze a variety of distributed systems. Performance-wise, ST2 surpasses its predecessor, making it apt for profiling any common dataflow computation.

The thesis is structured into six chapters. In \cref{background}, we provide background and introduce related work referenced by the rest of the thesis. In \cref{ch:impl}, we discuss ST2's architecture, design considerations, and implementation. We then evaluate the performance of ST2's centerpiece, the program activity graph creation, in \cref{pag-evaluation}: We benchmark offline and online settings, contrast an implementation written in Timely with Differential, and compare our results to SnailTrail~1. In \cref{func-evaluation}, we examine ST2 from a functional perspective. After introducing its command-line interface and real-time dashboard, we benchmark the former and evaluate the latter's effectiveness in a case study. Finally, we summarize our findings and provide some pointers to possible areas of future work in \cref{conclusion}.

\chapter{Background}\label{background}

In this chapter, we provide background on concepts, systems, and related work that we will refer to throughout this thesis. First, we introduce the dataflow programming model (\cref{dataflow}). Built on top of this model, we give a brief introduction of Timely Dataflow (\cref{timely-dataflow}) and its \enquote{extension} Differential Dataflow (\cref{differential-dataflow}). We then discuss common ways to analyze dataflows in offline and online settings (\cref{dataflow-analysis}). Lastly, we introduce SnailTrail~1, a tool to run online critical path analysis on various stream processors, and the predecessor of ST2, the system created for this thesis (\cref{snailtrail}).

\section{Dataflow}\label{dataflow}

In the dataflow programming model, data coordinates a computation's execution. It was originally introduced by \citeauthor{rumbaugh1975parallel} \cite{rumbaugh1975parallel} and presents an alternative to the classic von Neumann architecture, where control structures such as conditionals and loops are used to coordinate execution. Computations in the dataflow model are expressed as a directed graph. Nodes in this graph are called operators and represent a unit of computation. Edges describe the channels between operators along which data can flow.

Compared to more traditional computation models, dataflows are especially apt to run concurrent and data-parallel computations. Dataflow programming commonly happens in two separate steps. First, the dataflow structure is created by connecting operators through channels. In a second step, this structure is fed with data, which is then processed by the dataflow pipeline. This model makes it easy to treat logical decisions about a dataflow separately from its physical representation --- incoming streams of data can be distributed across multiple worker threads and machines using data sharding. For these reasons, the dataflow computation model is well-suited and often used in stream processing contexts. Systems such as Apache Flink \cite{zotero-1215, carbone2016}, Spark Streaming \cite{zotero-1217}, Apache Storm \cite{zotero-1219}, Tensorflow \cite{zotero-1221}, and Microsoft Dryad \cite{poulain} all make use of dataflows under the hood. From a higher-level perspective, event streaming platforms such as Apache Kafka \cite{zotero-1225} also follow similar ideas in their systems' architecture.

\section{Timely Dataflow}\label{timely-dataflow}

Timely Dataflow (\enquote{Timely}) is a low-latency cyclic dataflow computational model that can be used for stream processing. It is the framework upon which ST2 was built. Its model was originally proposed by the Microsoft Naiad \cite{murray2013} system. It is written in the Rust programming language \cite{zotero-1231} and published open source \cite{mcsherry2019e}. The code of an exemplary timely dataflow computation is listed in \cref{lst:dataflow}. It first maps over its input, keying it by each record's remainder when divided by 10, then filters out all uneven keys, computes the sum per key, and inspects the result. A timely dataflow is always written from the perspective of a single worker, even if it is distributed across multiple threads later on (l.~9). The code sample also shows the two stages of dataflow programming: dataflow creation (ll.~11--23) and input ingestion (ll.~25--30).

\begin{lstlisting}[float=t,language=Rust,caption={Basic Timely Dataflow example},label={lst:dataflow}]
use timely::dataflow::InputHandle;
use timely::dataflow::operators::Input;
use timely::dataflow::operators::inspect::Inspect;
use timely::dataflow::operators::map::Map;
use timely::dataflow::operators::filter::Filter;
use timely::dataflow::operators::aggregation::Aggregate;

fn main() {
  timely::execute_from_args(std::env::args(), |worker| {

    // create dataflow structure
    let mut input: InputHandle<u64,_> = InputHandle::new();
    worker.dataflow(|scope| {
      scope.input_from(&mut input)
        .map(|x| (x % 10, x))
        .filter(|(k, _v)| k % 2 == 0)
        .aggregate(
          |_key, val, agg| { *agg += val; },
          |key, agg: i32| (key, agg),
          |key| *key as u64
        )
        .inspect(|x| println!("{:?}", x));
    });

    // feed input
    for round in 0..100 {
      if worker.index() == 0 {
        input.send(round as i32);
      }
    }
  }).unwrap();
}
\end{lstlisting}

\begin{figure}[htb]
\centering
\includegraphics[width=0.6\textwidth]{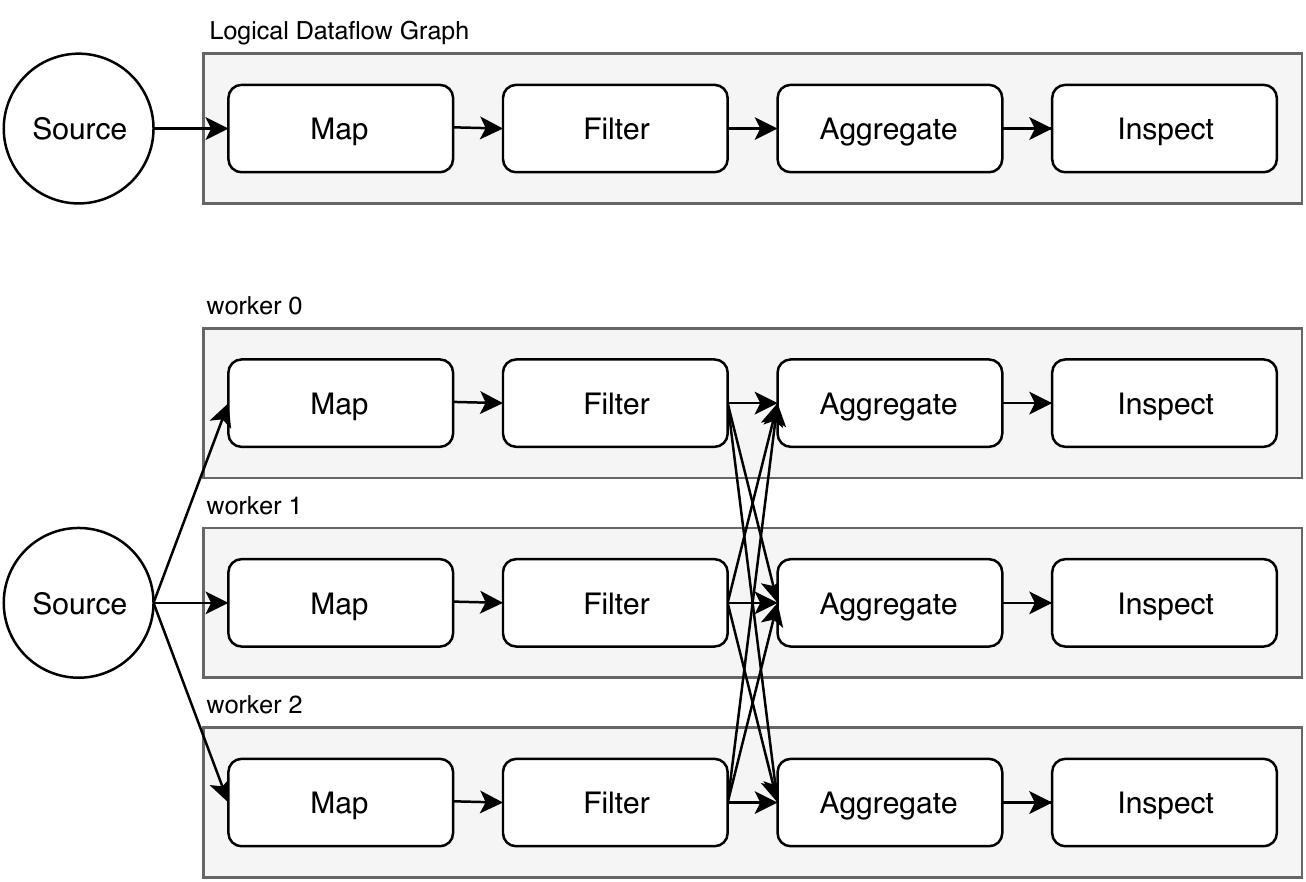}
\caption{Timely's logical and physical dataflow structure}
\label{fig:dataflow}
\end{figure}

\cref{fig:dataflow} depicts the logical and physical dataflow structure resulting from this computation. In Timely, every worker thread retains its own copy of the dataflow. Once the logical dataflow (shown at the top) has been created, it can freely be physically distributed to run across threads or whole clusters (shown at the bottom) --- as discussed in \cref{dataflow}, the computation's logical and physical structure are decoupled. While each worker processes a disjoint portion of the input data flowing in from the source, if there is more than a single worker, they must communicate with each other by exchanging data to arrive at correct overall results. This exchange is shown by the arrows between the \texttt{Filter} operator and the \texttt{Aggregate} operator.

Similarly to other stream processing systems, Timely also has to cope with unbounded streams of data that potentially arrive out-of-order at its dataflow operators. It has to determine times for which it can be sure that output results are correct (\enquote{safe} times), while still maintaining low latency: producing results too early will decrease latency at the cost of correctness, and vice versa. In the extreme case of not producing any results before all inputs have been seen, Timely \enquote{degrades} to a batch processor. What makes Timely unique compared to other stream processors is how it copes with this issue: It introduces a concept called \emph{Progress Tracking}. By attaching a logical timestamp to each datum in the system and imposing an order on them, Timely can control at what moment data becomes logically available (or \enquote{visible}) to the system. This visibility information is then propagated throughout the dataflow graph and across workers, allowing each operator to infer at which times it can safely produce results.

Under the hood, Timely uses an epoch-based progress tracking mechanism to achieve this. Each operator maintains a \emph{frontier} --- a set composed of mutually incomparable times\footnote{For totally-ordered timestamps, the frontier generally contains only a single element, since all times are mutually comparable. However, partially-ordered multidimensional timestamps are supported as well. In this case, the frontier might consist of multiple times.}. New inputs have to be expected for any times that are in advance of at least one element in the frontier: the upstream operator holds the \emph{capability} to still produce outputs for these open times. Conversely, stored results for data with times smaller than all elements in the frontier are safe to output, as no upstream operator will ever again produce data for these times. This progress tracking mechanism is the central way of coordination between worker threads and across processes. As a democratized peer-to-peer model, it is also a departure from other systems (e.g.\ Apache Spark \cite{zotero-1233}), which frequently make use of master-slave architectures for coordination. It allows Timely to communicate fine-grained progress information without requiring centrally coordinated control mechanisms. Lastly, progress tracking not only provides a way to effectively navigate the trade-off between low latency and correctness of results. It also allows Timely to express complex streaming computations such as (nested) iteration by using multidimensional timestamps. For example, a loop is able to keep track of its current iteration in its timestamp, which is then reconciled using Timely's progress tracking mechanism to prevent non-termination.   

\section{Differential Dataflow}\label{differential-dataflow}

Differential Dataflow (\enquote{Differential}) is a programming framework that implements \emph{differential computation} \cite{mcsherry2013}, a generalization of \emph{incremental computation}, on top of Timely. While ST2 is built with Timely, during its development we therefore also considered to what extent Differential can act as a stepping stone for ST2 to enable richer semantics, more efficient computation, and database-like relational and declarative operators to express its algorithms, and we also report on our findings over the course of this thesis. 

Like Timely, Differential is written in Rust, and also published open source \cite{mcsherry2019d}. It extends Timely with generalized incremental operators such as \texttt{map} and \texttt{filter}, and also more complex operators such as \texttt{reduce} and \texttt{join}. This way, it provides a novel way of writing complex streaming computations.

Differential achieves this by introducing the notion of \texttt{Collection}s, streams that contain a multiplicity coordinate in addition to the datum they carry. This makes it possible to think about streams as a collection of \emph{changes} (\enquote{differences}) with multiset semantics, where data is permanently added or retracted. \texttt{Collection}s also make use of partially-ordered, multidimensional timestamps to solve the issue of efficiently (i.e. in an incremental fashion) performing iterative computations in distributed environments. Imposing a partial order on multidimensional timestamps makes it possible to not only distinguish between fresh inputs and inputs from iterative feedback flowing into the iteration, but also to compute iterations for batches of input in parallel. Records belonging to related times are visible to each other, while unrelated records will not have an effect on each other and can therefore be processed independently.

Differential computation allows Differential to achieve very low latency characteristics, even in the face of data retractions for complex, iterative algorithms. Incremental computation makes this possible. Given a collection of records $X$, in a non-incremental setting,  we would compute a new collection $Y$ by applying a computation $f$ to all of $X$: $f(X)=Y$. To incrementalize this computation, we must find a $\delta f$ such that $f(X+\delta x) = f(X) + \delta f(\delta x) = Y + \delta y$ \cite{gobel2019}. $\delta f$ should then only have to do work in the order of the input change $\delta x$. This also works in practice, as long as an input change results in a similarly large output change. Thus, for many computations, Differential only has to do a small amount of work --- sometimes none at all --- for each update, while still being able to leverage Timely's efficient dataflow model.

Additionally, Differential also supports \emph{arrangements}, which maintain a compacted in-memory representation of historical \texttt{Collection} traces in the form of update batches. This indexed state can then be shared between independent operators, similarly to database indices.

As far as we know, Differential is unique in its ability to uphold a stream processor's performance characteristics even for complex computations involving relational operators, and also in its ability to provide a means to efficiently share historical inputs across the dataflow.  

\section{Dataflow Analysis}\label{dataflow-analysis}

Analyzing distributed dataflows is challenging. Similarly to general distributed systems profiling, many tasks, activities, and operators have to interact with each other on the logical and physical level to create a functioning system  in the face of partial failure and unpredictable delays at any level of the stack \cite{oldenburg, bailis2014, deutsch1994}. Especially the interdependencies between these components add complexity to any debugging or analysis endeavor --- often, performance issues such as bottlenecks appear in one part of the system, while their root cause is located elsewhere. Long-running, dynamic workloads further complicate this and make it challenging to define scalable and reusable metrics. Even if metrics have been defined, few systems are able to execute them while still keeping up with the profiled computation in an online setting.

On the other hand, successful dataflow analysis can provide viable insights into a distributed system. With effective metrics, the computation's behavior can be explained, provenance collected, and perhaps even future performance predicted. This allows for more efficient debugging of faulty computations, and performance tuning. It could also enable analysts to audit dataflow invariants to e.g.\ ensure that service-level-agreements are upheld by the system.

Prior work exists for debugging and profiling distributed systems and dataflows. \citeauthor{zhao2016} \cite{zhao2016} provide a comprehensive overview of existing approaches and assign them to three categories: using pre-defined event semantics (\enquote{intrusive} instrumentation), static analysis to infer the system model, or using machine learning. For example, similarity scores derived from graph matching networks \cite{li2019} might be used to deduce the differences between a healthy and a struggling dataflow computation. In dataflow analysis, intrusive instrumentation is most commonly used, as many stream processors (e.g.\ Apache Flink and Timely Dataflow) already output profiling information that can be used for further analyses. For example, \citeauthor{beschastnikh2014} present \emph{CSight} \cite{beschastnikh2014}, which allows them to infer models of distributed systems using log traces they emit.

 \citeauthor{hoffmann2018} \cite{hoffmann2018} compare dataflow performance analysis approaches. For example, \emph{Nagios} \cite{zotero-1248} and \emph{Ganglia} \cite{sacerdoti2003} provide aggregate metrics but cannot discover dependency-related performance issues, while \emph{Splunk} \cite{zotero-1250} and \emph{VMware vRealize} \cite{zotero-1252} ``can isolate specific instances of performance loss, but lack a big picture view of what really matters to performance [\ldots] on a varying workload'' \cite{hoffmann2018}. They propose SnailTrail, a system that is able to overcome other systems' limitations by using a modified critical path analysis that is able to capture worker dependencies of streaming computations in real-time (cf.~\cref{snailtrail}).

In the context of unbounded streaming data, many traditional analysis tools become inefficient or ineffective: They cannot run on all of the data, as there might never exist a state in which all data has been seen. Instead, analyses have to somehow present accurate results while only being applied to excerpts of the profiled data streams. To be of use, analyses also should happen in an online fashion: If they are either too slow to keep up with the monitored computation, or require the profiled system to be taken offline in order to extract information, results will be out-of-date, thereby diminishing their practical use. Two kinds of computations are well-suited even for these kinds of workloads.

\begin{description}
	\item [Graph algorithms and graph pattern matching] can be applied to subsets of a larger dataflow computation, if their log traces have been converted into a graph representation (cf.~\cref{snailtrail}). While \citeauthor{lattanzi2011} \cite{lattanzi2011} are able to express graph algorithms on top of the \emph{MapReduce} model \cite{dean2008} used by e.g.\ Hadoop \cite{zotero-1257}, \citeauthor{mcgregor2014} \cite{mcgregor2014} surveyed approaches to express similar algorithms in a streaming fashion. This opens the doors to use a sufficiently capable stream processor to analyze another stream processor's log event traces with graph algorithms. \citeauthor{oldenburg} \cite{oldenburg} make use of a query language built on top of lineage graphs to express debugging questions on traces of distributed executions. This way, similarly to the concept of \emph{data provenance} known from database systems \cite{buneman2001}, they hope to provide data-centric explanations of computation behavior. \citeauthor{khan2010} \cite{khan2010} introduce proximity patterns that can be used to mine even very large graphs, and also provide an overview of other common graph patterns and graph pattern mining algorithms.
	\item [Invariant checking] can be used to debug and optimize dataflows by verifying that they uphold correctness and performance guarantees. For example, \emph{SWIFT} \cite{yan2018} is a system to mine representative patterns from large event streams, and \emph{CSight} \cite{beschastnikh2014} can be used to check \enquote{always followed by}, \enquote{never followed by}, and \enquote{always preceded} event invariants on concurrent system logs, which could also be used to check a dataflow for its correctness. \citeauthor{beschastnikh2011} \cite{beschastnikh2011} also present algorithms to check temporal invariants on partially ordered logs, which is especially useful for performance monitoring of streams --- e.g., growing processing times at a specific dataflow operator hint at further optimization potential.
\end{description}

In addition, it is of course also possible to analyze dataflows using more conventional means: aggregate metrics can be used to provide an overview of a profiled computation, with the aforementioned caveat that they are unable to capture dependency information. Overall, the analysis of distributed dataflows presents similar opportunities and challenges as analysis of distributed systems in general --- in particular since their computations are potentially unbounded. A key to effectively extract insights from them is to make use of the log traces they publish. These log event streams can then be converted to more \enquote{traditional} representations such as graphs, which allows a sufficiently expressive stream processor to analyze them using familiar algorithms. 

\section{SnailTrail~1}\label{snailtrail}

SnailTrail~1 is a system built on top of Timely Dataflow that is able to analyze distributed dataflows in an online fashion (cf.~\cref{dataflow-analysis}). It was introduced by \citeauthor{hoffmann2018} \cite{hoffmann2018} and is ST2's predecessor, even though ST2 has been rewritten from the ground up, as will be discussed in \cref{ch:impl}. Using the stream of log events that are produced by a profiled computation (the \enquote{source computation}) as input, SnailTrail~1 constructs a graph representation called the \emph{Program Activity Graph} (\enquote{PAG}).

\begin{figure}[htb]
\centering
\includegraphics[width=0.6\textwidth]{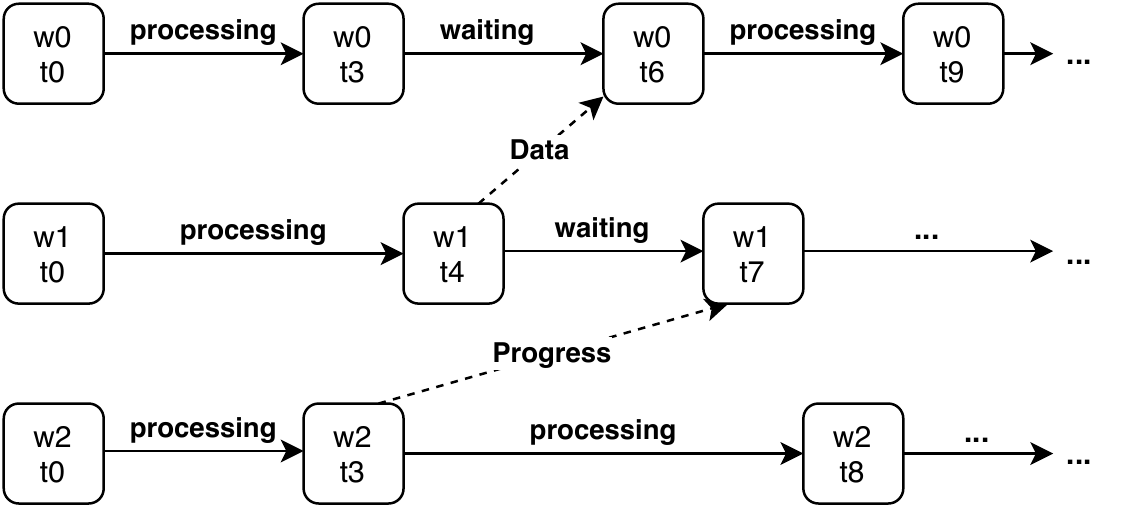}
\caption{Exemplary PAG}
\label{fig:pag}
\end{figure}

An exemplary PAG is depicted in \cref{fig:pag}. It consists of nodes and edges. PAG nodes represent the start or end of an activity in the dataflow. They belong to a specific worker ($w_0, w_1, w_2$ in the visualization) and are identified by the timestamp at which the log message was created ($t_0, t_1, ..., t_9$ in the visualization). PAG edges represent dataflow activities and can be either worker-local or remote (i.e., spanning multiple dataflow workers). Examples for local activities are scheduling and processing of operators, message (de)serialization when receiving or sending data across multiple processes, or waiting activities, where a worker cannot progress without external input from another worker. Waiting activities are particularly important, as they hint at optimization potential in the dataflow's execution --- if they occur, for some reason, a worker has been blocked. Remote activities are mostly data messages, i.e.\ the transfer of data from one operator to another (of course, this can also happen between operators in a local context), and control messages --- in Timely, these are the epoch messages that are propagated through the dataflow during its progress tracking mechanism. Additionally, PAG edges contain activity type-dependent metadata. For example, processing activities will typically include information on which operator was scheduled, and how many records were processed.

To create the PAG from a stream of log events, relatively little information is necessary: Most distributed systems and stream processors readily provide logging information on activity starts and ends, which is enough to deduce the corresponding starting and ending PAG nodes, and a connecting edge in-between. This makes the PAG suitable as a common abstraction of event traces stemming from a multitude of systems: SnailTrail~1 decouples its profiling algorithms from system-specific trace adapters through the common PAG representation. Thus, as long as an adapter that transforms the respective system's events into the PAG can be written, SnailTrail~1 is able to support multiple stream processors and dataflow systems with relatively low additional effort, such as Timely Dataflow itself, Apache Spark \cite{zotero-1233}, Tensorflow \cite{zotero-1221}, Apache Flink \cite{zotero-1215}, and Twitter's Apache Heron \cite{twitter}.

On top of the PAG, SnailTrail~1 runs the \emph{critical participation} metric, which captures the importance of an execution activity by estimating its contribution to the computation's critical path. As this happens in a possibly unbounded streaming context, the true critical path can only be estimated; \cref{sec:impl_considerations} discusses SnailTrail~1's fixed window semantics, which force the PAG to be created from arbitrarily cut-off execution snapshots. This not only complicates the critical path analysis, but also makes it difficult for SnailTrail~1 to support further analyses.

SnailTrail~1 is able to detect non-trivial bottlenecks in computations and demonstrates that a Timely Dataflow implementation is able to keep up with online streaming computations: In their benchmarks, \citeauthor{hoffmann2018} \cite{hoffmann2018} report that SnailTrail~1 is about two orders of magnitudes faster than it needs to be in order to handle the source computation's load. However, SnailTrail~1 does not support running in an online setting without further implementation work. In its evaluation, only offline log traces were used.

While SnailTrail~1 laid the groundwork for online dataflow analysis, it is not an \enquote{off-the-shelf} solution easily accessible by end users. As explained in \cref{introduction}, ST2's goal is to expand on SnailTrail~1's work and to overcome these limitations. ST2 starts from SnailTrail~1's conceptual ideas, in particular the creation of a graph representation from a stream of log events that is completely decoupled from the source computation. ST2 further aims to provide additional analyses using richer window semantics for the PAG construction, support running in offline and online settings, boost SnailTrail~1's performance in both modes, explore potential opportunities to use Differential, and present itself in an accessible way to end users. ST2's initial version focuses on profiling the Timely ecosystem. As the Timely stack is particularly demanding, support for further stream processors is simple to build later-on.

\section{Background Summary}

In this chapter, we introduced important concepts and related work to support understanding the \enquote{why} and \enquote{what} of ST2. We gave an overview of the dataflow model, which provides a scalable and data-centric model for streaming computations. We then presented Timely Dataflow, a streaming computational framework leveraging this model and solving common streaming challenges with its unique progress tracking mechanism. Further, we discussed Differential Dataflow, which, powered by differential computation, extends Timely Dataflow with incremental relational operators. In \cref{dataflow-analysis}, we introduced challenges and opportunities of dataflow analysis, and provided background on approaches to analyze dataflows and distributed systems in an online fashion and in the face of potentially unbounded data. In particular, this highlighted graph algorithms and invariant checking on top of graph representations of execution traces as a promising way to debug and analyze the performance of dataflow systems. Lastly, we gave an overview of SnailTrail~1, ST2's predecessor. SnailTrail~1 discovers bottlenecks in a source computation by estimating each activity's contribution to the computation's critical path. For this, it uses the PAG, a graph representation of the source computation's execution traces. However, it does not come without caveats, e.g.\ regarding its window semantics and real-world capabilities. ST2 aims to overcome these limitations by providing richer window semantics and new analyses that are readily available to end users, boosting performance in offline and online settings, and evaluating opportunities to use Differential.

\chapter{Implementation}\label{ch:impl}

In this chapter, we discuss ST2’s implementation. The chapter does not solely focus on ST2’s final state, but also discusses alternative implementations, even if they might not have been used in ST2’s final version.

While ST2's high-level architecture and application flow, which we discuss in \cref{sec:impl_architecture}, are similar to SnailTrail~1, its implementation is rewritten from the ground up. In \cref{sec:impl_considerations}, we present and compare three major computational considerations that led to this decision: the scaling model, which defines what kind of scaling can and should be expected from ST2, window semantics, which define how aggregate computations work and what one iteration of ST2’s PAG represents, and time semantics, which define how ST2’s time progresses and how Timely and Differential can interact with ST2.

We then explore each of the architecture flow’s elements in greater detail: The adapter between source computation and ST2 (\cref{sec:impl_adapter}), the \texttt{LogRecord} (\cref{sec:impl_logrec}) and PAG construction (\cref{sec:impl_pag}), and analytics based on the PAG (\cref{sec:impl_analysis}). Lastly, we summarize the chapter’s findings in \cref{sec:impl_summary}.

\section{High-Level Architecture}
\label{sec:impl_architecture}

\begin{figure}[htb]
\centering
\includegraphics[width=0.7\textwidth]{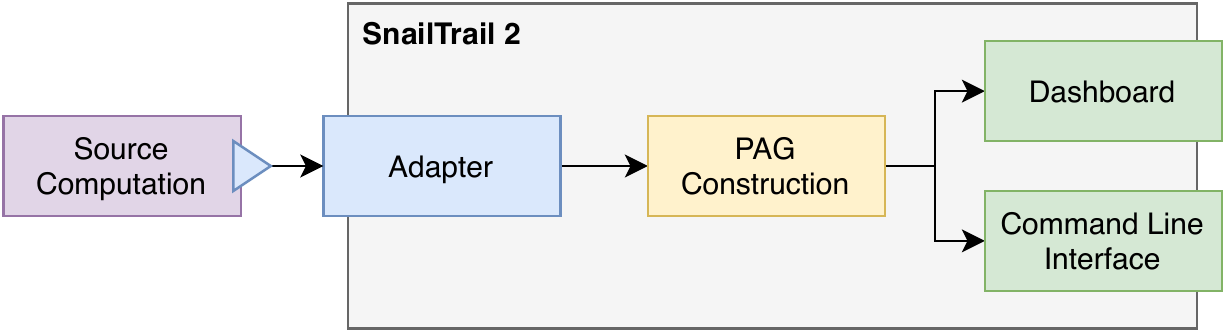}
\caption[ST2 flow]{ST2 flow. Event ingestion in blue, PAG construction in yellow, analyses and user-facing components in green.}
\label{fig:st2}
\end{figure}

ST2 internally works like a data pipeline. It ingests traces from a source computation running on a stream processor --- in its current version, only Timely and Differential are supported ---, then creates the PAG representation from them (cf.~\cref{snailtrail}), and finally runs algorithms on top of the PAG. These steps are visualized in \cref{fig:st2}.

First, the source computation that should be analyzed has to be enriched with calls to an adapter library. The calls trigger every time a new epoch in the computation ticks (represented as blue triangle in \cref{fig:st2}). The adapter then processes the raw event data it receives from the source computation and creates an event stream from it for further consumption (cf.\ \cref{sec:impl_adapter}). For the current ST2 version, only Timely and Differential computations are supported by the provided \texttt{st2-timely}, but, as the adapter is independent of computations commencing downstream, new adapters could be written for other stream processors\footnote{This is similar to SnailTrail~1, which supports Timely, Apache Flink, Tensorflow, Heron, and Spark.}, as long as they satisfy the same contract (cf.\ \cref{subsec:impl_contract}). Once ST2 has connected to the event stream --- either by reading from file, or from socket ---, it can replay the stream into its own dataflow (cf.\ \cref{subsec:impl_online_offline}). This dataflow creates the \texttt{LogRecord} representation from the replayed events, the PAG from the stream of \texttt{LogRecord}s (cf.\ \cref{sec:impl_pag}), and analysis output from the PAG (cf. \cref{sec:impl_analysis}).

\begin{figure}[htb]
\centering
\includegraphics[width=0.7\textwidth]{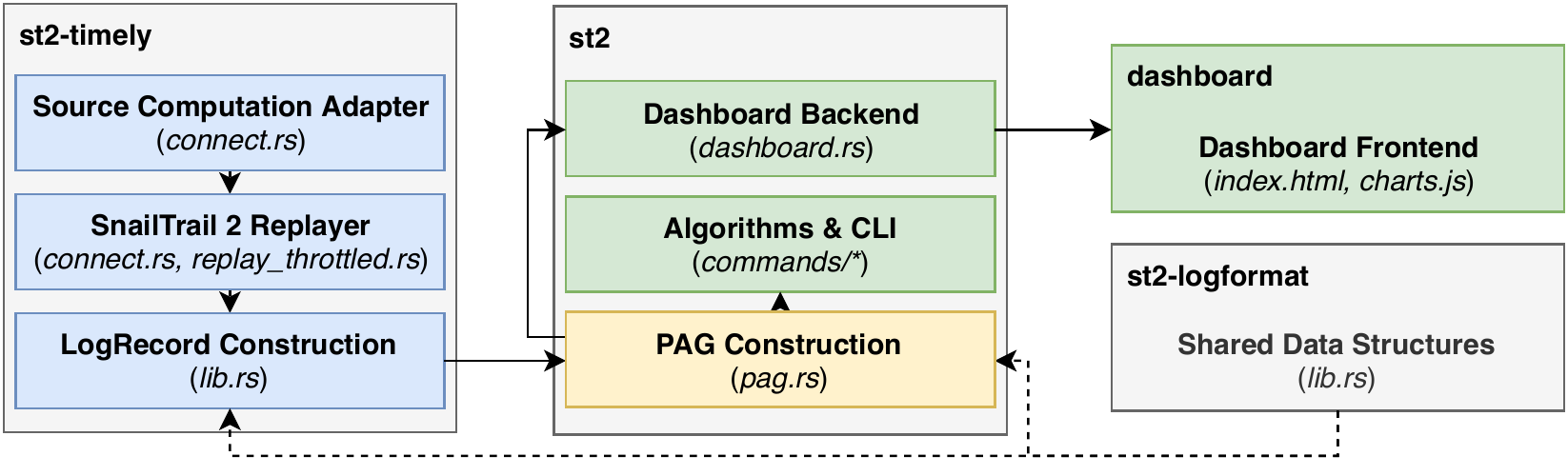}
\caption[ST2 code structure]{ST2 code structure. Color-coding matches \cref{fig:st2}.}
\label{fig:st2arch}
\end{figure}

In code, the pipeline is structured into multiple Rust crates, depicted in \cref{fig:st2arch}. \texttt{st2-timely} contains the implementation for the adapter between ST2 and a Timely or Differential source computation, the custom operator that replays the source computation log events into ST2, and the \texttt{LogRecord} creation. It relies on \texttt{st2-logformat}, which contains shared data structures such as the \texttt{LogRecord} struct and the \texttt{Pair} timestamp type. \texttt{st2} generates the PAG and analyzes it. Its results are either exposed through a user-facing command line interface, or fed into an interactive dashboard. \texttt{dashboard} contains this dashboard's frontend code.  

\section{Computational Considerations}
\label{sec:impl_considerations}

In the design of ST2, we had to make three major choices: what is the dataflow's scaling model, should it support some kind of windowing, and how should time be modelled. This section discusses these considerations.

\subsection{Scaling Model}

Similarly to many other stream processing jobs, the main performance guarantee that ST2 has to provide is that it is able to keep up with the incoming flow of data to avoid a growing amount of back-pressure; ST2 needs to be able to run live analyses on a batch of event log traces from the source computation before the next batch of traces has been generated. While we also evaluate ST2's performance characteristics (cf. \cref{pag-evaluation}), as long as it can keep up with an arbitrary source computation, it is \enquote{good enough} for practical use. Depending on the profiled computation, it is challenging to meet this requirement. There is no obvious relationship between a given source computation and the size of the log traces it generates --- sometimes there are more log events generated for ST2 to analyze than raw tuples processed by the source computation. In many cases, keeping up with such a computation is not a big issue, since Timely is able to outperform most stream processors, but for profiling a timely or differential dataflow, this poses a bigger challenge; all the more as ST2’s goal is to implement new analyses on top of the PAG that might be more computationally intensive than SnailTrail~1’s \emph{critical participation metric} (cf.~\cite{hoffmann2018}). These challenges can be tackled in two ways: The PAG construction and analysis can be sped up by implementing them in a more efficient way, or they can be parallelized across workers and processes. SnailTrail~1 mostly focused on the former, while ST2 also expands on the latter.

There are three hotspots for parallelization in both SnailTrail versions: Ingesting source computation events (\enquote{I/O}), creating the PAG, and analyzing the PAG (cf.~\cref{fig:st2}). Speeding up the event ingestion can either be achieved by feeding event batches to multiple \emph{instances} of ST2 in a round-robin fashion, such that each instance maintains a partial view on the source computation, or by feeding event batches to multiple \emph{worker threads and processes} of a single ST2 instance. The latter is discussed in this thesis. For the PAG construction, initial parallelism across workers can be achieved in the same vain as SnailTrail~1: By assigning one ST2 worker to one source computation worker, the \enquote{timeline-local} part of the PAG construction can happen in isolation. However, this limits scalability. Supporting a higher degree of parallelism on ST2's end than matching workers one-to-one requires additional implementation, as will be discussed in greater detail in \cref{sec:impl_adapter}, but opens up new performance opportunities that were not supported by SnailTrail~1. The control gained by the customized event transmission can then also be used to implement the PAG construction (cf. \cref{sec:impl_pag}) and analysis (\cref{sec:impl_analysis}) in a performant, non-blocking fashion.

In summary, ST2 should be able to scale with the demands of an online computation --- regardless of input batch size, generated event log trace density, or algorithmic complexity. To achieve this, we implement its hotspots as efficiently as possible: We parallelize beyond one-to-one worker matching, and implement the PAG construction and analyses in a non-blocking fashion.

\subsection{Window Semantics}
\label{subsec:impl_windows}

In stream processing, an important decision to make is how to treat aggregates. Streamed data might arrive late, which makes continually producing correct aggregate results while maintaining low latency challenging: A stream is a potentially unbounded flow of data --- operators cannot wait until they have seen all events before they make progress and produce results, as this would block the overall computation indefinitely. As discussed in \cref{timely-dataflow}, Timely uses its progress tracking mechanism to solve this issue. As long as time progresses, Timely can advance the computation and provide results that are consistent and correct as of that time. By using multidimensional timestamps, it is able to make progress even if only one time \emph{dimension} advances. Differential takes this one step further (cf.~\cref{differential-dataflow}): It relaxes the order imposed on time and implements its operators in the context of partially-ordered time, allowing it to process data belonging to unrelated times in parallel. Using differential computation, it is also able to efficiently maintain aggregate results over time. On input changes at a later time, Differential will incrementally compute the necessary additions and retractions to reflect the updated results.

Other stream processors collect records into \emph{windows} based on a timeframe (cf.~\cite{stonebraker2005, carbone2016}). Once a window has closed, aggregate results for that window are produced. This way, the computation continually makes progress. However, consistency needs to be traded off against latency in the face of late-arriving data. Results are also only valid for a single window and are not maintained over time.

\begin{figure}[htb]
\centering
\subfloat[Fixed window. Assigns activities based on disjoint time frames.]{\includegraphics[width=0.31\textwidth]{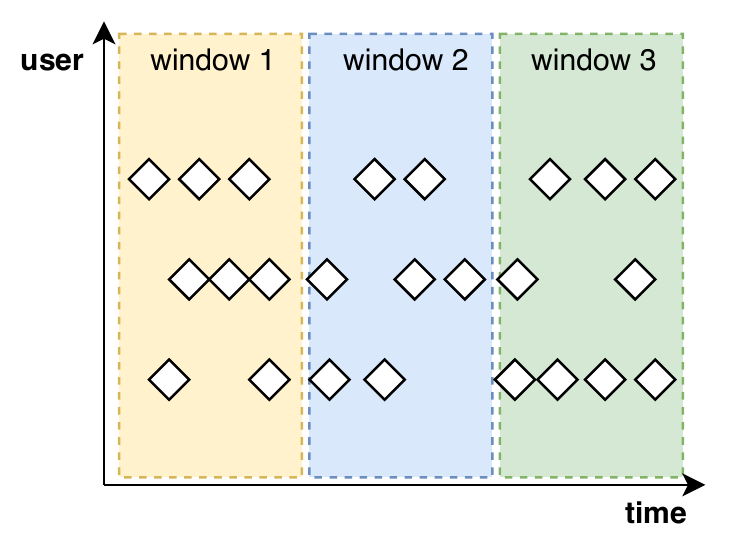}}\hfill
\subfloat[Sliding window. Assigns activities based on overlapping time frames.]{\includegraphics[width=0.31\textwidth]{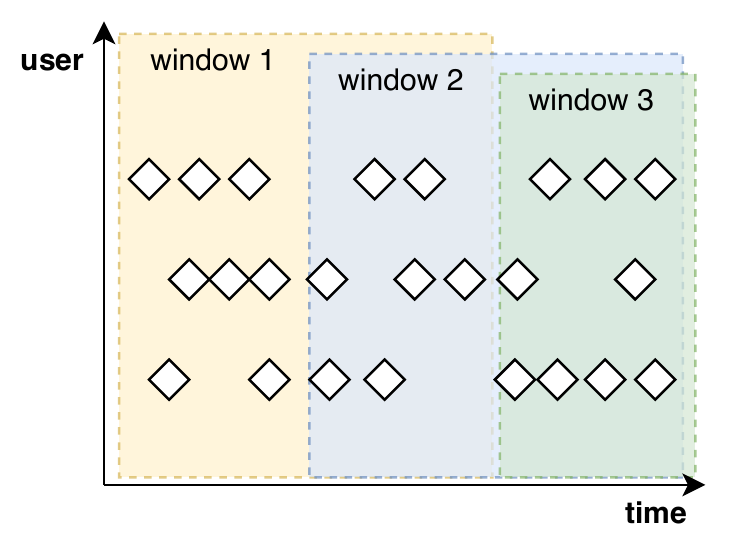}}\hfill
\subfloat[Session window. Assigns activities based on inactivity gaps.]{\includegraphics[width=0.31\textwidth]{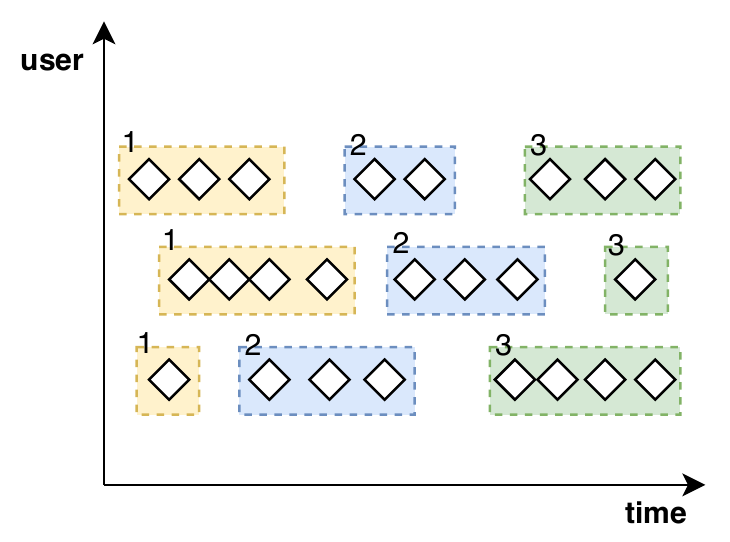}}
\caption[Window comparison]{Window comparison. Activities (diamonds) by multiple users (y-axis) are assigned to various windows types across time (x-axis).}
\label{fig:windows}
\end{figure}

Of course, there are cases where windowing is not a mere technical consideration, but used to express business logic. For example, an analyst might be interested in measuring sales statistics over the last 24 hours, or might like to create a daily sales report. In that case, using a sliding or fixed window \cite{akidau2018} is a simple way to produce the desired reports. Windows can be implemented using Timely and Differential as well\footnote{Implementing them in Differential is particularly interesting: By retracting records that fall out of the window, a changing view on the same data can be maintained over time.}. Both SnailTrail versions make use of window semantics to create and analyze the PAG. SnailTrail~1 uses a fixed window for creating the PAG. Below, we discuss the resulting implications of choosing this semantics and compare it to other window types depicted in \cref{fig:windows} that could be used in ST2.

\begin{description}
	\item [No Window] Both SnailTrail versions create a graph representation of the events they receive. In a no window setting, this graph continuously grows and accumulates state. Careful engineering is necessary to avoid buffer overflows. All operators need to be incrementalized (e.g.~using differential operators) to avoid computing from scratch for every new event. While overcoming these challenges might be possible, analyzing all events since the source computation's beginning is not really necessary: To analyze the current health of a streaming job, it mostly suffices to consider recent information. Thus, SnailTrail~1 and ST2 use windowing.

	\item [Fixed Window] In a fixed window (sometimes also called \emph{tumbling window}), the stream (for ST2, this is the stream of the source computation events) is split into windows at fixed processing time intervals (cf.~\cref{fig:windows}~\emph{(a)}). For example, a one-second fixed window would group events into the processing time (cf. \cref{subsec:impl_time}) chunks $[0s, 1s), [1s, 2s), ...$. Each window can contain a varying number of events, might split an epoch’s events at arbitrary points, or join multiple epochs’ events together. One side-effect of this is that every generated PAG’s timelines will have the same length, which complicates the computation of distance-based metrics such as critical path analysis (cf.~\cite{hoffmann2018}). Fixed windows also cannot make use of differential computation in any way: every window is completely different from the last, so no computational reuse is possible. Fixed-window semantics are easy to implement, but for ST2, this benefit does not outweigh its costs.
	
  \item [Sliding Window] Similarly to a fixed window, a sliding window also defines a processing time frame in which events are \enquote{collected} (cf.~\cref{fig:windows}~\emph{(b)}). Unlike fixed windows, sliding windows might overlap with each other, causing the impression that they \enquote{slide} across the data over time. They are a generalization of fixed windows. While sliding window semantics are useful for analyzing continuously changing graphs of data, in ST2’s case, they do not provide major benefits compared to a simple fixed window: The PAG is still cut off at arbitrary times, and all resulting analytical issues remain valid. A slight advantage might be that differential retractions happen in a smoother fashion (if Differential is used for the PAG construction). In a fixed window setting, events are inserted as they happen, and then have to be removed all at once as a window closes. Sliding windows allow the events to be retracted one by one as they fall out of the window.
  
  \item [Session Window] \enquote{An example of dynamic windows, sessions are composed of sequences of events terminated by a gap of inactivity greater than some timeout} \cite{akidau2018a} (cf.~\cref{fig:windows}~\emph{(c)}). In ST2's case, each of the source computation’s epochs can be interpreted as a tick in \enquote{event time}. Being able to group all events that result from one round of inputs into a PAG (\enquote{a session}) solves many of the issues with fixed and sliding windows. The PAG is not cut off at any arbitrary position --- distance-based metrics work out of the box. It contains all events that belong to one epoch --- statements about the whole dataflow can be made. In that way, it is \enquote{consistent}. Furthermore, it might be able to leverage differential features: Operators and their channels stay the same in a static dataflow. Thus, PAG epochs are structurally similar, and a single PAG representation could be maintained over time. As new epochs introduce or retract edges due to varying source computation runtime characteristics, the PAG is updated accordingly. PAG epochs should also be comparable, which allows analyzing the source computation’s performance development over time. An \enquote{epochal session window} provides the basis for such analysis.
  
  Of course, it is no silver bullet. Identifying which PAG edges are the same between PAG epochs is challenging, as their identity can only be established through metadata. The source computation needs to output additional profiling information to communicate when a round of inputs has closed. Outside of Timely, the concept of \enquote{rounds of input} might not even exist. In a Timely setting, the computation must not compute more than one epoch at a time; otherwise, events from multiple epochs might get mixed up again (while this issue might be solvable by constructing PAGs for multiple epochs in parallel, it is not trivial to implement such functionality). However, even with these caveats, sessions provide a basis that is richer and allows for more interesting and easier analyses than the original fixed-window semantics. They also provide an arguably more natural mental model for thinking about what one PAG iteration represents. Lastly, they open up opportunities for further research into using Differential to support ST2, whereas with the original time semantics, there are no such opportunities. For these reasons, we chose epochal session window semantics for ST2.
\end{description}

\subsection{Time Semantics}
\label{subsec:impl_time}

Timely and Differential offer richer time semantics than the majority of existing stream processors. Not only a monotemporal timestamp can be assigned to an event, but also bi- and multitemporal timestamps are possible. This is especially useful when event and processing time have to be modelled, or when loop iterations should be tracked in addition to the overall progress of a computation. The mathematical ordering between individual timestamp coordinates is adjustable as well. This enables use cases where bitemporal time using a total order (e.g.\ a lexicographical order) does not suffice, and instead partial orders with independently tracked times need to be used (cf.~\cref{timely-dataflow} and \cref{differential-dataflow}). One such example is the issue of late arrivals --- data belonging to an earlier time that might arrive late (e.g.\ due to network issues). In that scenario, common stream processors either have to sacrifice correctness by providing a result \enquote{early}, or they sacrifice latency and wait a set amount of time before they publish a result. However, this breaks when data can be infinitely delayed, and does not solve the underlying issue --- conflating processing and event time. With Differential, these times can be tracked independently of each other. Correct results can be continuously produced for a given tuple $(event\ time, processing\ time)$, and computation can continue even if some times have to be kept open.

Time is typically provided as part of the \texttt{Stream} (Timely) or \texttt{Collection} (Differential) type. It is then treated as a first-class citizen to Timely and Differential dataflows: e.g., multiple unrelated times can be processed in parallel, and data belonging to old times might get compacted (\enquote{consolidated}) to save space and computing overhead if distinguishing between old versions of the data is not necessary anymore. It is also possible to model an additional time dimension as part of the \emph{data} within a \texttt{Stream} or \texttt{Collection} and manually group by that time where records of each time should be treated in isolation. Furthermore, a Timely \texttt{Stream}'s time (which describes the physical time at which a record first appeared) is mostly independent of a Differential \texttt{Collection}'s time (which describes the logical time at which a record should be considered in the computation): The only requirement is that a record's Differential time should be equal or greater than the underlying \texttt{Stream}'s Timely time. Differential might even rebatch (i.e., regroup) multiple Timely times into their minimum time to speed up the computation. While physical times are modified that way, logical consistency and correctness guarantees are preserved.

To find the time semantics best suited for ST2, we compare several time representations. In all of these, \emph{processing time} describes the time at which an event happened in the source computation. Timely logs every event with a processing time timestamp. We assume that clock skew is negligible and processing time timestamps impose a global ordering on events across workers and processes. Within each worker, processing time is strictly monotonically increasing for each event. This way, it also acts as a globally unique identifier for each event down to nanosecond precision (although the sequence number abstraction presented in \cref{subsec:impl_pag_3}, together with each worker’s globally unique identifier, arguably provide an even better unique identifier). \emph{Epoch time} describes the source computation's round of input to which an event belongs. An epoch corresponds to a version (or iteration) of the PAG, so that algorithms requiring a complete PAG can produce final results only once no new events will be seen for that epoch. \cref{subsec:impl_contract} outlines the contract between source computation and ST2 that synchronizes input rounds and epoch time. The trade-off between time semantics is summarized in \cref{tab:time}.

\begin{table}[htb]
\centering
\begin{tabular}{l|llll}
 & \begin{tabular}[c]{@{}l@{}}distinct\\ epochs\end{tabular} & \begin{tabular}[c]{@{}l@{}}ordered\\ events\end{tabular} & expressive & non-blocking \\ \hline
mono (processing) & no & yes & no & yes \\
mono (epoch) & yes & no & yes & no \\
mono (processing, epoch in data) & yes & yes & no & yes \\
mono (epoch, processing in data) & yes & yes & yes & no \\
bi (product partial order) & yes & yes & yes & yes \\
bi (lexicographical total order) & yes & yes & yes & yes
\end{tabular}
\caption{Time semantics comparison}
\label{tab:time}
\end{table}

\begin{description}
  \item [Monotemporal (processing time)] SnailTrail~1 uses monotemporal time semantics with processing time as timestamp. While this suffices in a fixed-window use case, there is no way of distinguishing between epochs without some kind of epoch marker in time or data (\enquote{distinct epochs} in \cref{tab:time}); a session-based PAG construction is impossible. Thus, this time semantics does not suffice.
  \item [Monotemporal (epoch time)] Storing only the epoch as time information allows ST2 to decide on which event belongs to which PAG version. However, there is no ordering information on events available (\enquote{ordered events} in \cref{tab:time}). This prevents ST2 from constructing the PAG and from carrying out time-based analyses in any meaningful way. It is therefore not useful for ST2.
  \item [Monotemporal (processing time, epoch time in data)] Storing epoch time in data in addition to monotemporal time semantics using processing time timestamps remediates the disadvantage of using only a monotemporal processing time timestamp without any epoch information. A session-based PAG construction is now possible. This solution complicates the dataflow logic, since it now has to group by epoch for every within-epoch computation (for ST2's current version, this is always the case). Further, if Differential is used for the PAG construction, this complicates cleaning up the old epoch state discussed in \cref{subsec:impl_windows}, as it's now impossible to provide a compaction or state discard rule to Differential without inspecting the \texttt{Collection}'s records' \texttt{data} field. Any other Differential features that rely on rich time semantics meet a similar fate: For example, implementing a consistent differential PAG is further complicated, as event-level changes create inconsistent intermediate representations of the PAG where parts of it belong to the old, parts to the new version. In summary, this time semantics is able to successfully construct an epoch-based PAG, but complicates the dataflow logic and lacks support for future (differential) features (\enquote{expressive} in \cref{tab:time}).
  \item [Monotemporal (epoch time, processing time in data)] In a similar fashion, monotemporal epoch time with processing time in data remediates the disadvantages of monotemporal epoch time without processing time in data. Its main limitation becomes obvious when running ST2 online. As only epoch time acts as an indicator to Timely and Differential that it is free to advance its frontier, all dataflow operators that have to block until time has advanced to output correct results (e.g.\ aggregates, joins, \texttt{consolidate}) will block at least parts of the overall computation until a complete round of inputs has finished (\enquote{non-blocking} in \cref{tab:time}). While this might not be an issue for small dataflows, for large source computations it may become viable from a performance perspective to be able to continually make progress during a large round of input. Depending on the analytics used (cf. \cref{sec:impl_analysis}), it might also help in keeping a lower memory footprint by discarding parts of an epoch's records once they are no longer needed. Overall, this time semantics is able to successfully construct an epoch-based PAG, and also supports Differential features such as compaction and a differential PAG construction. However, its ability to scale in a large epoch source computation setting is questionable.
  \item [Bitemporal (product partial order)] With all possible combinations of monotemporal timestamps in time and data explored, bitemporal timestamps might mitigate monotemporal disadvantages. ST2 can use a partially-ordered $(epoch\ time, processing\ time)$ pair to provide PAG version information and record creation time simultaneously. This resolves the monotemporal disadvantages. Compared to keeping epoch time in data, PAG bounds are clear and all of Timely's and Differential's features can be used. Compared to keeping processing time in data, the computation can progress even if the epoch has not closed yet, as a within-epoch happened-before relation \cite{lamport1978} exists: e.g., $(2, 10ns)\ \text{\emph{happened-before}}\ (2, 13ns)$. While ST2 could be implemented using this semantics, maintaining a partial order is unnecessary. Defining a product partial order on a bitemporal timestamp $(event\ time, processing\ time)$ is reasonable, as some events are incomparable. In contrast, every event in a $(epoch\ time, processing\ time)$ pair \emph{is} comparable. In this case, a lexicographical total order is preferable.
  \item [Bitemporal (lexicographical total order)]\footnote{One could argue that a lexicographically ordered multi-dimensional timestamp is not bitemporal, since its coordinates directly depend on each other. Here, it is still treated as bitemporal to emphasize the semantical difference between epoch time and processing time, even if they are physically dependent.} By ordering the two-dimensional timestamp by epoch first, processing time second, all bitemporal benefits are upheld, without the additional complexity of paying attention to possibly incomparable timestamps. While in other dataflows (and even in ST2, once additional timestamp dimensions such as loop iterations occur), partially ordered timestamps are invaluable, a totally-ordered bitemporal timestamp is adequate for ST2. It supports splitting up epochs into smaller chunks for performance gains, while retaining the ability to compact old records and to extend ST2 to offer cross-PAG comparisons. Treating epoch time and processing time as first-class citizens also emphasizes their importance in a consistent and correct PAG construction. We therefore use these time semantics for ST2.
\end{description}

\section{Adapter}
\label{sec:impl_adapter}

As introduced in \cref{sec:impl_architecture}, adapters are the centerpiece to connecting a source computation to ST2. An adapter's task is two-fold. It filters incoming source computation events down to relevant events for the PAG construction and creates a stream from them that contains time information. It also provides functions to connect ST2 to this stream, either through an offline serialization layer that writes (reads) the generated event stream to (from) file, or online via TCP sockets.

\subsection{Profiling Contract}
\label{subsec:impl_contract}

\newtheorem{contract}{Term}

To ensure that the downstream computation works correctly, the source computation has to fulfill the following contract's terms.

\begin{contract}
The source computation logs at least processing events, data messages, and control messages. It provides logging information to uniquely identify them.
\end{contract}

\begin{description}
 \item [Processing] occurs when a dataflow node does some work, e.g.\ when it executes a \texttt{map} operator. This is a thread-local event. For Timely and Differential, this is represented through a \texttt{ScheduleEvent}\footnote{\url{https://docs.rs/timely/0.10.0/timely/logging/struct.ScheduleEvent.html}}. Timely and Differential do not distinguish between \emph{scheduling} an operator and an operator actually \emph{processing} data. However, Timely version 0.9 introduced an event-driven execution model\footnote{\url{https://github.com/TimelyDataflow/timely-dataflow/commit/6d6896c}} that allows to schedule operators only if they have work to do, and observing the data messages that surround a \texttt{ScheduleEvent} allows to infer whether the operator really processed data. Therefore, we ingest every \texttt{ScheduleEvent} and decide in the PAG construction (cf.~\cref{sec:impl_pag}) whether it denotes empty spinning or actual data processing.
 \item [Data Messaging] occurs when a dataflow node sends data (e.g.\ records created by applying a \texttt{map} operator) to another node, i.e.\ when communication on the data plane commences. This can happen thread-local or across workers. In Timely and Differential, data messages are logged as \texttt{MessagesEvent}s\footnote{\url{https://docs.rs/timely/0.10.0/timely/logging/struct.MessagesEvent.html}}.
 \item [Control Messaging] occurs when the dataflow has to coordinate on the control plane. In Timely and Differential, progress messages (logged as \texttt{ProgressEvent}s\footnote{\url{https://docs.rs/timely/0.10.0/timely/logging/struct.ProgressEvent.html}}) are the most common control messages. They are broadcasted via channels within and across worker threads and signify that the sending worker is ready to advance the computation (e.g.\ due to an advancing frontier).
\end{description}

While additional event types could be used in the PAG construction and subsequent PAG analysis, these three suffice to detect major bottleneck causes and identify undesirable patterns in dataflow execution. However, for Timely and Differential, the source computation should also log \texttt{Operates} and \texttt{Channels} events to obtain additional information about the dataflow's structure. This is used to strip outer operator scopes before constructing the PAG (cf.~\cref{sec:impl_logrec}), and to match up broadcasted control messages (cf.~\cref{sec:impl_pag}). The ST2 adapter can discard all other events. Additionally, it removes worker-local data and control messages. While they can delay subsequent cross-worker messages and therefore indirectly cause cross-worker bottlenecks, they also prolong \texttt{Schedules} events. Therefore, these can serve as a proxy. By filtering out logged events that are non-essential to the PAG construction, around 40\% less events have to be serialized or sent over the TCP socket and processed in the downstream computation, which in turn speeds up ST2.

\begin{contract}
    The source computation computes one epoch at a time.
\end{contract}

For SnailTrail~1, fixed window semantics were used for assigning events to a PAG \autocite{hoffmann2018}. While this simplified PAG construction, extracting viable information from the PAG became challenging. A PAG was neither guaranteed to be consistent --- e.g., it could contain dangling edges ---, nor did the PAG's epoch correspond to a source computation's epoch in a meaningful way. The original PAG was using system time semantics to express computations where event time information is preferable. This resulted in a limited number of possible applications and the need to use heuristics, e.g.\ the critical participation metric, to estimate global critical path participation from local information.  In the presented version of ST2, one PAG epoch represents the activity graph of one source computation epoch. This way, the PAG is guaranteed to be consistent, and any algorithm that relies on complete dataflow execution should work out of the box. It also allows establishing similarity between epochs of PAGs, which could then be used to analyze a dataflow's development over time.

For this to work, the source computation has to run with only one epoch in flight at a time. Otherwise, log messages from multiple epochs might get mixed up in one epoch of the PAG, and one PAG epoch would no longer correspond to one source computation epoch. While this is a limitation of the current ST2 version, it might be possible to extend ST2 to support multiple epochs in flight, as long as they remain distinguishable downstream. Then, PAGs for multiple epochs could be constructed in parallel, further speeding up the overall online log analysis.

\begin{contract}
    The source computation logs the end of each epoch.
\end{contract}

The source computation's epoch bounds have to be communicated so that they can be synchronized with the event stream that is used in the PAG construction. Thus, the source computation has to output a custom log event that acts as a marker every time an epoch ends. This marker is then used by the adapter to discern epochs. In theory, this can even be used to simulate SnailTrail~1's fixed window semantics by sending out an epoch end at a fixed interval.

\begin{lstlisting}[float=t,language=Rust,caption={Source computation adapter setup},label={st2minimal}]
use timely::dataflow::InputHandle;
use timely::dataflow::operators::{Input, Inspect, Probe};

// (A) Include adapter library
use st2_timely::connect::Adapter;

fn main() {
  timely::execute_from_args(std::env::args(), |worker| {
    // (B) Create ST2 adapter instance
    //     at the beginning of the worker closure
    let adapter = Adapter::attach(worker);

    // Dataflow Structure
    let mut input = InputHandle::new();
    let probe = worker.dataflow(|scope|
      scope.input_from(&mut input)
           .inspect(|x| println!("{}", x))
           .probe()
    );

    // Dataflow Input
    for round in 0..10 {
      input.send(round);
      input.advance_to(round + 1);
      while probe.less_than(input.time()) {
        worker.step();
      }

      // (C) Communicate epoch completion
      adapter.tick_epoch();
    }
  }).unwrap();
}
\end{lstlisting}

\cref{st2minimal} shows a Timely source computation that adheres to the contract. At \texttt{(A)}, the adapter library is added to the computation. At \texttt{(B)}, it starts listening for all relevant events as defined by Term~1. The computation is written in a way that only one epoch is in flight at a time (cf.\ ll.\ 25--27). Lastly, it logs each epoch end at \texttt{(C)}.

\subsection{Event Stream Consumption}
\label{subsec:impl_online_offline}

Once the source computation's event stream has been created, it has to be consumed by ST2. Timely already provides a \texttt{Replay} operator\footnote{\url{https://docs.rs/timely/0.10.0/timely/dataflow/operators/capture/replay/trait.Replay.html}} that can be used to import a stream serialized using a \texttt{BatchLogger}\footnote{\url{https://docs.rs/timely/0.10.0/timely/logging/struct.BatchLogger.html}} into another dataflow. To comply with ST2's custom time semantics, the \texttt{BatchLogger} had to be reimplemented so that epoch ticks can be provided manually using the marker events discussed in \cref{subsec:impl_contract}. This way, the event stream's frontier can be manipulated by manually adding and retracting capabilities. For example, a custom \texttt{drop} implementation cleans up all remaining open capabilities if the source computation ends prematurely.

To be able to use ST2 in a multitude of settings, the event stream should be consumable both offline and online. \texttt{st2-timely} provides offline serialization by writing the generated stream of events to file. Alternatively, it pushes them on a TCP socket that is connected online to a running ST2 instance. In both cases, the source computation writes to one file or socket per source computation worker. As ST2 is decoupled from the upstream computation, it can then use an arbitrary number of workers to consume the event streams it is provided with.

\begin{figure}[htb]
\centering
\subfloat[Progress notifications. Progress notifications of any epoch are broadcasted to all writers.]{\includegraphics[width=0.3\textwidth]{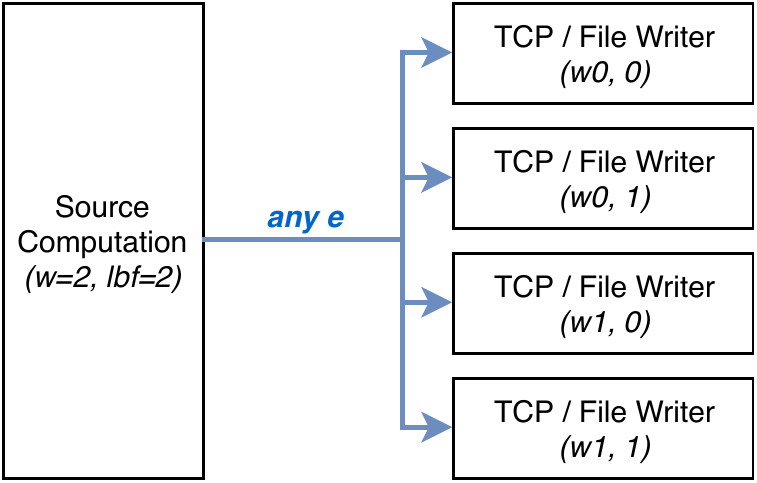}}\hfill
\subfloat[Dataflow setup. Dataflow setup (\texttt{Operates} and \texttt{Channels} events) at epoch 0 is broadcasted to all writers.]{\includegraphics[width=0.3\textwidth]{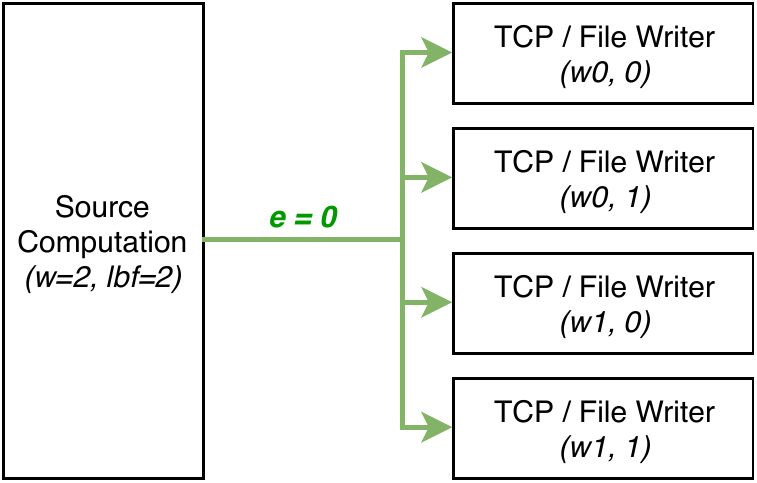}}\hfill
\subfloat[Log events. Each worker's log events are sent to the worker's respective writers in a round-robin fashion that cycles on each epoch.]{\includegraphics[width=0.3\textwidth]{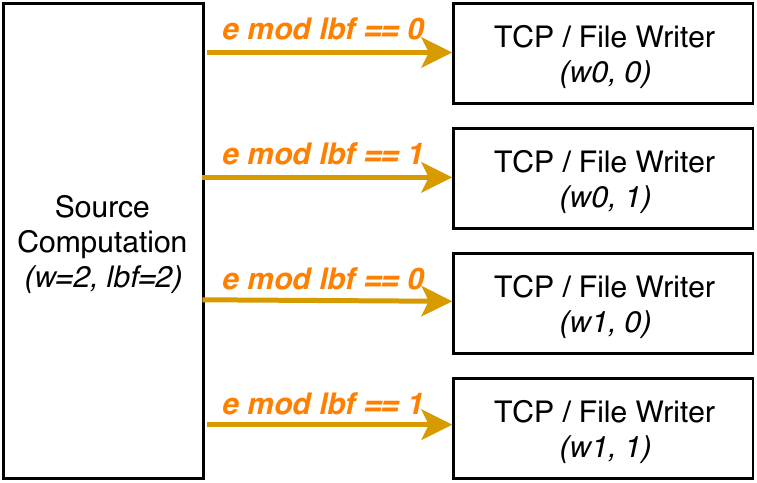}}
\caption[Load balance factor writer assignment]{Load balance factor writer assignment over epochs (denoted as \emph{e}) for various event types. Writers are spawned from a 2-worker, 2-load-balance-factor source computation: for every worker, two writers are created.}
\label{fig:lbf-impl}
\end{figure}

Similarly to database systems, I/O can also become a bottleneck in streaming computations. While a four-worker source computation generates four event streams that can be read by up to four ST2 workers in parallel, if I/O is the bottleneck, a higher degree of parallelization on ST2's end will not speed up the overall computation, as ST2 is limited by the four workers that replay the event streams. To alleviate this effect, \texttt{st2-timely}'s custom \texttt{BatchLogger} takes a \emph{load balance factor} as parameter. This factor is used to route incoming events to multiple files or sockets. \cref{fig:lbf-impl} visualizes this. Progress notifications are broadcasted to all of the respective worker thread's writers (cf.~\emph{(a)}): each ST2 worker needs this information to advance its worker-local computation. Note that these are not \texttt{ProgressEvent} log events derived from the source computation to be analyzed by ST2, but the progress information of the log event stream \emph{itself}. The dataflow setup commencing in the first epoch is also broadcasted to all writers (cf.~\emph{(b)}); each ST2 worker needs this information to execute the \texttt{peel\_ops} operator (cf.~\cref{sec:impl_logrec}). Any other log event batches are routed to each worker's writers in a round-robin fashion (cf.~\emph{(c)}). This is the majority of events that make up the stream each ST2 worker has to replay. Thus, spreading them out across ST2 workers lessens the load on individual workers. As a side effect, this forces the user to determine the maximum degree of I/O parallelism before running the source computation. However, since the source computation has to also actively opt in to profiling by ST2 before being run either way --- by implementing the contract discussed in \cref{subsec:impl_contract} ---, we believe that this is a reasonable tradeoff.

Lastly, we had to reimplement ST2's \texttt{Replayer} for offline replay, to ensure that it yields (i.e., allows other scheduled operators to run) after reading a batch from the event stream. Otherwise, in the worst case ST2's streaming computation could turn into a single batch computation, forfeiting all benefits of streaming. The custom operator also prevents ST2 from introducing too many epochs to the computation at the same time when reading multiple epochs from file, as this might slow down the computation due to additional worker communication overhead. Limiting the number of epochs in flight also has the positive side effect of making it easier to measure each epoch's performance independently for the performance evaluation described in \cref{pag-evaluation}, as no processing of any other epoch commences at the same time.

\section{\texttt{LogRecord} Construction}
\label{sec:impl_logrec}

Once the event stream has been replayed by ST2, an intermediate representation is created: the \texttt{LogRecord}. \texttt{LogRecord}s are a shared representation of a PAG event, so that distinct adapters' outputs are abstracted from the PAG construction and further downstream computations.

\begin{figure}[htb]
\centering
\includegraphics[width=0.5\textwidth]{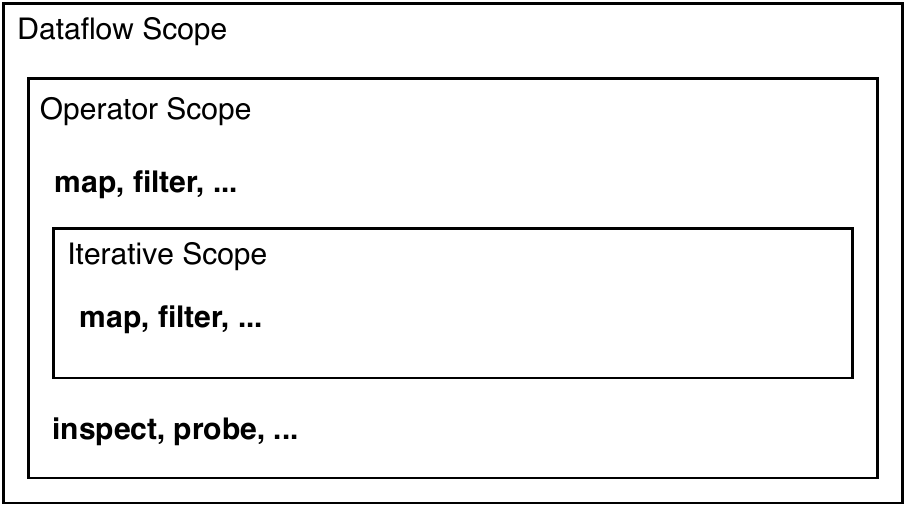}
\caption[Nested scopes in Timely Dataflow]{Nested scopes in Timely Dataflow. The outermost scope defines the overall dataflow. Additional scopes are created e.g.\ when iterative computations are executed.}
\label{fig:scopes}
\end{figure}

Before the \texttt{LogRecord} representation itself is created, we perform one last step of pre-processing. Timely uses the concept of nested \emph{scopes} to abstract dataflow details of inner timely dataflow graphs that other parts of the outer timely dataflow graph do not need to know about \autocite{mcsherry2014}. For this reason, Timely log traces might contain log messages from several levels of abstraction, which complicates the creation of a consistent PAG. To analyze the run-time behavior of Timely, only the log events of the currently inner-most \enquote{layers} of the nested log are useful. For example, if we create a PAG from the log events generated in the \enquote{iterative} scope depicted in \cref{fig:scopes}, we do not want to intermingle these events with messages from the surrounding \enquote{operator} and \enquote{dataflow} scopes. Thus, we want to filter out any events that belong to outer scopes.

\begin{lstlisting}[float=t,language=Rust,caption={Timely \texttt{peel\_ops} pseudocode},label={lst:peel}]
fn peel_ops(&self) -> Stream {
  let mut blacklist = BTreeSet::new();

  self.unary(Pipeline, move |input, output| {
    input.for_each(|epoch, event| {
      match event {
        StructureEvent(s) => {
          assert!(epoch == 0);
          blacklist.insert(s.addr.pop());
        }
        OtherEvent(o)  => {
          assert!(epoch > 0);
          if !blacklist.contains(o.addr) {
            output.give(event);
          }
        }
      }
    })
  })
}
\end{lstlisting}

To strip the outer layers away, we exploit that the nesting level of an operator is represented as a vector of address identifiers. By \texttt{pop}ping the vector, outer nesting layer identifiers can be constructed. We store these in a set and use them to filter out any events that belong to outer scopes --- basically, we antijoin the stream of log events on their operator addresses against a relation of blacklisted addresses. Since the dataflow structure that contains the operator addresses is part of the stream itself, we need to implement this antijoin either as a custom timely operator or using Differential's join primitives. We do not need to consider time in this operator, as the ST2 adapter ensures that the dataflow structure containing all events necessary for the \enquote{blacklist} is located at the stream's beginning. We use a hand-rolled Timely solution for its added performance. \cref{lst:peel} shows the pseudocode for this operator.

\section{PAG Construction}
\label{sec:impl_pag}

The PAG can be constructed from the stream of \texttt{LogRecord}s. It is represented as a stream of \texttt{PagEdge}s, each containing a source and destination node (cf.~\cref{snailtrail}). The PAG can be traversed by joining \texttt{PagEdge} source with destination nodes. To construct all PAG edges,
\begin{enumerate*}
    \item worker-local \texttt{LogRecord}s,
    \item cross-worker control edges, and
    \item cross-worker data edges
\end{enumerate*}
have to be connected.

There are multiple possible implementations to create PAG edges, which come with various performance and usability trade-offs. Here, we exemplarily discuss and compare different implementations for the worker-local edge construction; the construction of cross-worker edges commences in similar fashion.

\subsection{Option 1: Differential \texttt{Reduce}}
\label{subsec:impl_pag_2}

We first assess the local edge construction from a functional programming perspective. Using Differential's \texttt{reduce} operator \autocite{mcsherry2019a}, we can fold all local \texttt{LogEvent}s into a single PAG timeline. It allows us to write a stream transformation as if on a finite input vector, similar to Rust's \texttt{fold} method\footnote{https://doc.rust-lang.org/std/iter/trait.Iterator.html}. This makes it straightforward to implement the local edge construction logic, and, as the logic is wrapped into a differential operator, ensures that the input is sorted and additions and retractions are handled correctly for any downstream differential computations. However, as will be explained in \cref{subsec:impl_pag_3}, entering the \enquote{world} of Differential also has some disadvantages.

There is another major downside in using \texttt{reduce}: Differential treats its logic as blackbox, so it cannot incrementalize the local edge construction efficiently. Instead, for every new batch of events introduced at a new processing time (cf. \cref{subsec:impl_time}), \texttt{reduce} recomputes the whole epoch's PAG edges again. This forfeits the major benefit of constructing the PAG in a scalable manner through the use of bitemporal time semantics. Thus, the \texttt{reduce} operator is not suited for the local edge construction.

\subsection{Option 2: Differential \texttt{Join}}
\label{subsec:impl_pag_3}

Next, we evaluate a \emph{relational} approach. Similarly to databases, Differential provides various join operators \autocite{mcsherry2019a}: They (semi-/anti-)join records formatted as \texttt{(key, value)} tuples based on their key coordinate. The local edge construction can also be expressed as a join: By assigning a worker-local sequence number to each event during the event stream creation, a correct edge can be created by joining record $n$ with record $n + 1$. Note that this still requires some care with regards to the \texttt{peel\_ops} operator discussed in \cref{sec:impl_logrec}, as \texttt{peel\_ops} might leave sequence number gaps between events. Differential's join operators are also incrementalized, which solves the \texttt{reduce} operator's black box problem. Lastly, a differential implementation of the local edge join is very concise and arguably easy to understand, as \cref{lst:local} highlights.

\begin{lstlisting}[float=t,language=Rust,caption={Differential local edge construction},label={lst:local}]
fn make_local_edges(&self) -> Stream {
  let edge_start = self.map(|x| (x.seq_no, x));
  let edge_end = self.map(|x| (x.seq_no + 1, x));
  edge_start.join_map(&edge_end,
    |_key, curr, prev| build_local_edge(&prev, &curr))
}
\end{lstlisting}

However, using joins to construct the PAG is not efficient enough in our setup, as we cannot make use of many of Differential's unique and performance-boosting features. We discuss this in depth in \cref{diffpag}.

Introducing differential semantics in ST2 also complicates our downstream logic if we plan to drop down to Timely Dataflow again for two reasons. 

First, Differential is built around change-based semantics. Every \texttt{Collection} consists of triples that are either \emph{additions} or \emph{retractions} of computation input. If some data is retracted upstream, Differential is able to compute downstream effects efficiently. To maintain these benefits with a custom Timely operator, this complex behavior has to be implemented manually. Even for seemingly simple computations, a correct implementation is not trivial. As long as every custom Timely operator is situated before differential computation in the dataflow, this issue does not apply. However, as soon as we drop down to timely operators, we need to be careful to avoid correctness issues.

Secondly, Differential aims to provide a relational view on a stream of data via its \texttt{Collection} interface. Since the end user interacts with \texttt{Collection}s rather than \texttt{Stream}s, Differential does not guarantee that the underlying Timely times do not change. Differential respects the \emph{logical} times at which an event is allowed to be \enquote{seen} by the computation --- the \texttt{time} part of each triple in a \texttt{Collection}. While it might already process an event that logically happens in the future for performance gains, only events that are older than the computation's notion of \enquote{now} are considered in results that are issued by the computation. Differential does \emph{not} consider the \emph{physical} time at which an event is introduced to the computation\footnote{Obviously, Differential still considers physical time to the extent that it cannot process an event that has not yet been physically introduced to the computation.} --- the \texttt{time} part of each event in a Timely \texttt{Stream}. Instead, it might rebatch multiple distinct Timely times into a single time for performance gains. Further, some Differential operators such as \texttt{consolidate} will exchange the data they received based on a hash function. If the data is then rebatched and exchanged back, it still maintains its intra-batch order, but its inter-batch order might be off. Thus, if we want to continue writing order-dependent timely operators downstream, we either need to insert an additional \enquote{translation step} (e.g., using a \texttt{Delay} operator to synchronize Timely and Differential times), or we have to reorder across batches within the operator, which greatly hurts performance: Outputs now have to be buffered until the frontier has advanced past their time to ensure their order is correct. 

In both cases, there is a trade-off in switching between Differential and Timely mid-computation. As we might not want to implement downstream computations using Differential, and taken together with the unfavorable benchmark results discussed in \cref{diffpag}, we therefore refrain from constructing the PAG using differential joins.

\subsection{Option 3: Timely Operator}
\label{subsec:impl_pag_1}

When using only Timely and no Differential, writing a custom Timely operator is the common way of implementing custom dataflow behavior \autocite{mcsherry2019b}. Since we control the replay operator, we can ensure that all events of one source computation worker also arrive in the order they happened in at one ST2 worker. Local edges can then be created by iterating over the incoming events once, repeatedly matching the current event with the previous one using a one-element operator state (in practice, we actually need a two-element operator state, as we also need to access the previous event's predecessor to deduce waiting edges). Pseudocode for the resulting operator is shown in \cref{lst:localt}. As a downside, the custom Timely operator imposes requirements on the structure of the computation, while Differential enabled us to specify the edge construction in a declarative manner. Still, the Timely implementation does not restrict downstream analyses and is much more efficient than constructing the PAG using differential joins (cf.~\cref{pag-evaluation} for the PAG construction evaluation). We therefore use the Timely PAG implementation for ST2's current version.

\begin{lstlisting}[float=t,language=Rust,caption={Timely local edge construction},label={lst:localt}]
fn make_local_edges(&self) -> Stream {
  let mut prev = None;
  let mut prev2 = None;

  self.unary_frontier(Pipeline, move |input, output| {
    input.for_each(|time, logrecords| {
      for lr in logrecords {
        if let Some(prev) = prev {
          if let Some(prev2) = prev2 {
            output.give(build_local_edge(prev2, prev, lr));
          }

          // prev -> prev2
          prev2 = Some(prev);
        }

        // lr -> prev
        prev = Some(lr);
      }
    });
  })
}
\end{lstlisting}

%

\section{Data Analytics}
\label{sec:impl_analysis}

In this section, we discuss how ST2's analyses are implemented. For details on their functionality, cf.~\cref{func-evaluation}. We first thematize ST2's \emph{commands} (\cref{commands}) --- tools accessible via a command-line interface ---, then examine the frontend (\cref{frontend}), which connects to the \texttt{dashboard} command to visualize many of ST2's analytics.

\subsection{Commands}
\label{commands}

ST2's command-line interface is built using the Rust \emph{clap} library \cite{kevin2019}, which groups multiple user-facing tools into subcommands. All of these subcommands --- their functionality is explained in-depth in \cref{func-evaluation} ---, even those that only provide simple aggregate metrics, are implemented on top of the PAG. This is not strictly necessary --- the PAG's main advantage, compared to a simple event stream, is that it takes worker dependencies into consideration. However, thinking in graphs makes many analyses straightforward to implement --- the epochal PAG provides a semantically useful abstraction on top of a mere log. This comes at the performance cost of creating the PAG for every command, but, as discussed in \cref{pag-evaluation}, the PAG construction is efficient enough to give leeway for commands sitting on top of it. Furthermore, some commands such as the \texttt{dashboard} use multiple analyses under the hood; the PAG representation \emph{has} to be used for some of them, but can then be shared with all other commands.

Taken together with the subcommand architecture, the PAG abstraction makes it easy to extend ST2's functionality --- new subcommands (e.g.~complex graph algorithms) can be implemented on top of it as a Timely job, or even in Differential, if the performance trade-off is worth it (cf.~\cref{diffpag}). All subcommands existing in the current ST2 version are implemented as timely dataflows. We'll briefly discuss the implementation of three of them, each increasing in complexity.

\subsubsection{Invariants}

The invariants command allows to define rules that are monitored across the dataflow execution. If any rule is violated, alerts are thrown (e.g., in the command-line interface they are logged to \texttt{stdout}, in the frontend they are shown in the dashboard). One of the invariants tracks whether a control or data message takes longer than a specified amount of time. Its code is depicted in \cref{lst:maxmsg}. It highlights how simple it is to implement basic data analytics as Timely operators on top of the PAG: After filtering out the relevant edges, their duration is compared to the maximum allowed duration to filter out all violating edges. These can then be displayed to the user.  

\begin{lstlisting}[float=t,language=Rust,caption={Maximum message duration invariant operator},label={lst:maxmsg}]
fn max_message(&self, max: Duration) -> Stream<PagEdge> {
  self.filter(move |edge|
    (edge.edge_type == ControlMessage ||
     edge.edge_type == DataMessage) &&
    edge.duration() > max
  )
}
\end{lstlisting}

\subsubsection{Metrics}

The metrics command allows exporting aggregate statistics about the analyzed source computation to file. Pseudocode of it is shown in \cref{lst:metrics}. First, as the statistics should be aggregated per-epoch, all edges are forwarded to the next epoch using \texttt{delay\_batch}. Then, Timely's \texttt{aggregate} operator\footnote{https://docs.rs/timely/0.10.0/timely/dataflow/operators/aggregation/aggregate/trait.Aggregate.html} accumulates count, total duration, and total data processed (ll.~6--10) for each unique combination of edge source worker, edge destination worker, and edge type (l.~4, hashed in l.~12). Lastly, the results are formatted and produced (l.~11). They can then be written to file (not shown). While this is a slightly more complicated operator than before, it is still able to express a complex analysis in less than 15 lines of code.

\begin{lstlisting}[float=t,language=Rust,caption={Metrics operator},label={lst:metrics}]
fn metrics(&self) -> Stream {
  self
    .delay_batch(|time| Pair::new(time.epoch + 1, 0))
    .map(|e| ((e.src.w_id, e.dst.w_id, e.edge_type), e))
    .aggregate(
      |_key, edge, acc| {
        *acc = (acc.0 + 1,
                acc.1 + edge.duration(),
                acc.2 + edge.data_count);
      },
      |key, acc| (...key, ...acc),
      |key| calculate_hash(key))
}
\end{lstlisting}

\subsubsection{K-Hops}\label{impl-khops}

The k-hops command implements a graph pattern that traverses the PAG backwards for \emph{k}~hops, starting from waiting activities, to highlight potential bottleneck causes. \cref{lst:khops} shows its pseudocode. The most important operator to traverse the PAG is the \texttt{hop} operator, which is little more than a stream-stream join implemented in Timely. We want to find the activities that have resulted in the end of a waiting activity. Therefore, we join (\enquote{hop}) on each edge destination's timestamp (l.~2 maps the stream accordingly). The first hop (l.~16) consists of all non-local edges ending into a waiting activity. As data messages belonging to a waiting activity only become visible in a subsequent processing activity, ll.~6--8 first find these processing activities (l.~7) and then traverse the data edges from there (l.~8). Ll.~10--14 traverse the rest of non-local edges, starting from the waiting edges (l.~13): By definition, all remaining local edges are the waiting activities themselves, so they are filtered out (ll.~10--11) before traversal (l.~14). After the first step, no special logic has to be applied; every subsequent hop is just joining the newest step against the \texttt{edges} stream (ll.~17--18). Since this generates separate streams for each hop depth, all streams are concatenated in the end (ll.~21--23).

\begin{lstlisting}[float=t,language=Rust,caption={K-hops operator},label={lst:khops}]
fn khops(&self) -> Stream {
  let edges = self.map(|e| (e.dst.timestamp, e));

  let data_edges = edges
    .filter(|(_, e)| e.type == DataMessage);
  let step_1_datamsgs = edges
    .processing_ends()
    .hop(&data_edges);

  let nowaiting_edges = edges
    .filter(|(_, e)| e.type != Waiting);
  let step_1_nowaiting = edges
    .filter(|(_, e)| e.type == Waiting)
    .hop(&nowaiting_edges);

  let step_1 = step_1_nowaiting.concat(&step_1_datamsgs);
  let step_2 = step_1.hop(&edges);
  let step_3 = step_2.hop(&edges);
  //...

  step_1
    .concat(&step_2)
    .concat(&step_3)
  //...
}
\end{lstlisting}

Note that while \cref{lst:khops} unrolls subsequent hops (ll.~17--19, 22--24), in practice, they can also be implemented as a more general loop to support an arbitrary number of hops.

In summary, even complex graph traversals can be expressed as simple timely dataflow operators and surfaced as a user-facing command-line interface by \texttt{inspect}ing the dataflows' results and then printing them to \texttt{stdout} or writing to file.

\subsection{Frontend}
\label{frontend}

While using the command-line interface is a simple way to analyze a source computation, ST2 also provides a real-time frontend that visualizes information from multiple analyses in a single dashboard. The dashboard itself does not introduce new analyses but only aggregates them. It is partially implemented in ST2 as the \texttt{dashboard} subcommand, which serves as a backend, and as a frontend application (cf.~\cref{sec:impl_architecture}).

In the backend, the dashboard runs most analyses from other subcommands --- aggregate metrics, multiple invariants, and the k-hop graph pattern. It then uses Timely's \texttt{inspect} operators\footnote{https://docs.rs/timely/0.10.0/timely/dataflow/operators/inspect/trait.Inspect.html} to push the results on a thread-safe, asynchronous, infinitely buffered Rust \texttt{mpsc} channel\footnote{https://doc.rust-lang.org/std/sync/mpsc/}. On a separate thread, a stateful WebSocket server implemented on top of Rust's \emph{ws-rs} library \cite{housley2019} takes the results off of the channel and buffers them by epoch. This allows it to promptly respond to requests without sacrificing performance. If state grows too large, older epochs can also be discarded, as they are less relevant for the online analysis of a source computation. A frontend can then connect to the WebSocket server and request specific analysis results for a specific epoch, which are then JSON-encoded using \emph{serde-rs} \cite{tolnay2019} and sent over the socket. As an exception, invariants are not accessed by epoch: they should be treated as alerts. Therefore, they are pushed to the frontend as soon as they pop up, independently of requested epoch.  

On the frontend, a single-page application implemented with HTML and JavaScript uses React.js \cite{facebookinc.} to display a stateful, dynamic dashboard. It also makes use of D3.js \cite{d3js} to visualize the PAG, and Vega \& Vega-Lite \cite{heer2013} to create interactive visualizations of the analysis results using declarative graph specifications. Since the dashboard only requests one epoch at a time, the WebSocket connection is still able to handle the amount of data that has to be transferred from ST2 to the dashboard for most source computations. Thus, it is also able to satisfy the online requirement.

\section{Implementation Summary}
\label{sec:impl_summary}

In this chapter, we discussed the implementation details of ST2 and compared them to other possible implementation approaches.

ST2's scaling model should allow it to keep up with any online source computation while retaining a high degree of expressivity. It uses epochal session windows to subdivide the PAG into semantically sensible units. Its time semantics support this interpretation of the PAG and allow the computation to scale, especially when run online. These computational considerations influence ST2’s overall architecture and flow. By defining a clear contract between source computation and ST2, writing adapters for various stream processors becomes possible. Making use of a custom adapter for timely and differential dataflows, we achieve performance gains and control how events are ingested. For example, this lets us efficiently discard events belonging to outer scopes. It also allows us to create the PAG with quasi-stateless timely operators, instead of using comparatively inefficient differential joins. We also use custom timely operators to implement the downstream analytics that power the command line interface. The PAG and analysis results are fed one epoch at-a-time to a frontend using a buffered socket connection. There, the source computation's behavior is visualized in a real-time dashboard powered by React.js, D3.js and Vega.

Overall, this allows ST2 to support rich and simple-to-implement analytics, while still remaining scalable and open to further research in how Differential’s unique characteristics might be used in the future for an online monitoring setting.

\chapter{PAG Evaluation}\label{pag-evaluation}

In this chapter, we evaluate the PAG construction. As presented in \cref{ch:impl}, the PAG acts as the foundation of consecutive downstream analytics. We evaluate its performance to ensure that it can power analyses on top of it.

In \cref{experimental-setup}, we present the general benchmark setup used for this evaluation.

In \cref{sec:pageval_offline-experiment}, we benchmark a timely PAG implementation in an offline setting.

We compare this implementation with a partially differential PAG in \cref{diffpag}, where \texttt{LogRecord} construction and operator peeling (cf. \cref{sec:impl_logrec}) happen in Timely, but edge creation and joins are implemented in Differential (cf.\ \cref{subsec:impl_pag_3}). In addition to evaluating the PAG construction itself, this allows us to draw conclusions on the performance impact that comes with the high abstraction Differential provides compared to hand-rolled dataflow operators.

In \cref{online-experiment}, we benchmark the timely PAG implementation in an online setting to spot \enquote{production environment} peculiarities.

Lastly, we compare ST2's PAG construction performance with SnailTrail~1 \cref{st1comp}.

\section{Experimental Setup}\label{experimental-setup}

In this section, we discuss the experimental setup used to evaluate ST2: our choice of hardware and software (\cref{hardware-and-software-specifications}), the source computation and its configuration from which the log traces to analyze are generated (\cref{source-computation}), and the modes in which we run the experiments (\cref{subsec:pageval_settings}).

\subsection{Hardware and Software Specifications}\label{hardware-and-software-specifications}

All experiments ran on an Intel Xeon E5-4650 v2 2.40 GHz machine with 4~x~10 cores (4~x~20 threads) and 512GB RAM, running on Debian 8 (\enquote{Jessie}). For multi-machine experiments, all machines use identical specifications. They are connected through InfiniBand networking with a bandwidth of 13~Gbit/sec and a mean latency of \numprint{0,085}ms.

To compare with SnailTrail~1, we also ran the experiments on an Intel Xeon E5-4640 2.40 GHz machine with 4 x 8 cores (4 x 16 threads) and 512GB RAM, running on Debian 7.8 (\enquote{Wheezy}).

\subsection{Source Computation}\label{source-computation}

ST2 can analyze any timely or differential source computation. To test its capabilities, we chose an implementation of triangle counting from the  Differential Dataflow examples provided in the source code repository \cite{mcsherry2019c}. To achieve high performance, it is incrementalized and uses worst-case optimal join algorithms \cite{ngo2012}. In our experiments, we distinguish between the edges we feed to the triangles computation and the edges that are created as result of the PAG construction. We call the former \enquote{tuples} (they represent one row of input to the source computation), the latter \enquote{edges}.

\subsubsection{Configuration}\label{configuration}

\begin{figure}[htb]
\centering
\includegraphics[width=0.5\textwidth]{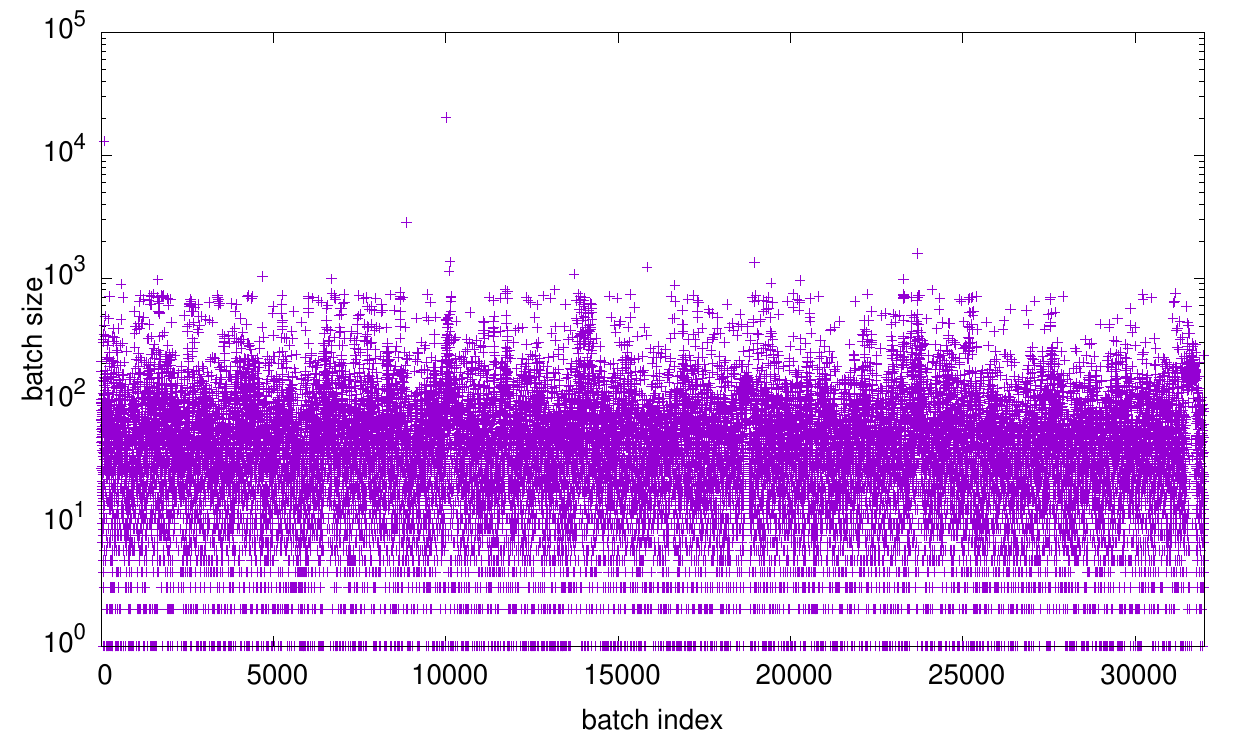}
\caption{Triangles edge batch-size distribution}
\label{fig:edge_distribution}
\end{figure}

The triangles computation is run over \numprint{1000} rounds of input (\enquote{epochs}) from a large \emph{LiveJournal} \cite{zotero-1158} social network dataset \cite{backstrom2006, leskovec2008} published in the \emph{Stanford Large Network Dataset Collection} \cite{snapnets} and converted to binary using \emph{frankmcsherry/GraphMap} \cite{mcsherry2019d}. Every round, each worker inserts a batch of new edges with varying size into a shared graph. The batch sizes are visualized for a 32-worker configuration in \cref{fig:edge_distribution}. Worker 0 inserts the edge batches at 0, 32, 64, ..., worker 1 at 1, 33, 65, ..., and so on. Notable outliers are batch \numprint{10009} and 87, which contain more than \numprint{20000} and \numprint{13000} edges, respectively, as compared to the next-largest batch \numprint{8859}, which contains less than \numprint{3000} edges. The computation is monoidic, i.e., no edges are retracted in later rounds of input. After providing input, all triangles in the graph are counted. Differential incrementalizes all operators; the graph does not need to be recomputed from scratch after each round.

While the source computation is closely aligned to real-world use cases such as social network analysis \cite{scott1988, wasserman1994} to provide a realistic test setting, it is also a highly performant differential computation (cf.~\cite[Section~5]{mcsherry2013}), which makes it a challenging computation for ST2 to keep up with in an online setting. In addition, we artificially blow up the number of events per epoch the source computation generates and which ST2 has to analyze in two ways. First, we increase the source computation's number of workers. More workers induce a greater coordination and communication overhead between the workers, e.g.~due to progress tracking: each progress message is also logged as an event which has to be considered in the PAG construction. Secondly, we insert edges multiple times. This leads to more data messages between operators and workers, and, consecutively, more log events for ST2 to analyze. A caveat of blowing up log traces like this is that it causes meaningless work for the source computation and thereby trades off some of its performance for more events to analyze.

\begin{table}[]
\centering
\begin{tabular}{@{}rrrrr@{}}
\toprule
workers& batch factor & \begin{tabular}[c]{@{}r@{}}events per epoch\\ (rounded)\end{tabular} & \begin{tabular}[c]{@{}r@{}}time per epoch\\ (milliseconds)\end{tabular} & \begin{tabular}[c]{@{}r@{}}PAG-relevant events \\ (generated gigabytes)\end{tabular} \\ \midrule
16 & 1 & \numprint{40000} & 9 & 3.8 \\
16 & 5 & \numprint{45000} & 12 & 4.4 \\
16 & 10 & \numprint{50000} & 14 & 4.9 \\
16 & 50 & \numprint{75000} & 34 & 8.1 \\
16 & 100 & \numprint{100000} & 60 & 12 \\
16 & 200 & \numprint{175000} & 114 & 19 \\
16 & 300 & \numprint{210000} & 171 & 25 \\
16 & 400 & \numprint{290000} & 231 & 32 \\
16 & 500 & \numprint{320000} & 293 & 37 \\
32 & 1 & \numprint{140000} & 22 & 14 \\
32 & 5 & \numprint{150000} & 25 & 18 \\
32 & 10 & \numprint{175000} & 31 & 19 \\
32 & 50 & \numprint{250000} & 67 & 30 \\
32 & 100 & \numprint{350000} & 108 & 41 \\
32 & 200 & \numprint{500000} & 194 & 61 \\
32 & 300 & \numprint{700000} & 286 & 83 \\
32 & 400 & \numprint{900000} & 384 & 103 \\
32 & 500 & \numprint{1000000} & 490 & 122 \\ \bottomrule
\end{tabular}
\caption[Triangles configuration impact]{Triangles configuration impact. Impact of source computation worker count and batch factor on generated log events. One epoch corresponds to one round of differential inputs, where each round of input consists of $edges(round) \times batch\ factor$ edges.}
\label{tab:config_impact}
\end{table}

\cref{tab:config_impact} summarizes the effect various worker configurations have on the source computation and its log traces. Increasing the number of workers has the largest effect on log size, while increasing the batch factor --- how often every edge is inserted --- has a large effect on time and source computation throughput. Due to space restrictions imposed by the in-memory partition used to store serialized traces (cf. \cref{subsec:pageval_settings}), the maximum number of events per epoch we can generate is 1 million, which takes 122 GB to store in-memory after filtering out all non-relevant events (cf.~\cref{sec:impl_adapter}).

To scale out the benchmark further, we artificially combine multiple epochs into one. For example, by combining 10 epochs into one, we can generate a log trace that contains 10 million events per epoch using 32 workers and a batch factor of 500. As a downside, this sacrifices sample size; we now measure ST2's performance over 100 instead of \numprint{1000} epochs.

Using the results from \cref{tab:config_impact}, we run all experiments with 32 source computation workers. We run with batch factors of 5, 50, 100, 200, and 500, which translate into around \numprint{150000}, \numprint{250000}, \numprint{350000}, \numprint{500000}, and \numprint{1000000} log events per epoch to analyze, respectively. For some experiments (cf. \cref{sec:pageval_offline-experiment}), we also merge epochs to measure performance with up to \numprint{100000000} events.

The source computation's source code can be found in the \texttt{triangles} example, the ST2 inspector in the \texttt{inspect} command, both available in the ST2 source code repository \cite{sandstede2019}.

\subsubsection{Effect on PAG Construction}

\begin{figure}[htb]
\centering
\subfloat[tuple throughput]{\includegraphics[width=0.5\textwidth]{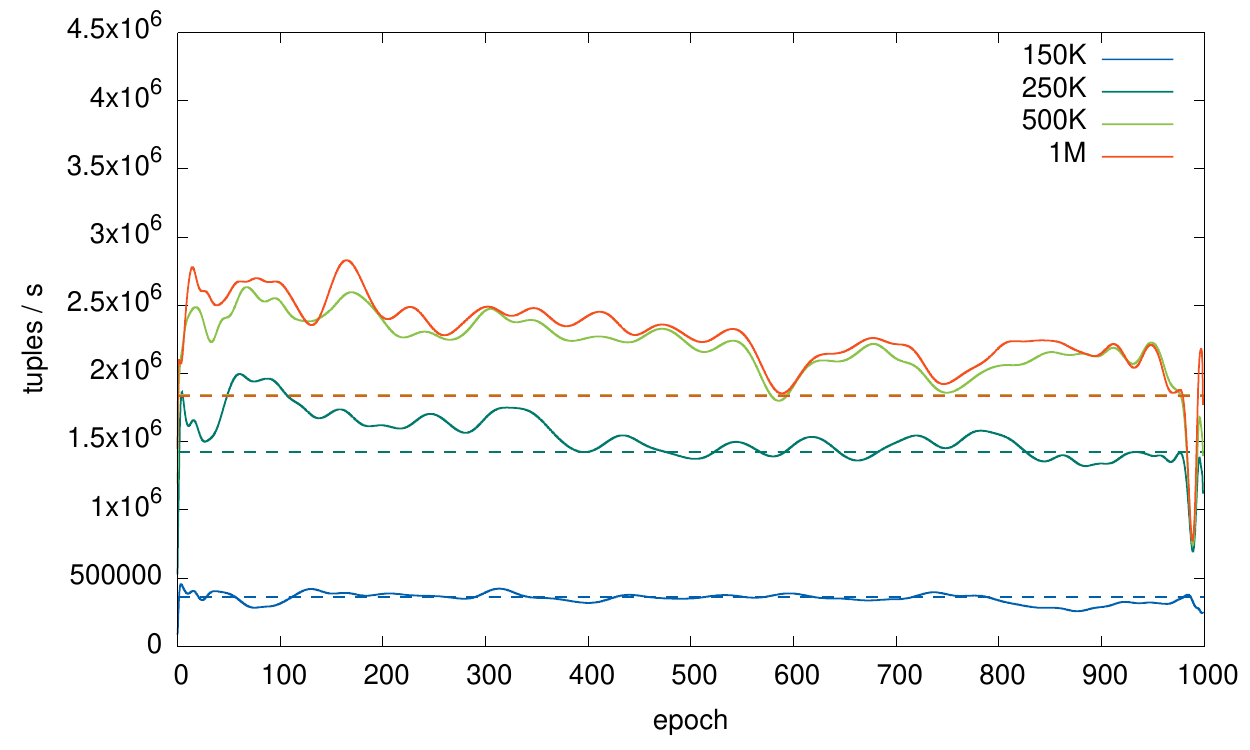}} 
\subfloat[event throughput]{\includegraphics[width=0.5\textwidth]{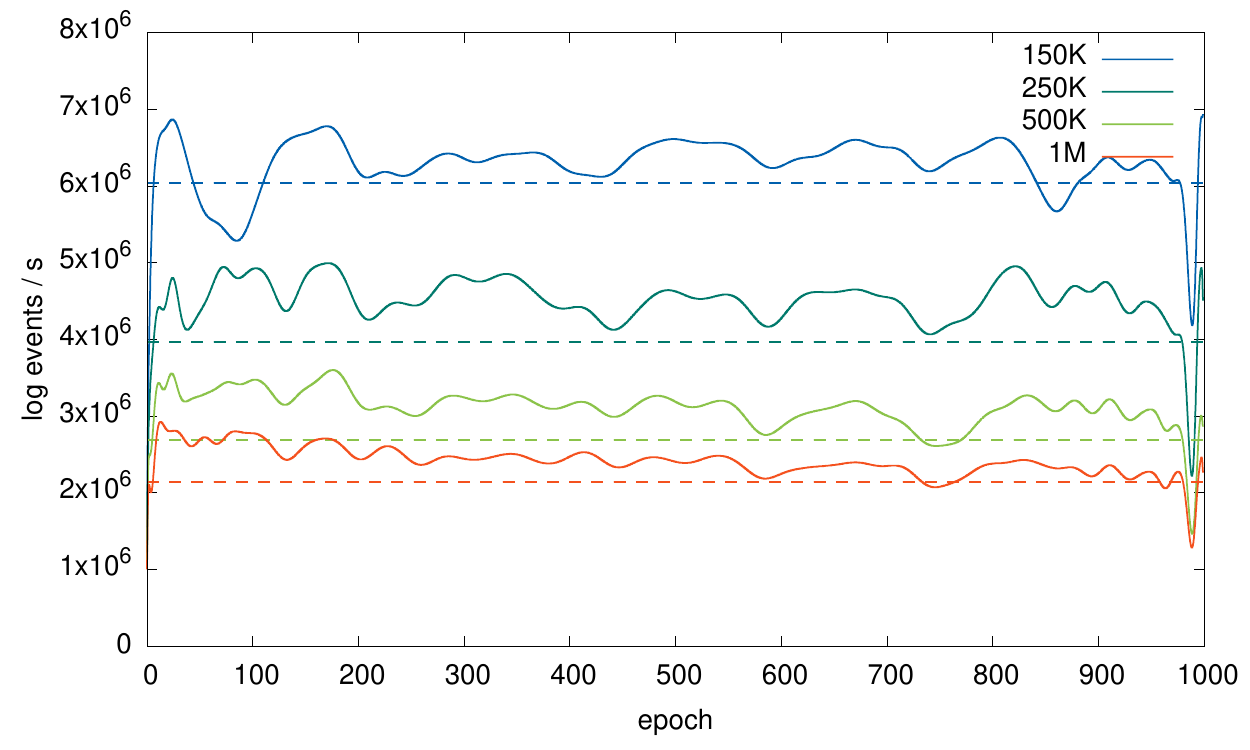}}
\caption[Triangles throughput]{Triangles throughput per epoch for various events per epoch settings. Dashed lines indicate mean across all epochs.}
\label{fig:tc_throughput}
\end{figure}

\begin{figure}[htb]
\centering
\includegraphics[width=0.5\textwidth]{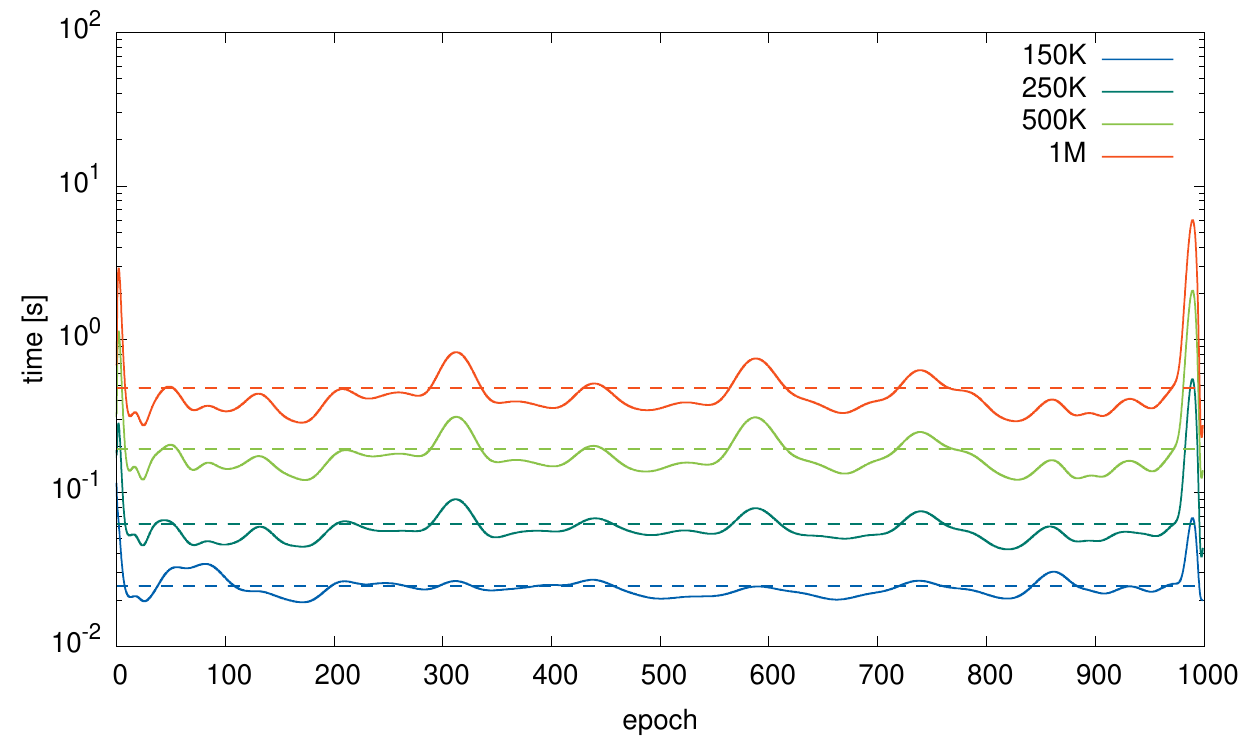}
\caption[Triangles latency]{Triangles latency per epoch for various events per epoch settings. Dashed lines indicate mean across all epochs.}
\label{fig:tc_latency}
\end{figure}

\cref{fig:tc_throughput} depicts the source computation's tuple and event throughput\footnote{Event throughput is the throughput at which log events to be analyzed by ST2 are generated.} during each epoch for batch factors up to 500, and \cref{fig:tc_latency} shows the triangle computation's per-epoch latency. As discussed, the triangles computation should represent a real-world use case: Even at the smallest batch factor, it is able to process an average of around \numprint{400000} tuples per second on a single machine with 32 workers, which should suffice for many real-world applications. With higher batch factors, tuple throughput increases: Each redundant edge insertion causes less-than-proportional overhead due to Differential's incremental computation model. For log events, the relationship between throughput and batch factor is inverted. With more tuples to process each epoch, epoch latency increases; furthermore, additional edge insertions generate a diminishing number of log events (e.g.\ in the form of data messages).

\begin{figure}[p]
\centering
\includegraphics[width=.9\textwidth]{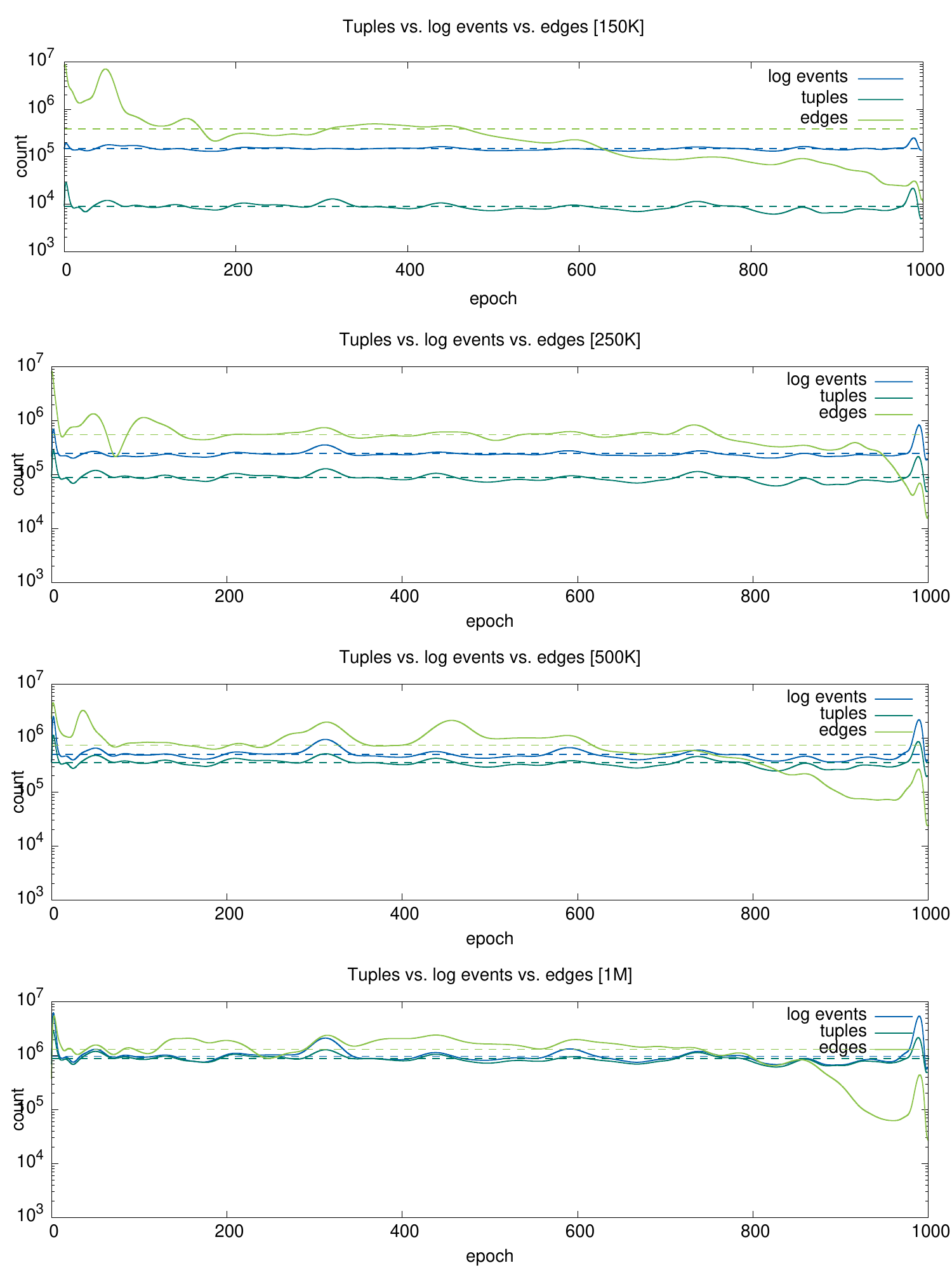}
\caption[Tuples vs.\ log events vs.\ edges]{Tuples vs.\ log events vs.\ edges. Epoch size provided in brackets. Comparison between tuples processed by the source computation, generated log events to be analyzed by the PAG construction, and resulting PAG edges for each epoch. Dashed lines indicate mean across all epochs.}
\label{fig:tuples_vs_events_vs_edges}
\end{figure}

The relationship between tuples processed by the source computation, generated log events to be analyzed by the PAG construction, and resulting PAG edges for each epoch is also shown in \cref{fig:tuples_vs_events_vs_edges}, combining information from \cref{tab:config_impact}, \cref{fig:tc_throughput}, and \cref{fig:tc_latency}. With a batch factor of 5, the triangles computation has to process around \numprint{9000} tuples per epoch. This results in approximately \numprint{150000} log events per epoch. All PAG-relevant log events (cf. \cref{subsec:impl_contract}) are forwarded to the PAG construction, which creates close to \numprint{40000} edges each epoch. Regardless of epoch size, the average PAG edge count stays higher than the mean log events count. In addition to creating local edges for consecutive log events, additional edges are inserted for cross-worker activities. With higher epoch sizes, the delta between tuples and log events dwindles: As discussed above, additional edge insertions generate a smaller number of log events in higher batch factor configurations.

It is noteworthy that in many cases, ST2 has to process more records than the source computation did. Again, this makes it challenging for ST2 to complete its analyses faster than the source computation, in particular since both programs run on top of the same underlying stream processing framework.

\subsection{Modes}
\label{subsec:pageval_settings}

We run the experiments in two modes: offline and online.

The offline environment lets us measure ST2's maximum performance. In this setting, the source computation writes generated events to file. Running offline allows us to measure ST2's performance independent of network influences and source computation limitations. To avoid measuring I/O performance, which might quickly become the computation's bottleneck when reading from file, we store the intermediate files on an in-memory partition.

The online environment allows us to simulate a real-world use case for ST2. The source computation runs on one physical machine and is connected to another physical machine running an instance of ST2 via a TCP socket. Generated log events are written to and read from this socket instead of to and from file. While this does not allow us to measure ST2's peek throughput --- which is limited by the source computation's speed ---, it might surface ``production environment'' issues, e.g.~network limitations.

\section{Offline Experiment}
\label{sec:pageval_offline-experiment}

In the offline experiment, we want to benchmark ST2's maximum performance under load. We first evaluate how ST2's timely PAG construction's mean latency and throughput scales across various epoch sizes. We then analyze per-epoch throughput and latency distributions for each epoch size using cumulative distribution functions (CDFs). 

\subsection{Setup}

We measure ST2 when it is not limited by the source computation, I/O, or network. We run the triangles source computation in multiple configurations over \numprint{1000} epochs (cf. \cref{source-computation}), then use ST2 to analyze the generated log traces with 1, 2, 4, 8, 16, and 32 workers.

In many benchmarks, it is interesting to find a tipping point at which the benchmarked computation \enquote{breaks down}. To arrive at such a tipping point for ST2, we would need to consider the profiled source computation as well. As soon as ST2's latency exceeds the source computation's latency, it will not be able to run indefinitely in an online setting on this source computation. As each source computation behaves differently from a performance standpoint, and the relationship between the source computation and generated PAG-relevant log events is not straightforward, we mostly focus on the overall scaling behavior, result stability, and absolute performance of ST2 in the following benchmarks, using the triangles computation as a \enquote{typical} non-trivial streaming task for comparison. Where appropriate, we extend this comparison to other common streaming jobs.

\subsection{Scaling Results}\label{scaling}

\begin{figure}[htb]
\centering
\subfloat[latency]{\includegraphics[width=0.5\textwidth]{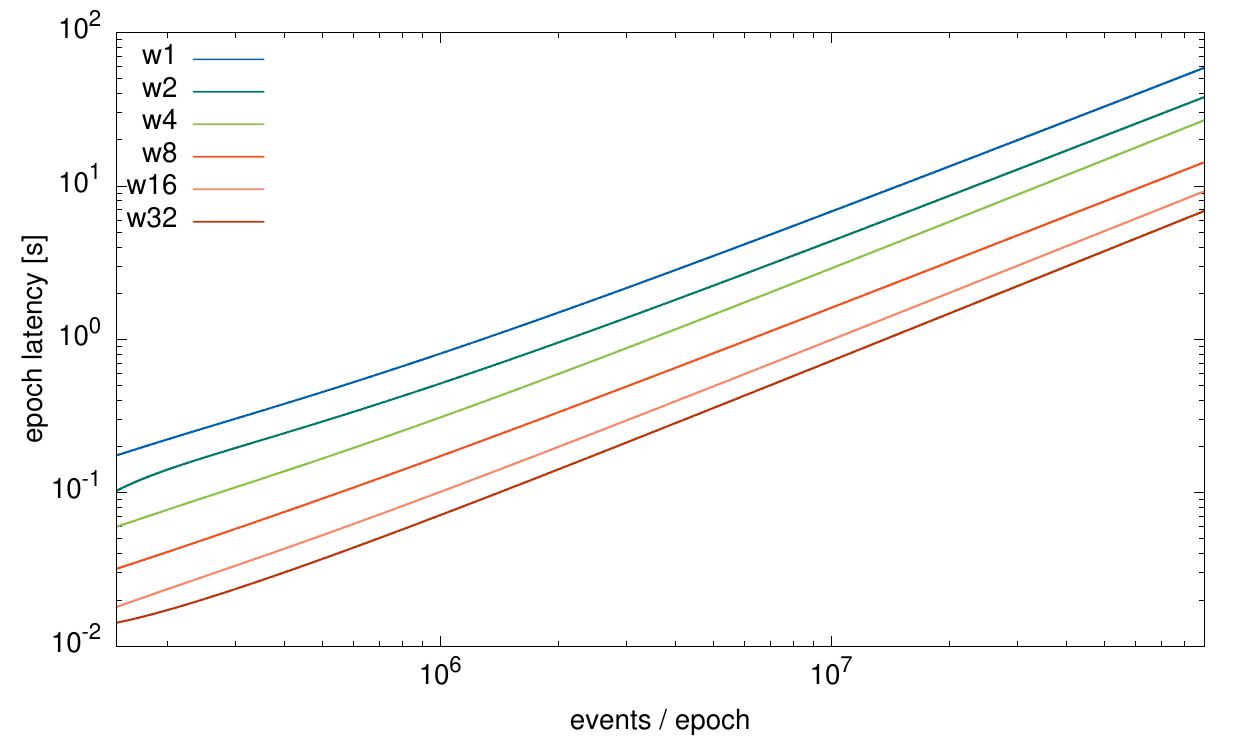}} 
\subfloat[throughput]{\includegraphics[width=0.5\textwidth]{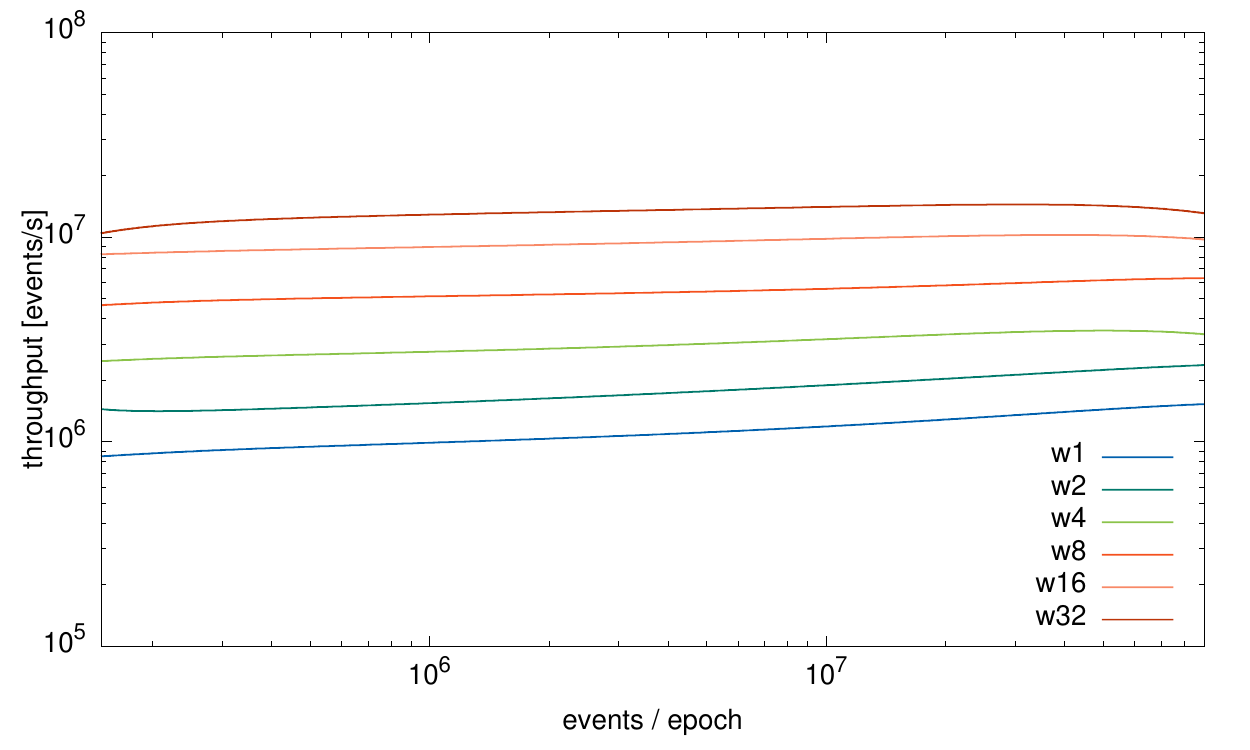}}
\caption{Timely PAG scaling}
\label{fig:st_scaling}
\end{figure}

ST2's mean epoch latency and throughput over varying epoch sizes for each worker configuration are shown in \cref{fig:st_scaling}. As would be expected, higher epoch density leads to longer time per epoch, as more events have to be processed each epoch. Increasing the worker count reduces latency --- the PAG construction parallelizes well. It also boosts throughput. Throughput remains stable across all epoch densities. We were not able to determine a cut-off point where latency spikes or throughput drops. One reason for this is the in-memory partition limitation mentioned in \cref{configuration}. The other is that we deem epoch sizes exceeding \numprint{1000000} unlikely, let alone the close to \numprint{100000000} we tested with. For all common use cases, ST2 scales out latency- and throughput-wise.

\begin{figure}[htb]
\centering
\includegraphics[width=0.6\textwidth]{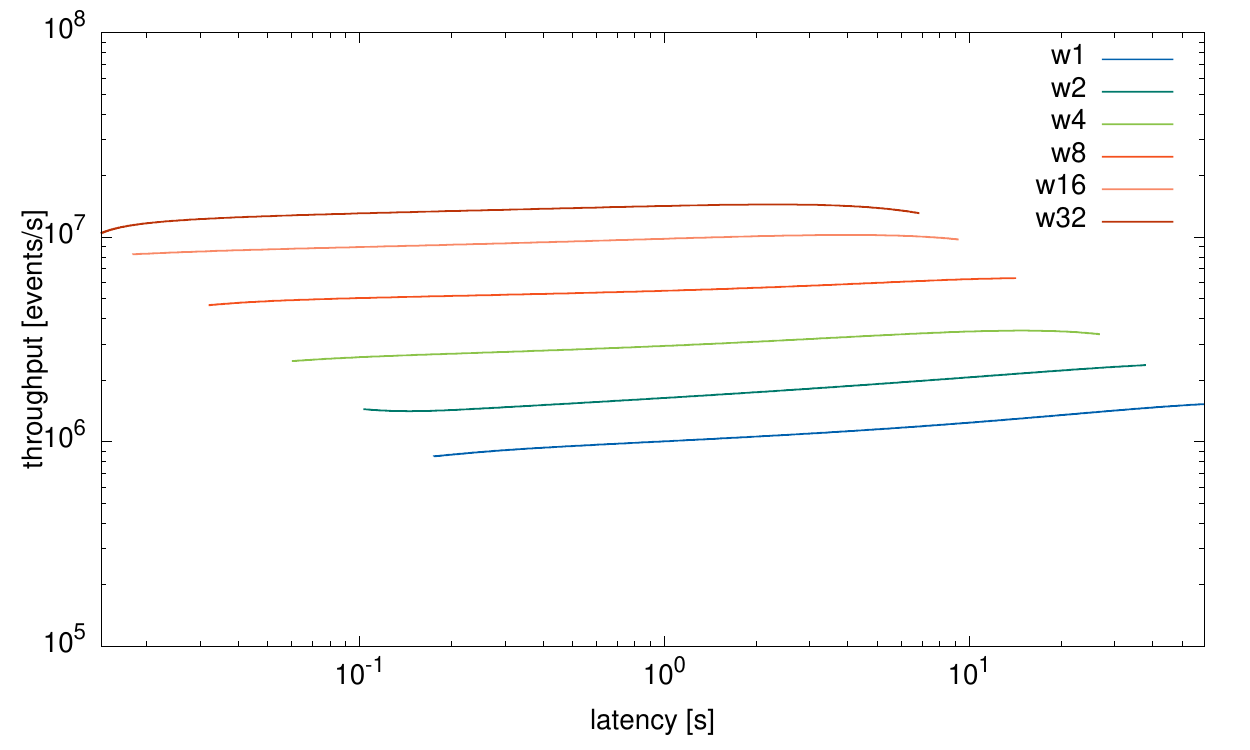}
\caption{Timely PAG throughput vs.\ latency}
\label{fig:st_latvsthrough}
\end{figure}

Compared with other latency and throughput benchmarks, it might come as a surprise that ST2's throughput is not affected by higher latencies --- \cref{fig:st_latvsthrough} illustrates their relationship. Throughput is mostly unaffected by latency due to the way ST2's PAG construction works: It consists of largely stateless, non-blocking operators. There are no benefits to be had from e.g.\ storing and arranging data (actions that would potentially increase latency) and then processing it as batch. The maximum throughput mostly depends on hardware configuration, and, to some extent, data skew. Latency, at least for the observed epoch sizes, thus becomes a function of throughput and events per epoch. For an alternative visualization of this, refer to \cref{fig:st_latvsthroughcloud} in the Appendix, which plots latency against throughput for each epoch and epoch size. 

In absolute terms, ST2 running on 32 workers is able to process more than \numprint{10000000} events per second on average, and even when running with only one worker, \numprint{1000000} events can be handled each second. This translates to latencies where the PAG for a \numprint{1000000}-event epoch can be constructed on average in \numprint{0.07} seconds with 32 workers. For comparison, as discussed in \citeauthor{hoffmann2018} \cite{hoffmann2018}, for ST2 to keep up with e.g.\ Spark traces from the big data \cite{zotero-1171} and TPC-DS benchmarks \cite{zotero-1169} generated with the tool \cite{zotero-1164} used by \citeauthor{ousterhout2015making} to analyze performance in data analytics frameworks \cite{ousterhout2015making}, a throughput of only around \numprint{12000} events per second would have been necessary. Another real-world comparison is LinkedIn \cite{zotero-1174}, who are using \enquote{the largest deployment of Apache Kafka in production at any company} \cite{narkhede2015}. According to \citeauthor{narkhede2015}, LinkedIn processes more than \numprint{1.1} trillion messages per day (around \numprint{13000000} messages per second). For this, LinkedIn \enquote{runs over 1100 Kafka brokers organized into more than 60 clusters} \cite{zotero-1178}. Thus, an average throughput of \numprint{10000000} events per second at an epoch latency of \numprint{0.7} seconds on a single 32-worker machine that could be scaled out further should suffice for most use cases.

\subsection{CDF Results}\label{cdfs}

\begin{figure}[p]
\centering
\includegraphics[width=.9\textwidth]{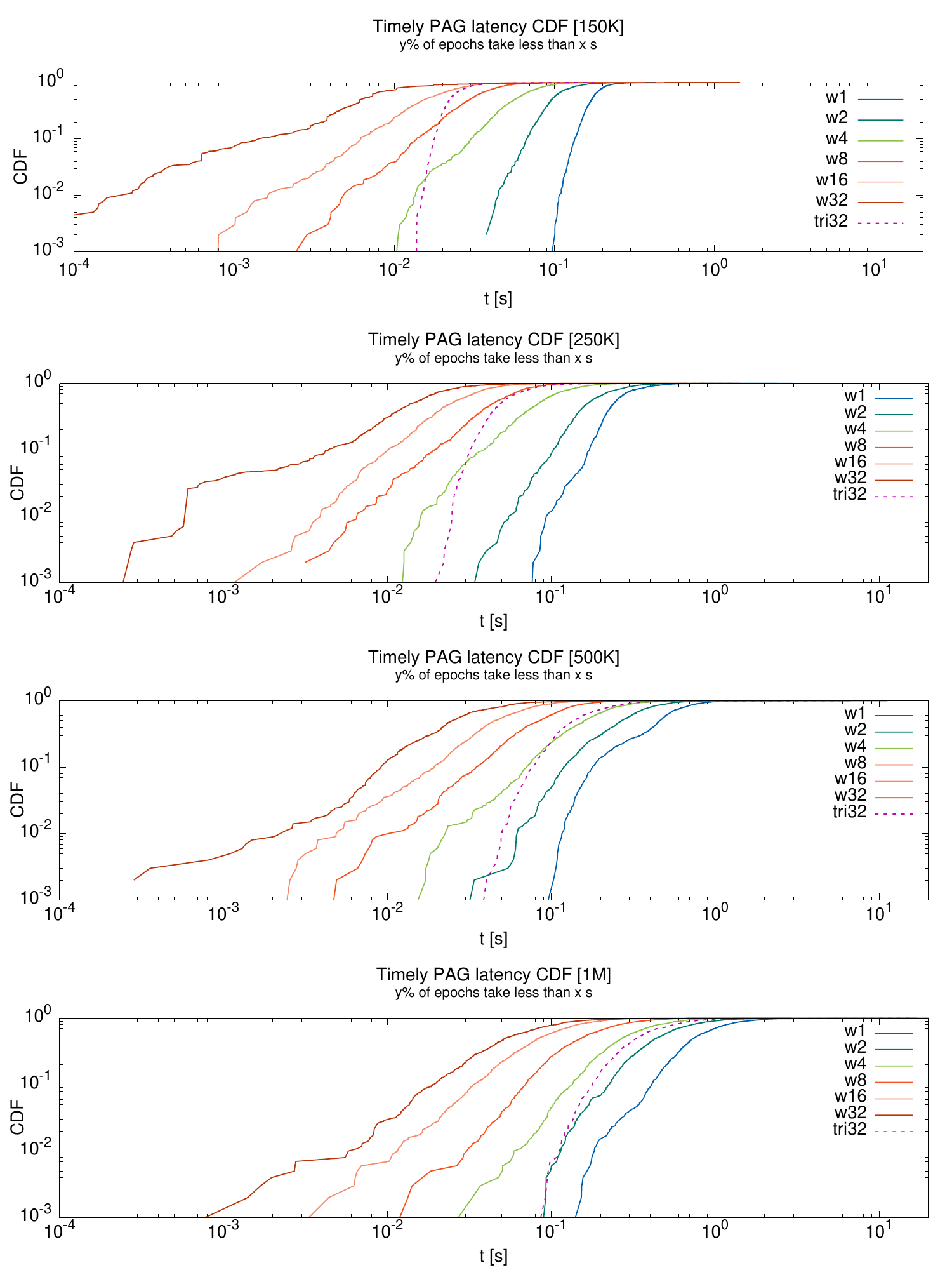}
\caption[Timely PAG latency CDFs]{Timely PAG latency CDFs for multiple epoch sizes and worker configurations. Epoch size provided in brackets. Dashed line denotes the triangles source computation. For a given point, its y coordinate denotes the percentage of epochs that take less time than the corresponding x coordinate in seconds.}
\label{fig:latency_cdf}
\end{figure}

\cref{fig:latency_cdf} shows the timely PAG construction's epoch latency for each epoch size and worker configuration as CDF. An alternative visualization is available in the Appendix (\cref{fig:latency}). Again, the CDFs highlight that increasing ST2's worker count decreases latency significantly: On a one-worker configuration with \numprint{1000000} log events to process each epoch, 90\% of epochs take \numprint{0.4} to two seconds to process. In comparison, on a 32-worker configuration, 90\% of epochs take only \numprint{0.02} to \numprint{0.2} seconds to process. Across all measured epoch sizes, ST2 running with 32 workers is the most \enquote{unstable} configuration: about 10\% of epochs run significantly faster than the rest. For all epoch densities, a 16-worker configuration is enough to keep up with the 32-worker source computation.

Note that this comparison is slightly lacking for two reasons. First, every source computation might behave differently, such that the triangles computation cannot serve as a general benchmark (although it is competitive with many streaming jobs from a performance perspective). Secondly, higher epoch densities are achieved by forcing the source computation to do empty work, artificially slowing it down --- this is also the reason why four ST2 workers are enough to keep up with the source computation in the \numprint{1000000} event size setting. Thus, the source computation should only serve as a rough indication and not as a conclusive comparison.

\begin{figure}[p]
\centering
\includegraphics[width=.9\textwidth]{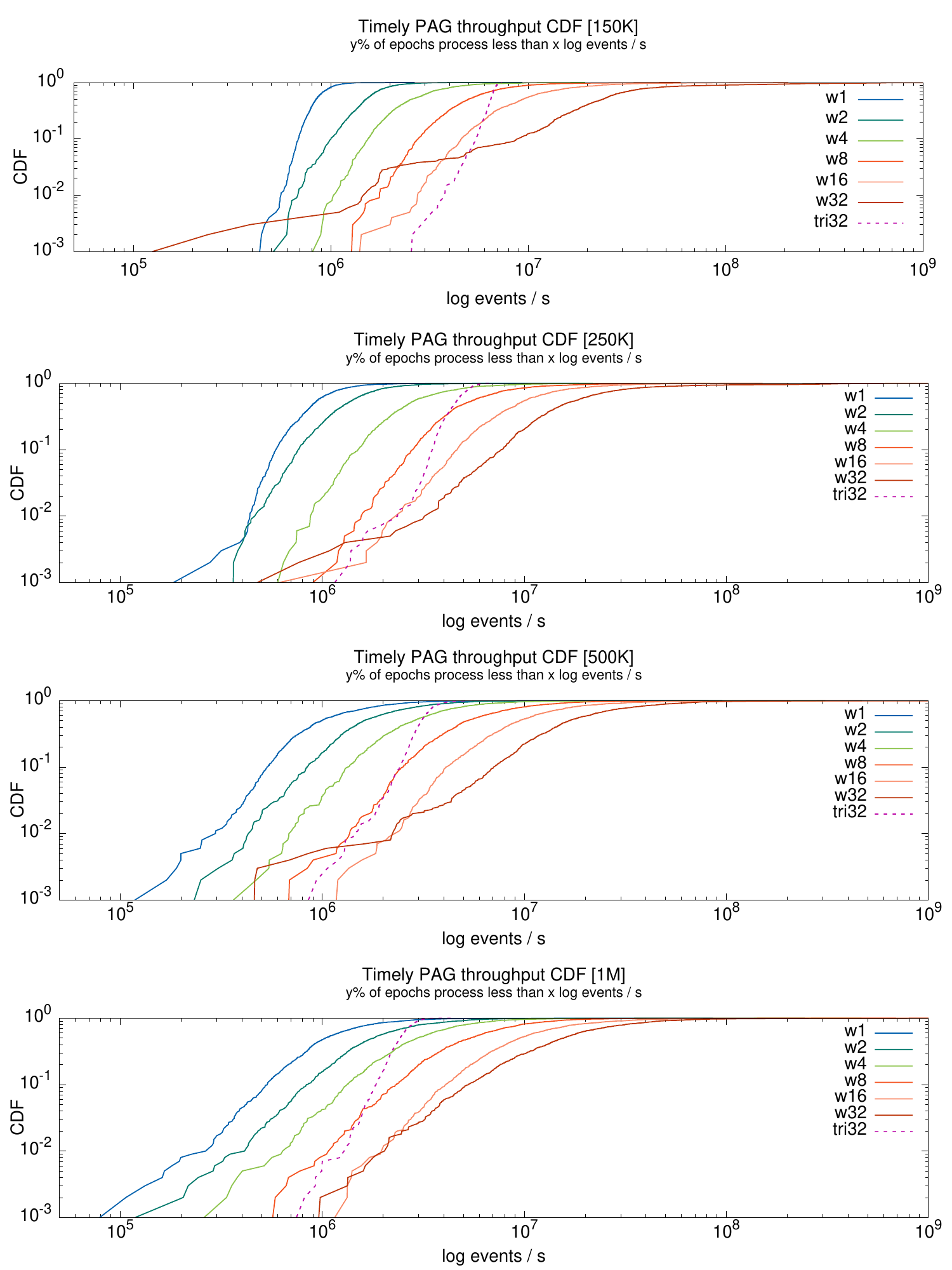}
\caption[Timely PAG throughput CDFs]{Timely PAG throughput CDFs for multiple epoch sizes and worker configurations. Epoch size provided in brackets. Dashed line denotes the triangles source computation. For a given point, its y coordinate denotes the percentage of epochs that process less events per second than the corresponding x coordinate.}
\label{fig:throughput_cdf}
\end{figure}

\cref{fig:throughput_cdf} shows the timely PAG construction's epoch throughput for each epoch size and worker configuration as CDF. An alternative visualization is available in the Appendix (\cref{fig:throughput}). Similarly to the latency CDFs, higher worker count has a positive effect on throughput: On a one-worker configuration with an epoch size of \numprint{1000000} log events, 90\% of epochs are processed with at least \numprint{500000}, and 70\% with more than \numprint{800000} event per second. A 32-worker configuration is able to process 90\% of epochs with at least \numprint{5000000}, and 70\% with more than \numprint{10000000} events per second. Again, the 32-worker configuration is the most unstable configuration, with some epochs achieving a significantly smaller throughput, especially on smaller epoch sizes. Similarly to the latency CDFs, 16 workers should suffice to keep up with the 32-worker source computation's log event throughput --- especially since not all of the source computation's generated log events have to be considered for the PAG construction (cf.~\cref{subsec:impl_contract}). Of course, the same caveats as for the latency comparison apply. 

Overall, a small number of epochs' performance degrades for both latency and throughput CDFs, especially at high worker counts and low epoch densities. Still, as discussed in \cref{scaling}, the overall latency and throughput results are well within margin to keep up with most streaming computations online.

\subsection{Increasing the Load Balance Factor}\label{increasing-the-load-balance-factor}

We previously benchmarked ST2's performance by running it with at most the same number of workers as the source computation. For particularly taxing workloads, ST2 is also able to scale beyond the source computation's worker count by modifying the \emph{load balancing factor} (cf.~\cref{subsec:impl_online_offline}). This allows it to run with more workers than the source computation.

\begin{figure}[htb]
\centering
\subfloat[Throughput]{\includegraphics[width=0.5\textwidth]{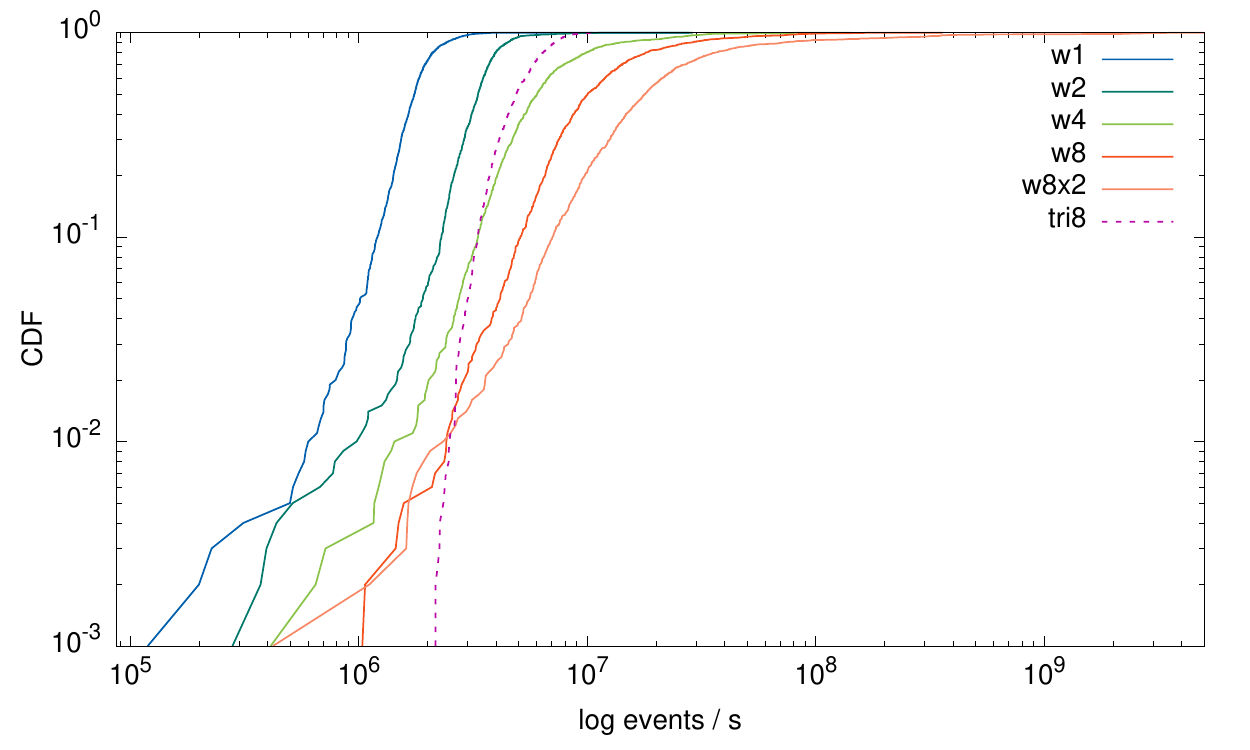}} 
\subfloat[Latency]{\includegraphics[width=0.5\textwidth]{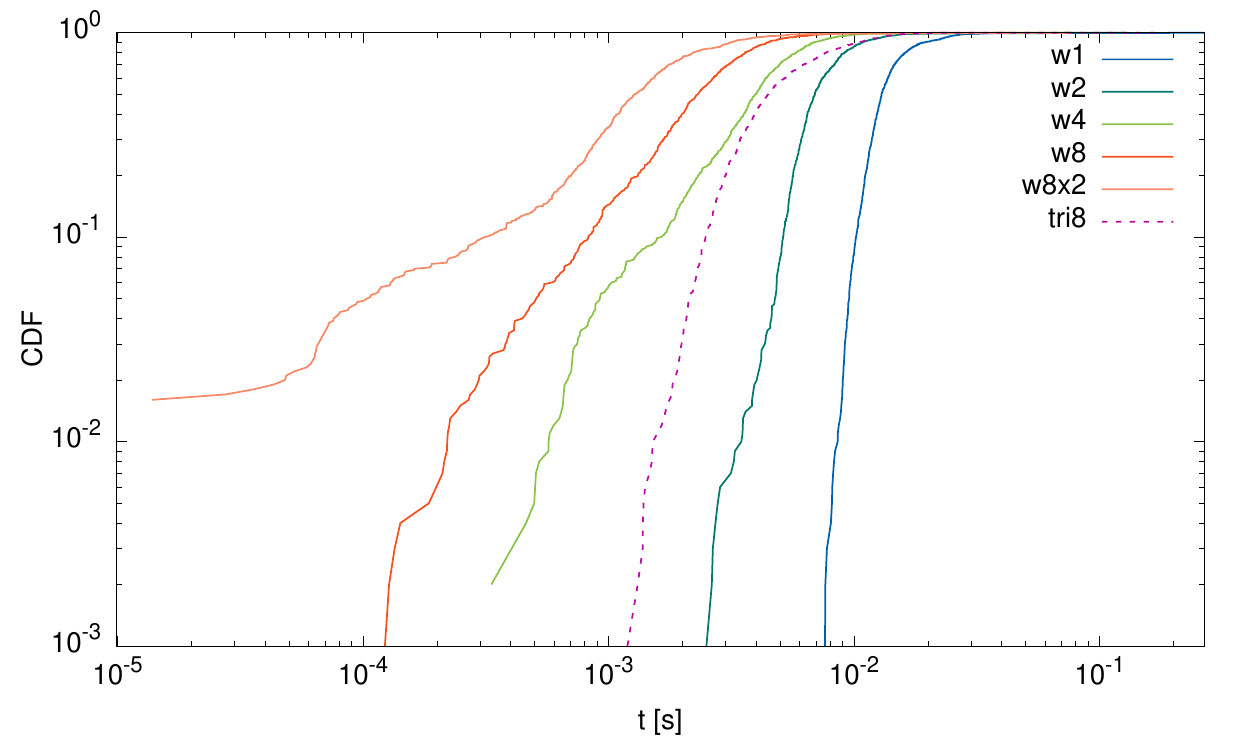}}
\caption[Timely PAG load balance factor CDFs]{Timely PAG load balance factor CDFs. Epoch size: \numprint{250000}. Source computation ran with 8 workers (\enquote{tri8}). It is profiled with up to 16 ST2 workers (\enquote{w8x2}).}
\label{fig:lbf}
\end{figure}

We evaluate the load balance factor by re-running the offline experiment with an 8-worker source computation profiled with up to 16 ST2 workers. Exemplary throughput and latency CDFs for an epoch size of \numprint{250000} are depicted in \cref{fig:lbf}. As shown in the figure, running ST2 this way is significantly faster than profiling the same source computation configuration with only 8 ST2 workers.

The load balance factor has to be set programmatically on the source computation's side before running it. Therefore, it is a little less flexible to use. Considering the previous benchmarks' positive results, in most cases, running ST2 with at most a one-to-one worker match should suffice. However, for demanding source computations, the load balance factor provides additional performance gains.

\subsection{Summary}

In the offline experiment, we benchmarked ST2's maximum performance under load. After evaluating its scaling behavior and its latency and throughput CDFs, we conclude that ST2's timely PAG construction is efficient enough to keep up with all common streaming computations, and still leaves room for the implementation of further analyses on top of it. For particularly challenging workloads, it can also be run with a higher worker count than the source computation to achieve even higher throughputs.

\section{Differential PAG Comparison}\label{diffpag}

Differential encourages a \enquote{relational} mindset when thinking about data transformations that is similar to interacting with traditional databases: E.g., joins and other relational algebra operators are preferred to writing custom streaming operators that rely on the data's underlying order. Such a high-abstraction language for expressing operations on streams is a powerful tool.

However, using Differential for the PAG implementation comes at a performance cost, as we are unable to make use of many of its advanced features. We cannot (yet) update a single PAG over time to reflect advancing epochs, which might allow us to incrementalize its construction --- instead, we have to create a fresh self-contained PAG for each epoch. We also cannot use Differential's support for partially ordered timestamps to parallelize and speed up the computation, as epochs are handed to ST2 in order and one at a time. Lastly, we would have to sacrifice hand-optimized dataflow operators, which exploit the log event order and custom replay behavior to increase performance (e.g., they avoid exchanging data between workers for the local edge construction), in favor of generalized differential operators. With these limitations, the differential PAG implementation is not able to attain similar latency, throughput, or cross-worker scaling characteristics as a Timely-based PAG.

To gain insights on the extent of this performance cost, we now compare the previously benchmarked timely PAG implementation with a PAG implemented partially on top of Differential --- while \texttt{LogRecord} construction and operator peeling still happen in Timely, the PAG edge creation is implemented using Differential \texttt{join} operators. This will help us  to decide when to drop down to custom Timely operators, trading off expressivity against performance; it allows us to \emph{quantify} the performance impact of Differential's high-level relational operators.

\subsection{Setup}

We use the same setup for the differential PAG comparison as for the offline experiment, such that a comparison is straightforward: We run the source computation in multiple configurations over \numprint{1000} epochs, then use ST2 to analyze the generated log traces with 1, 2, 4, 8, 16, and 32 workers.

\subsection{Scaling Results}\label{diff_scaling}

\begin{figure}[htb]
\centering
\subfloat[Differential PAG latency scaling]{\includegraphics[width=0.5\textwidth]{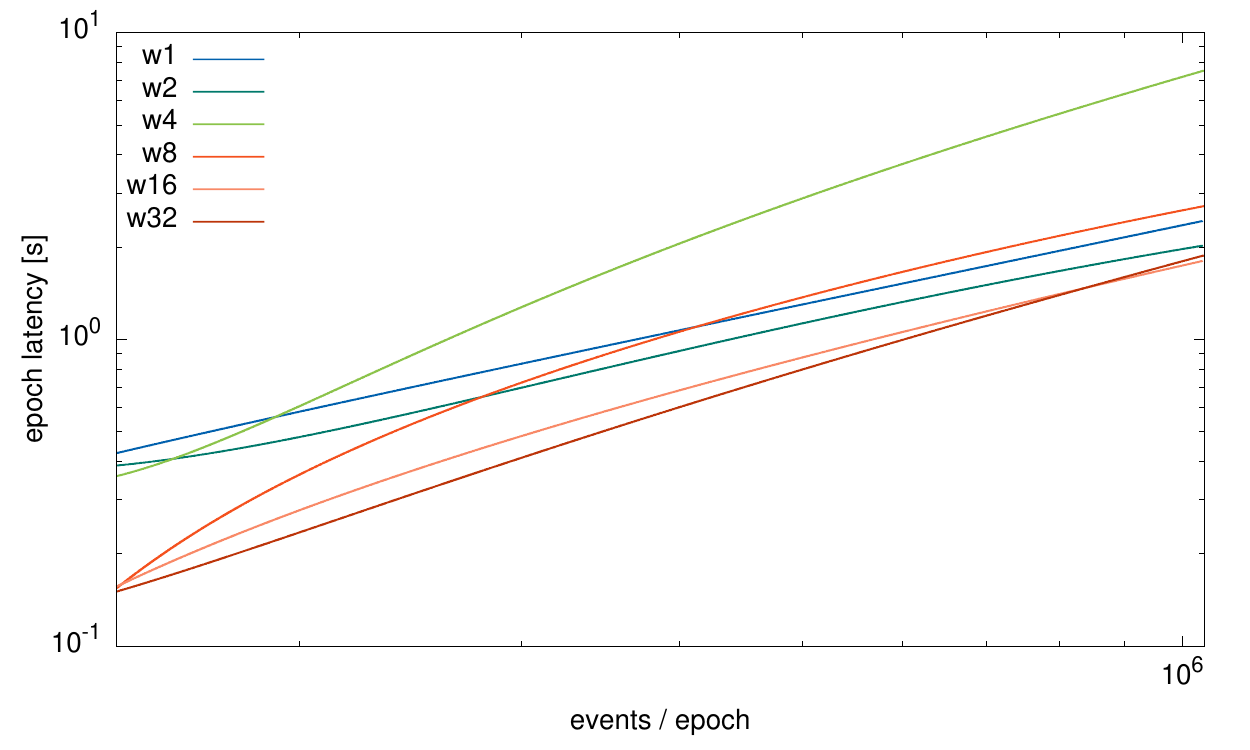}} 
\subfloat[Timely PAG latency scaling]{\includegraphics[width=0.5\textwidth]{img/st_scaling1.pdf}}

\subfloat[Differential PAG throughput scaling]{\includegraphics[width=0.5\textwidth]{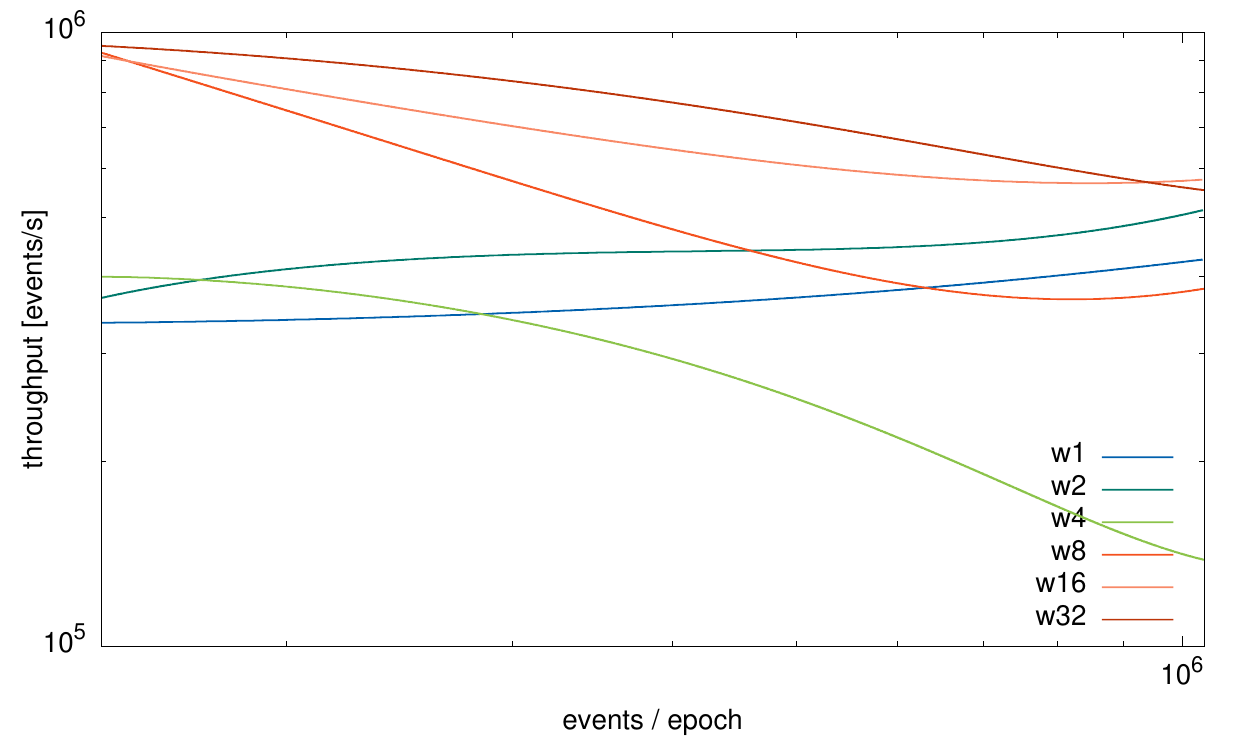}} 
\subfloat[Timely PAG throughput scaling]{\includegraphics[width=0.5\textwidth]{img/st_scaling2.pdf}}
\caption{Differential vs.\ timely PAG scaling}
\label{fig:diff_st_scaling}
\end{figure}

\cref{fig:diff_st_scaling} compares the differential and timely PAG's latency and throughput scaling characteristics. For a visualization of the combined scaling behavior, refer to \cref{fig:diff_st_latvsthroughcloud} in the Appendix, which plots latency against throughput for each epoch and epoch size. Compared to Timely, we capped the epoch size at \numprint{1000000} events per epoch for Differential. Like the timely PAG implementation, the differential PAG's mean latency increases with epoch size. Generally, increasing the worker count decreases the mean latency. However, this effect is smaller than on the timely PAG, and deteriorates with increasing epoch sizes for some worker configurations (e.g.\ four and eight workers). Unlike the timely PAG implementation, differential PAG throughput does not remain stable across epoch densities; 4, 8, 16, and 32 worker configurations' throughput all decrease (similarly to the latency results, the 4-worker configuration takes the largest hit), while 1 and 2-worker configurations slightly increase. 

It is not completely clear why the 4-worker configuration degrades. The differential PAG implementation uses joins, such that most of the log events have to be exchanged between workers and re-indexed multiple times until they have been processed completely. It is possible that these data exchanges are particularly unfavorable for the 4-worker configuration. They might also explain why throughput is somewhat stable for one and two workers: The probability that an event has to be exchanged is smaller with fewer workers, and exchanges might be expensive enough that they outshine the computing power of additional workers. With higher epoch densities, the differential joins grow more complex and involve much larger relations. This might cause the decline in performance at higher batch sizes. To alleviate the issue of frequent re-indexing, Differential provides \emph{arrangements} as a primitive to share indexed state \cite{mcsherry2018a}, but ST2 cannot make much use of them, as the index keys change for each join and previously arranged data will not match again in joins of later epochs. There is no simple way of circumventing frequent data exchanges in joins, apart from devising custom exchange pacts that can exploit existing data locality --- however, this might somewhat damage Differential's value proposition of writing dataflows without having to care for the underlying data representation.   

In absolute terms, the differential PAG processes log events with a throughput about an order of magnitude smaller, and a latency about an order of magnitude higher than the timely implementation. Since the performance difference is so large, we sanity-checked it with a custom dataflow. In this dataflow, we compare the performance of a simple \texttt{map} operator with a differential \texttt{join} operator --- as discussed in \cref{sec:impl_pag}, the timely PAG mostly consists of custom-built \texttt{map} operators, while the differential PAG employs differential \texttt{join}s under the hood. Of course, the test's results can only serve as a rough estimate. Still, it suffices for basic sanity checking. Its code can be found in the \texttt{join\_vs\_map} example available in the ST2 repository \cite{sandstede2019}. The test arrived at similar results as the PAG benchmark comparison: The \texttt{map} dataflow is around eight times faster than the \texttt{join} dataflow. Thus, an order-of-magnitude speed-up of the timely PAG implementation over the differential PAG implementation seems reasonable.

\subsection{CDF Results}

\begin{figure}[p]
\centering
\includegraphics[width=.9\textwidth]{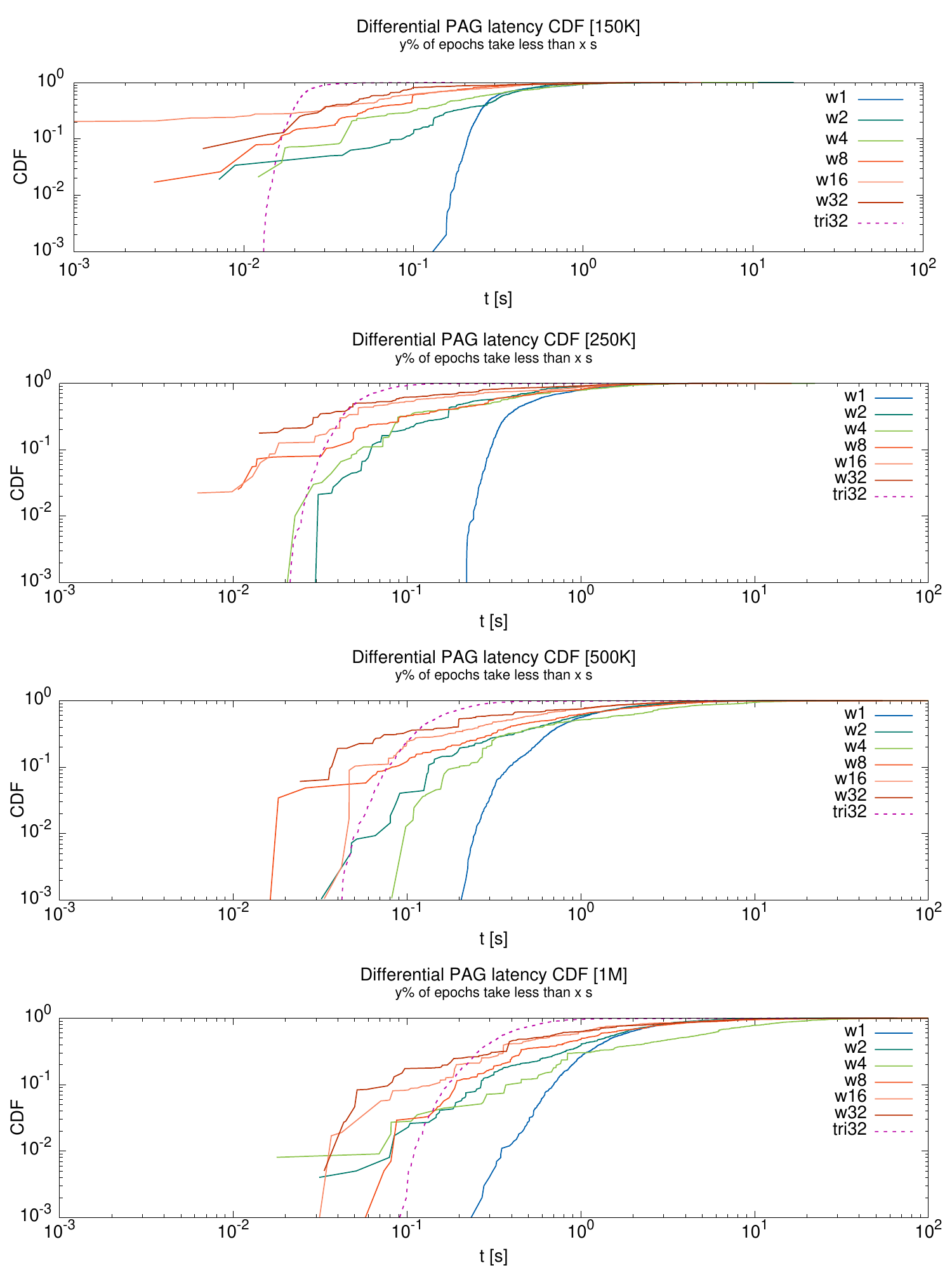}
\caption[Differential PAG latency CDFs]{Differential PAG latency CDFs for multiple epoch sizes and worker configurations. Epoch size provided in brackets. Dashed line denotes the triangles source computation. For a given point, its y coordinate denotes the percentage of epochs that take less time than the corresponding x coordinate in seconds.}
\label{fig:diff_latency_cdf}
\end{figure}

\begin{figure}[p]
\centering
\includegraphics[width=.9\textwidth]{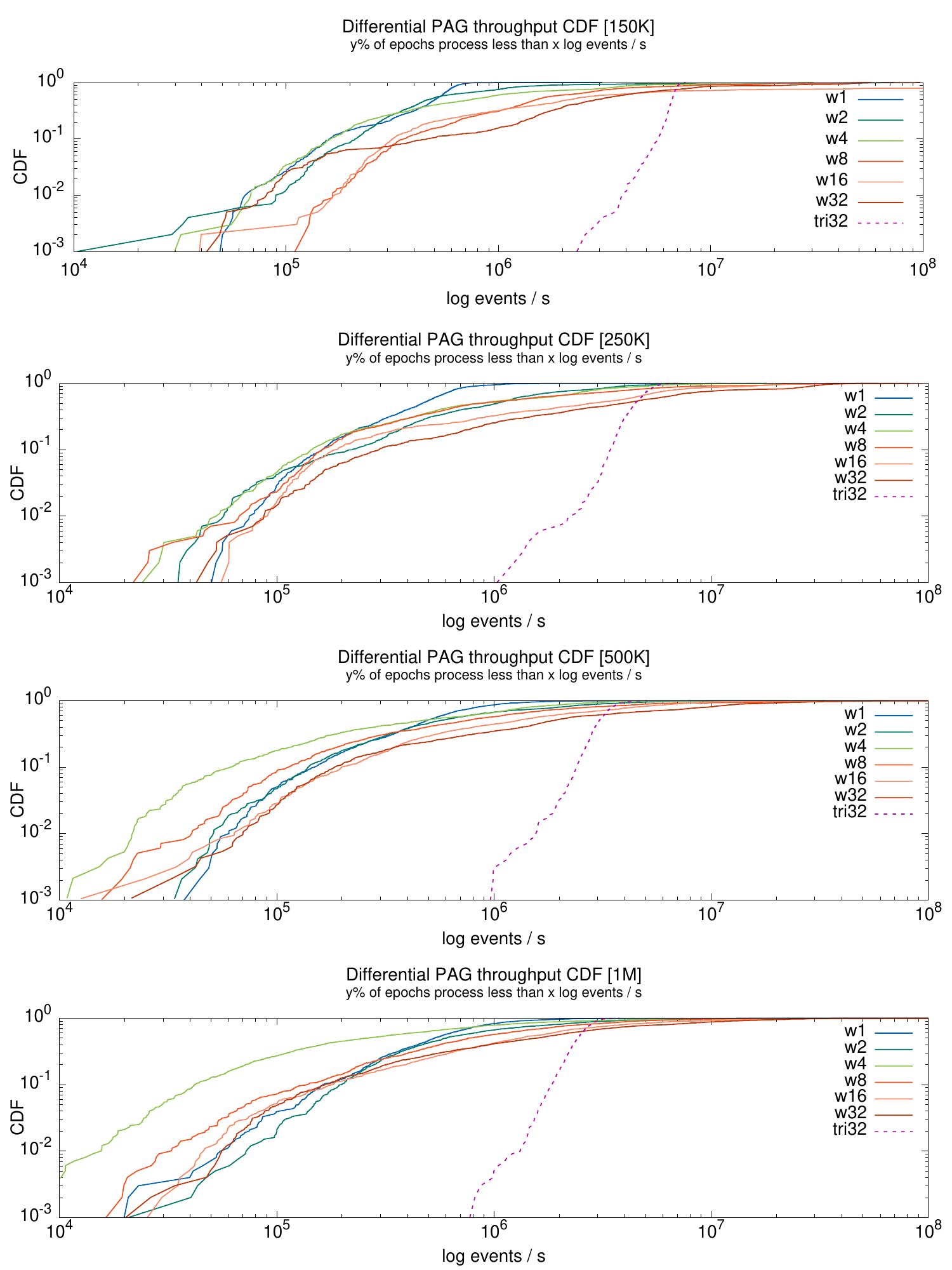}
\caption[Differential PAG throughput CDFs]{Differential PAG throughput CDFs for multiple epoch sizes and worker configurations. Epoch size provided in brackets. Dashed line denotes the triangles source computation. For a given point, its y coordinate denotes the percentage of epochs that process less events per second than the corresponding x coordinate.}
\label{fig:diff_throughput_cdf}
\end{figure}

\cref{fig:diff_latency_cdf} and \cref{fig:diff_throughput_cdf} show the differential PAG construction's latency and throughput CDFs for multiple epoch sizes and worker configurations. Alternative visualizations are available in the Appendix (\cref{fig:diff_latency} and \cref{fig:diff_throughput}). As discussed in \cref{diff_scaling}, latencies for the differential PAG are about an order of magnitude worse than the timely PAG latencies. Compared to the source computation's latency (dashed), this becomes particularly clear: No ST2 worker configuration will be able to keep up with the triangles computation when running online. Similarly, the differential PAG's throughput is about an order of magnitude slower than the timely PAG implementation. It also highlights the differential PAG's worse worker scaling characteristics: Especially the CDFs' lower probability \enquote{tails} mix without a clear pattern. Again, the comparison caveats from \cref{cdfs} apply --- for many computations (especially those running on other stream processors), a differential PAG construction should be able to keep up in an online setting.

For both plots, perhaps the most prominent features are their long tails, which highlight an uneven epoch performance distribution. Latency and throughput across all epoch densities and worker configurations (with the exception of the 1-worker latency CDFs) are affected. For example, 90\% of \numprint{1000000}-event epochs for a 32-worker configuration take at least \numprint{0.08} seconds, while 50\% already take more than half a second --- more than a six-fold increase, compared to a \numprint{2.5}-fold increase for the same experiment running on a timely PAG. At least for latency, the 1-worker outlier hints at these issues being rooted in the additional data exchanges required at higher worker numbers that were previously discussed in \cref{diff_scaling}.

The scaling and CDFs plots highlight that the differential PAG construction cannot compete with a PAG written purely in Timely. While this is true in case of ST2, it would be premature to generalize these results. It is obvious that, when comparing a hand-rolled implementation that avoids exchanging data wherever possible, with a join-heavy, generalized implementation, the former will fare better. Thus, order-of-magnitude performance losses sound worse than they are. This is especially true  when considering that this brings the differential PAG construction down to SnailTrail~1's performance levels, which still comfortably exceeded the requirements of any tested source computation; the triangles computation we used in our benchmarks can be seen as a \enquote{worst-case}. Profiling an optimized Differential Dataflow computation with another Differential Dataflow computation is bound to be challenging performance-wise. While the differential PAG implementation at hand did not achieve the scaling characteristics we would have liked to see, in many cases, more favorable exchange pacts can be employed, extremely large-scale joins avoided, and data pre-arranged for more efficient reuse. In particular, the computation at hand did not benefit from Differential's central feature --- incremental differential computation ---, and we also could not take advantage of its rich time semantics to parallelize our workload further.

\subsection{Summary}

In the differential PAG comparison, we compared a PAG implementation written partly using Differential Dataflow operators to the Timely PAG implementation we benchmarked in the offline experiment. As expected, the differential implementation was not able to match the timely implementation's scalability, latency, or throughput --- it performed about an order of magnitude worse. We attribute the extent of this performance loss to our use of Differential in a setting where we only benefit from its declarative nature, and cannot make use of its performance-related features. Taken together, we believe that for the problem at hand, dropping down to a timely dataflow is worth trading off some elegance of Differential's relational operators --- especially since our problem, at least in its current phrasing, is not well-suited to a differential formulation.   

\section{Online Experiment}\label{online-experiment}

In the previous sections, we evaluated ST2's performance in isolation, ignoring external influences such as limitations of the source computation, I/O, or network. However, in practice, ST2 runs in an online setting, where both the source computation and the network stack have a direct influence on its performance. The source computation's performance is the limit of ST2's performance: Events can only be ingested as fast as they are generated on the source computation's end. The network stack might become a bottleneck if too many log events have to be transferred. Thus, in this section, we evaluate ST2 in an online, \enquote{production environment}-setting.

Furthermore, we want to make sure that the source computation behaves identically in offline and online mode --- otherwise, profiling results might get distorted. We evaluate its behavior in \cref{online-sc}.

\subsection{Setup}

In the online experiment, we run the source computation on one machine and connect it to another machine running ST2 with the timely PAG implementation. Similarly to the offline experiment, we run the triangles source computation over \numprint{1000} epochs with varying epoch densities (cf.~\cref{source-computation}). We analyze the generated log traces with ST2 running on 16 and 32 workers, as according to the offline experiments, these worker counts should be able to keep up with the source computation.

\subsection{Source Computation Results}\label{online-sc}

\begin{figure}[htb]
\centering
\subfloat[latency (16w-profiled)]{\includegraphics[width=0.5\textwidth]{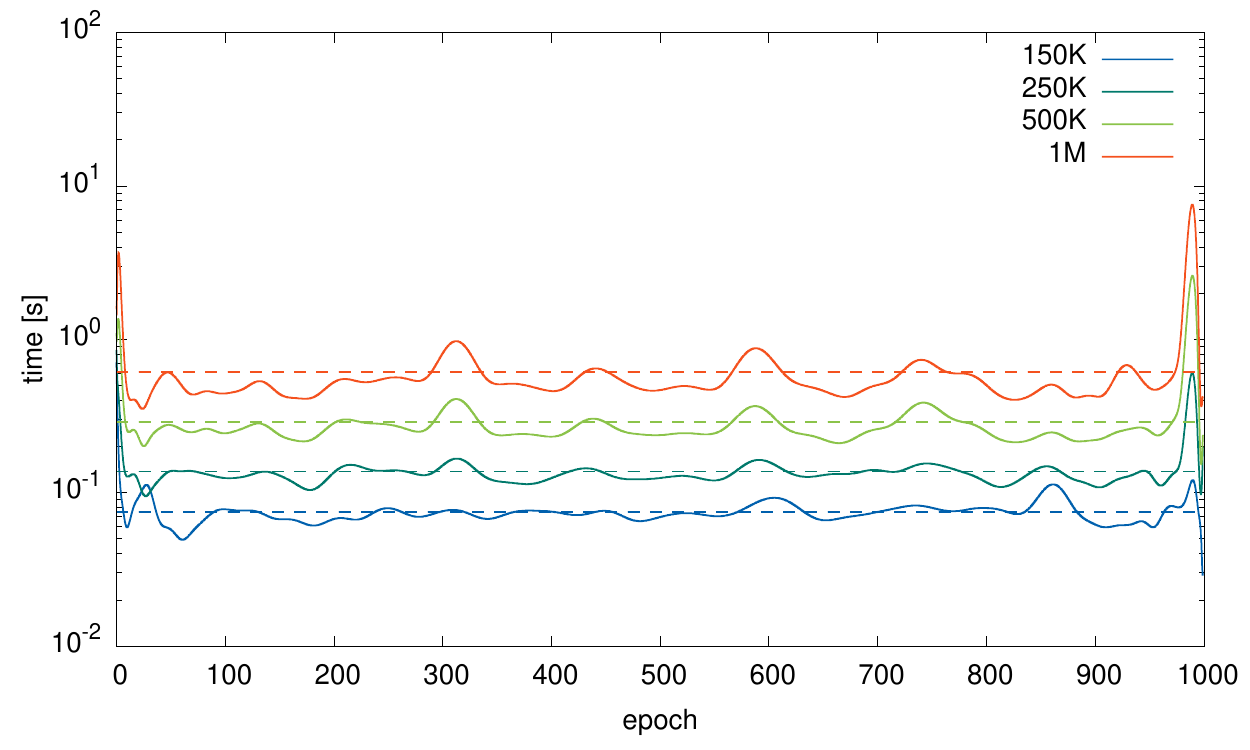}} 
\subfloat[latency (32w-profiled)]{\includegraphics[width=0.5\textwidth]{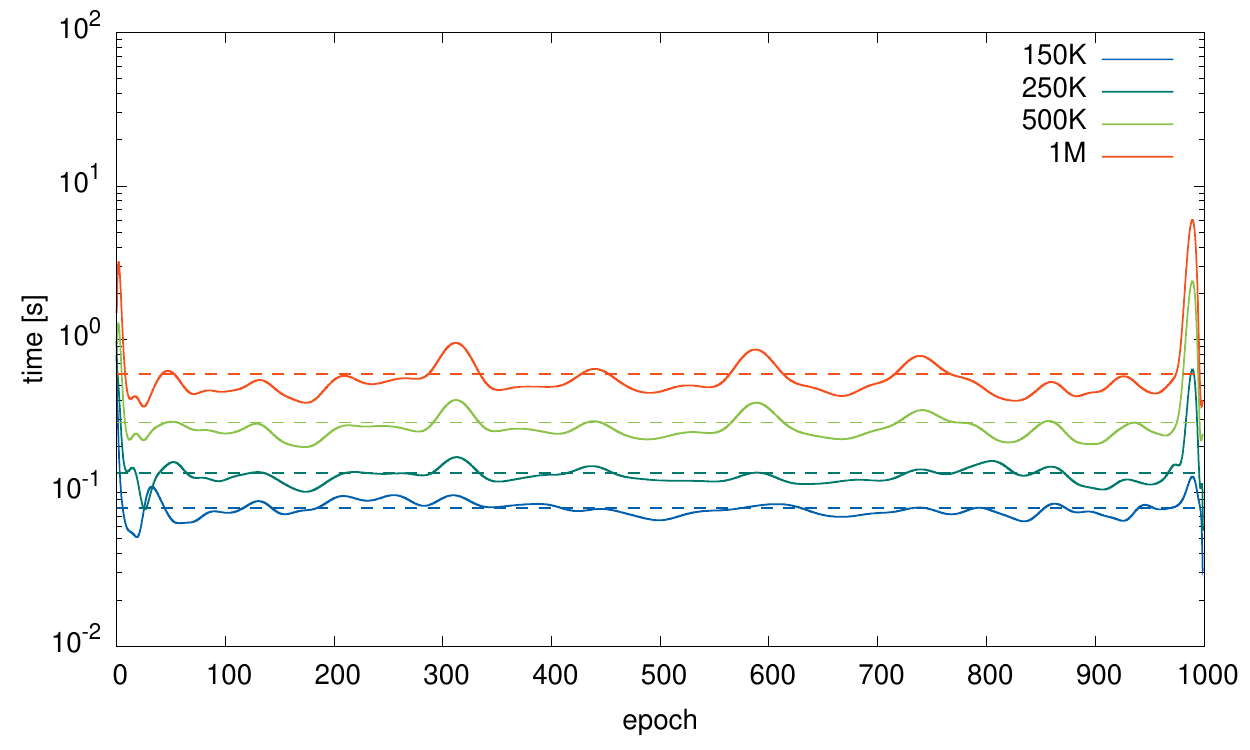}}

\subfloat[latency (16w-profiled)]{\includegraphics[width=0.5\textwidth]{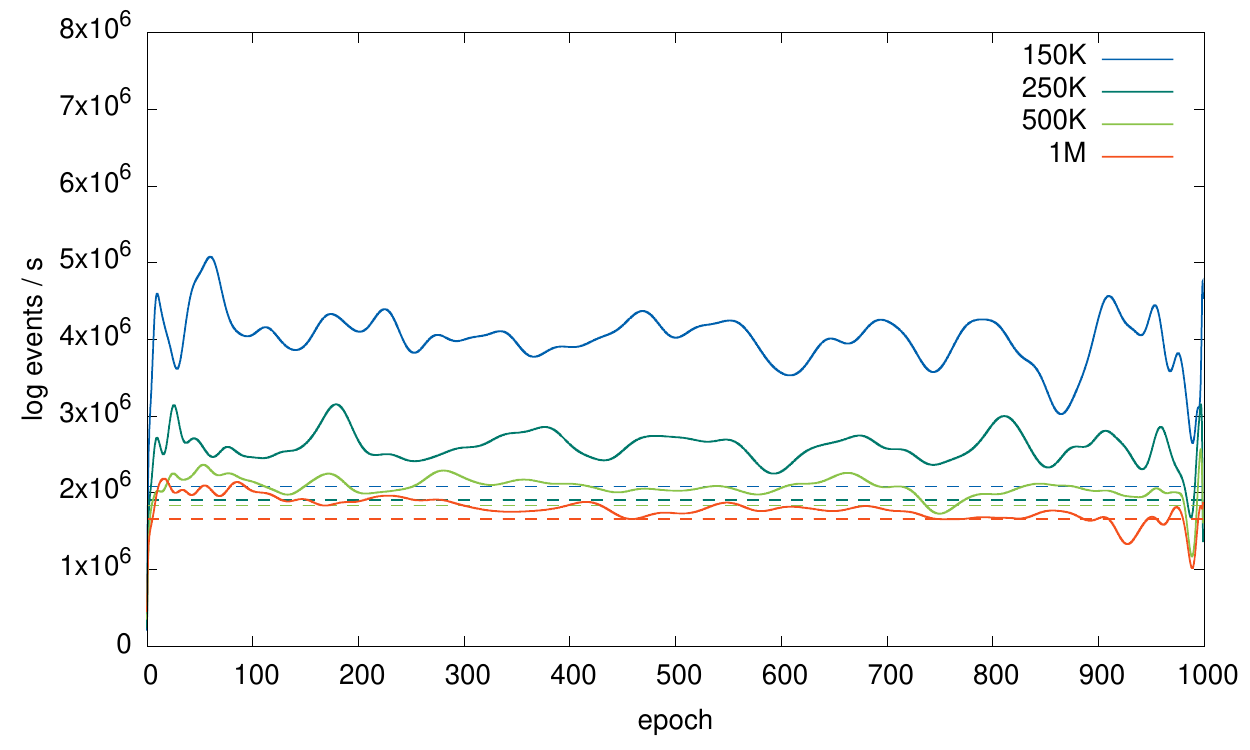}} 
\subfloat[latency (32w-profiled)]{\includegraphics[width=0.5\textwidth]{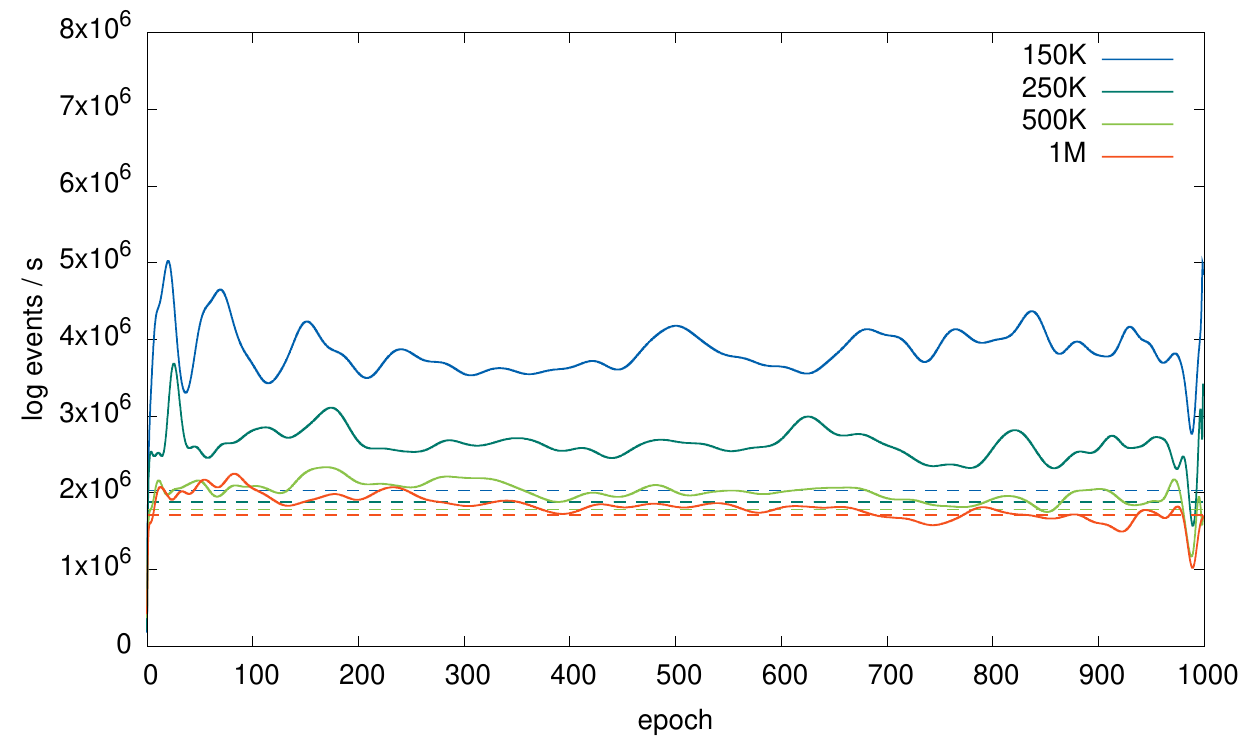}}
\caption[Triangles online performance]{Triangles online performance per epoch for various epoch densities while profiled by ST2. Dashed lines indicate mean across all epochs.}
\label{fig:tc_online_performance}
\end{figure}

To find out whether the source computation is affected by ST2's worker configuration, we measure and compare its latency and throughput separately for each configuration. \cref{fig:tc_online_performance} depicts the source computation's latency and throughput across each epoch when being profiled with 16 ST2-workers and 32 ST2-workers, respectively. As expected, there is no significant difference between the measured latencies --- ST2's worker configuration does not appear to have an impact. Comparing the online throughput measurements highlights minor differences: for example, in the 32-worker ST2 configuration with an epoch size of \numprint{1000000} log events, mean throughput slightly increases compared to the 16-worker configuration. We assume that this is owed to the generally more volatile throughput behavior across measurements, and not necessarily a statistically significant result. Therefore, we argue that neither latency nor throughput are affected in a major way by varying ST2's worker configuration in an online setting.

While these results appear to match our expectations, issues arise when comparing them to the offline experiment (cf.\ \cref{fig:tc_throughput} and \cref{fig:tc_latency}). Regardless of epoch size, in absolute terms, the source computation's online latency is worse. For example, for \numprint{1000000} events per epoch, each epoch on average takes \numprint{0.1} seconds longer; and for \numprint{150000}-event epochs, the mean latency undergoes a four-fold increase from \numprint{0.02} to \numprint{0.08} seconds. Online throughput is affected as well: While the source computation run offline scaled from around \numprint{2000000} events per second for an epoch size of \numprint{1000000} to \numprint{6000000} events per second for an epoch size of \numprint{150000}, in an online environment, it stagnates at around \numprint{2000000} events per second regardless of epoch size. To explain the disparity in behavior, we examined the major differences between running ST2 online and offline.

A first such difference is ST2's \texttt{replay} operator. In the offline experiment, it gets handed all log events at once and then feeds them to ST2 one epoch at a time. In comparison, events in the online experiment arrive at the operator piece by piece. To simulate this without a network roundtrip, we ran the source computation and ST2 offline concurrently, such that ST2 could only read new events from file after they had been written by the source computation. In this setting, the source computation's performance returns to old form. Thus, the \texttt{replay} operator is not the agitator --- instead, networking-related issues seem to be the cause.

We are able to rule out network bandwidth limitations: An offline triangles computation running on 32 workers and a batch factor of 5 creates ca.\ 18GB of PAG-relevant events in 25 seconds (cf.~\cref{tab:config_impact}). This translates to around 720MB/s that have to be sent over the network, while the cluster is able to handle around 13Gbit/s, i.e.\ \numprint{1625}MB/s. Instead, we examined the TCP socket configuration used to transmit log events. When we ran the TCP socket's \texttt{send} method in non-blocking mode, the application repeatedly crashed with Rust's \texttt{WouldBlock} error message, hinting at a TCP buffer overflow. This finally explains the source computation's slowdown: In blocking mode, it has to wait until the receive buffer is cleared on ST2's end before it can push additional events on the buffer. This waiting activity blocks the computation.

The offline experiment revealed that ST2's latency and throughput should be able to handle the source computation in an online setting with ease. To check whether unfavorable dataflow scheduling on Timely's was the main issue, we created a minimal Rust program that establishes a socket connection with the source computation, repeatedly reads from the socket and discards received data right away. This way, we simulated the fastest \enquote{dataflow} possible --- one that does nothing in-between reads. The source code for this example can be found in the repository's \texttt{pipe} subcrate \cite[commit: \texttt{cec8ca}]{sandstede2019}. While this led to better performance, the source computation still did not reach the offline experiment's throughput. Therefore, instead of tuning Timely's scheduling mechanism or introducing a separate networking thread, the TCP configuration itself has to be revised.

To that end, two promising optimizations exist. If the server allows it, the network buffer's size can be increased by modifying the TCP receive window size, e.g.\ by setting a higher window scaling factor, such that a single socket can handle higher source computation throughputs. Alternatively, instead of sharing a single high-throughput socket across all workers, each ST2 worker could retain its own socket. While not difficult per se, we did not implement the multi-socket solution due to the thesis' time restrictions, and instead require appropriately configured TCP window sizes in the current ST2 version. If that is the case, the source computation behaves as expected: there is no difference between offline and online profiling.

\subsection{Scaling Results}

\begin{figure}[htb]
\centering
\subfloat[latency]{\includegraphics[width=0.5\textwidth]{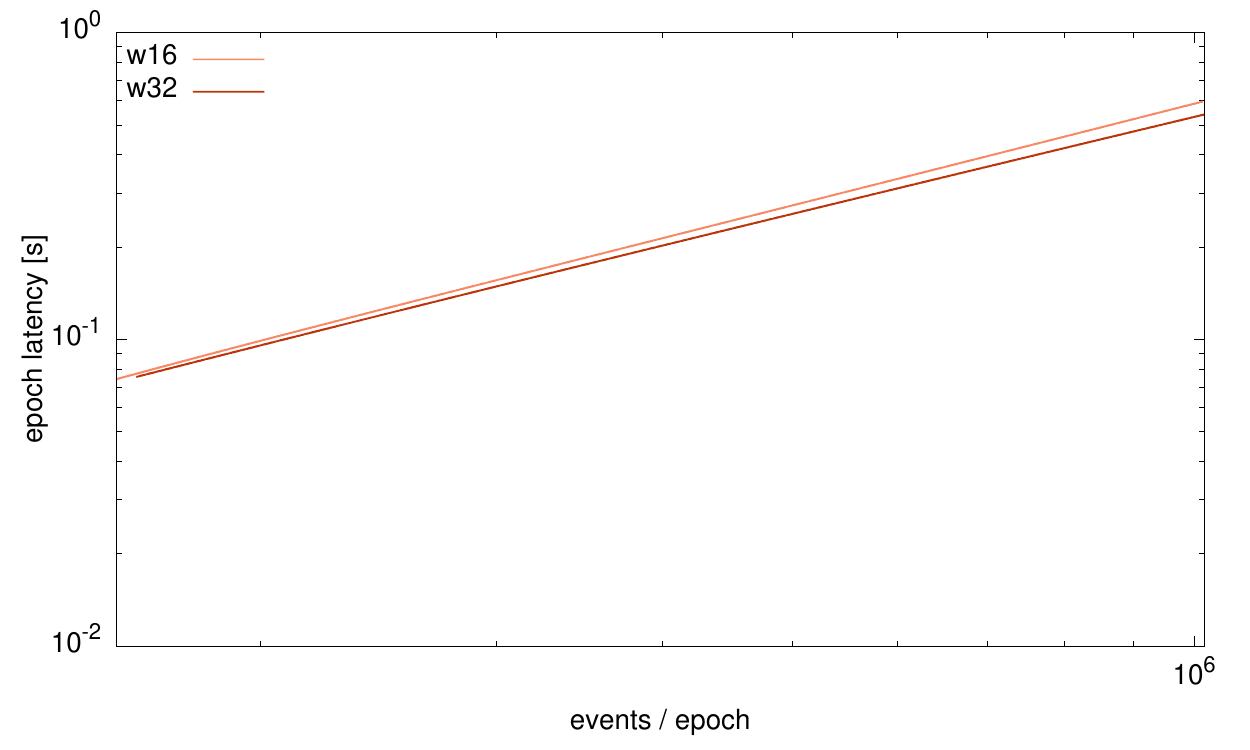}} 
\subfloat[throughput]{\includegraphics[width=0.5\textwidth]{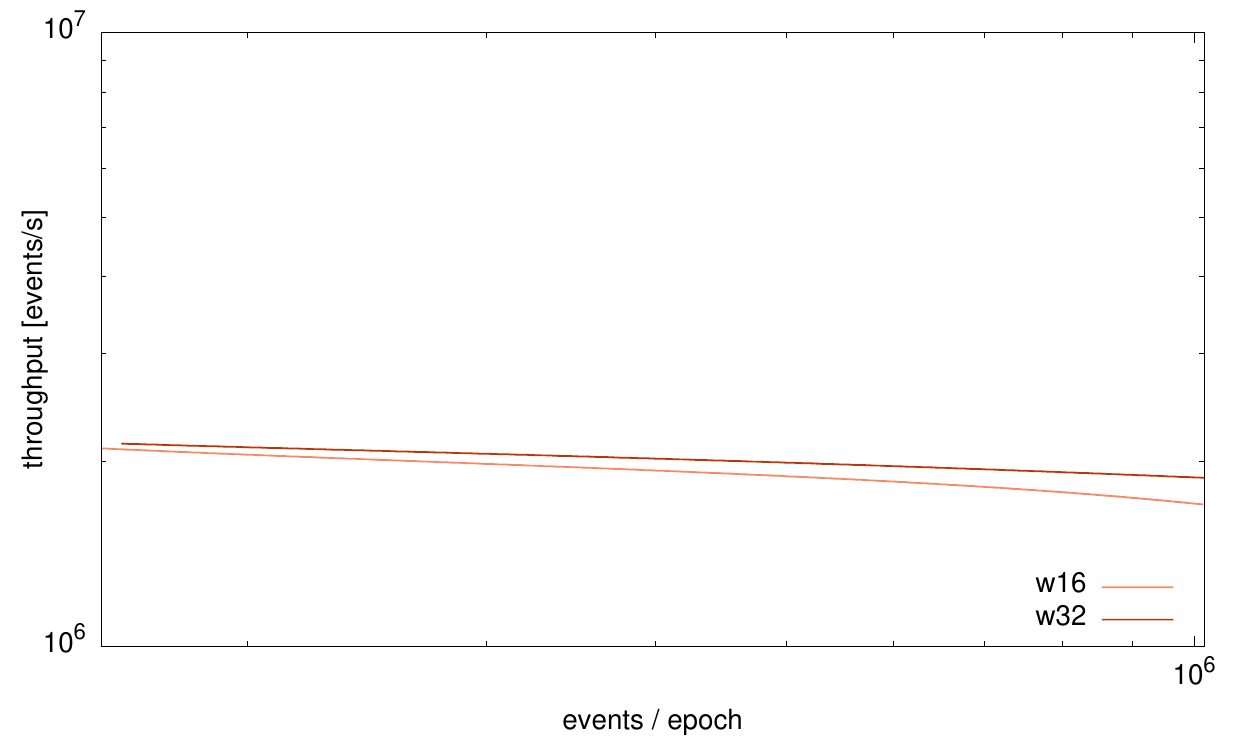}}
\caption{Timely PAG online scaling}
\label{fig:online_scaling}
\end{figure}

Keeping in mind the caveats discussed in \cref{online-sc}, we now analyze ST2's online scaling characteristics. \cref{fig:online_scaling} shows how epoch latency and throughput scale across increasing epoch densities. Similarly to the offline experiment, latency increases with higher epoch sizes. Both worker configurations display very similar latencies. A possible reason for this --- which we can further validate in \cref{online-cdfs} by examining the CDFs --- is that the measured ST2 worker configurations are able to keep up with the maximum rate at which the source computation can provide them with the events of an epoch. This also matches our results in \cref{sec:pageval_offline-experiment}, where a 16-worker configuration was always able to keep up with the source computation. While latency scales similarly to the offline experiment, the throughput scaling results deviate from the offline setting. With higher epoch densities, throughput slightly decreases. We suspect a combination of the TCP buffer limit and the source computation performance limit to be the cause of this --- compared to the offline setting, ST2 incurs additional overhead due to the networking at higher epoch densities that negatively affect throughput.

%

As expected, in absolute terms, ST2's mean latency and throughput are worse than in the offline setting. We hypothesize that this is due to the performance restrictions imposed by the source computation and consult the CDFs to validate this claim.

\subsection{CDF Results}\label{online-cdfs}

\begin{figure}[p]
\centering
\includegraphics[width=.9\textwidth]{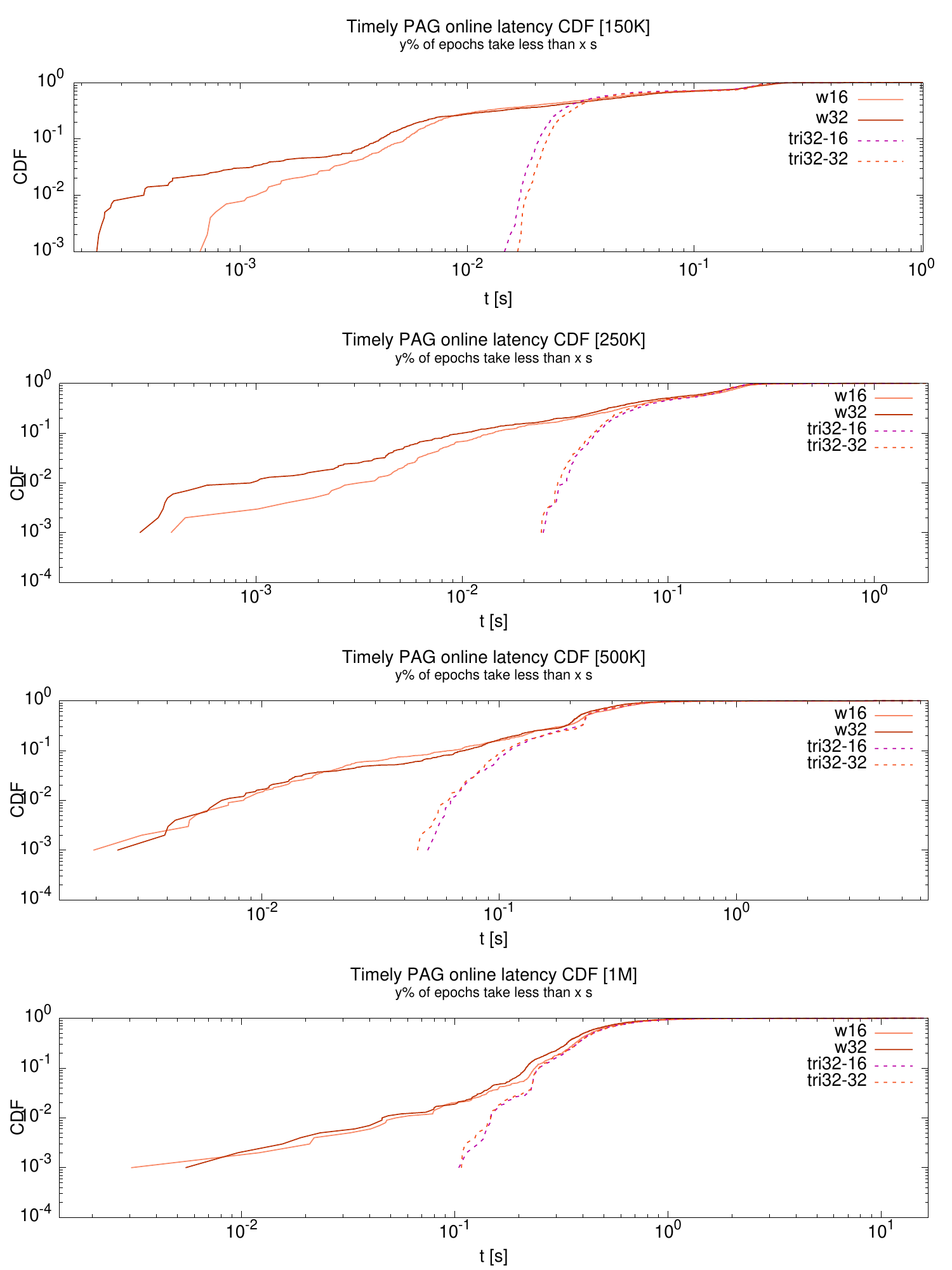}
\caption[Timely PAG online latency CDFs]{Timely PAG online latency CDFs for multiple epoch sizes and worker configurations. Epoch size provided in brackets. Dashed lines denote the corresponding triangles source computation that was profiled. For a given point, its y coordinate denotes the percentage of epochs that take less time than the corresponding x coordinate in seconds.}
\label{fig:online_latency_cdf}
\end{figure}

\begin{figure}[p]
\centering
\includegraphics[width=.9\textwidth]{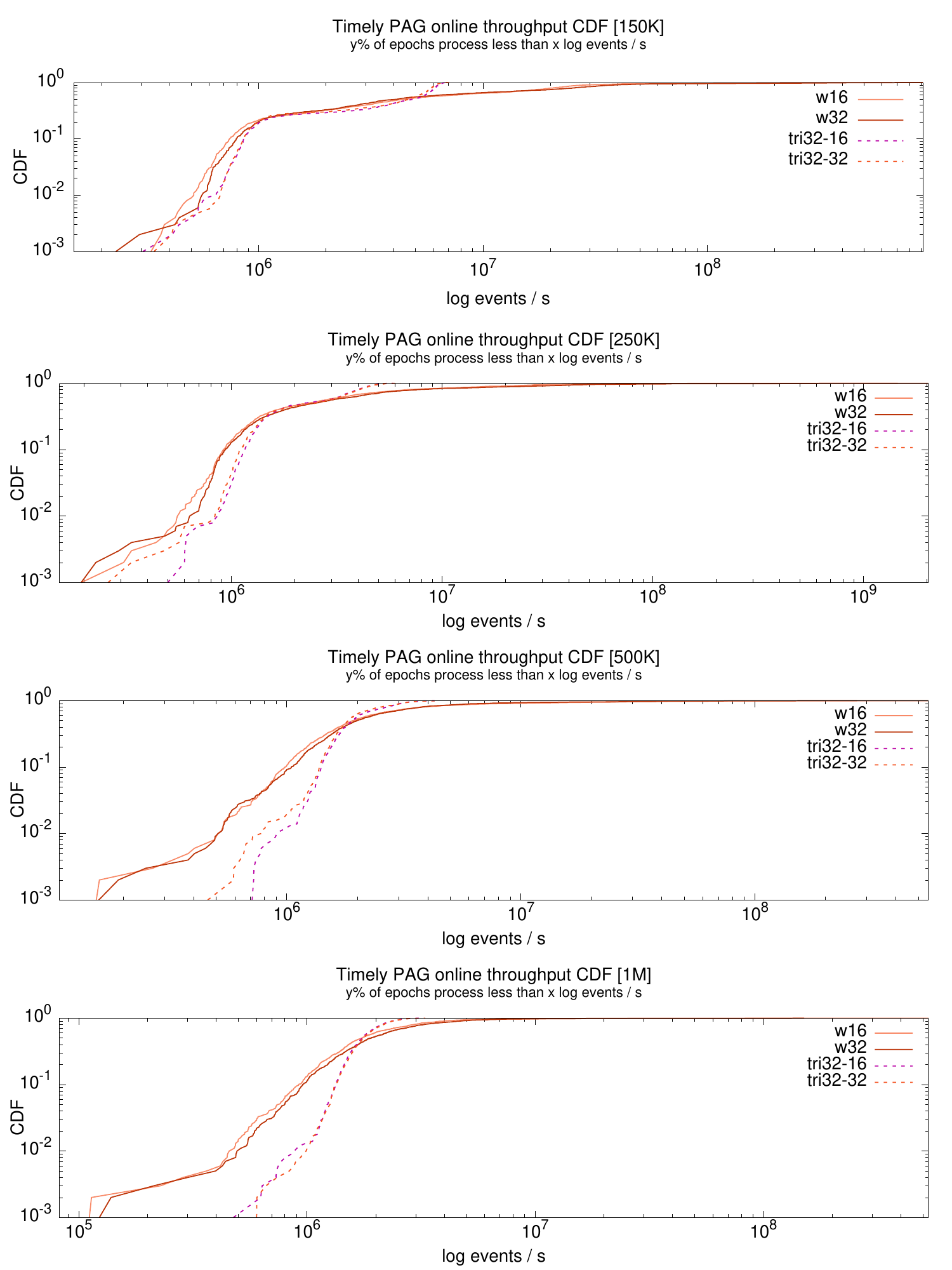}
\caption[Timely PAG online throughput CDFs]{Timely PAG online throughput CDFs for multiple epoch sizes and worker configurations. Epoch size provided in brackets. Dashed lines denote the corresponding triangles source computation that was profiled. For a given point, its y coordinate denotes the percentage of epochs that process less events per second than the corresponding x coordinate.}
\label{fig:online_throughput_cdf}
\end{figure}

\cref{fig:online_latency_cdf} depicts the latency CDFs for the online PAG construction. Compared to previous CDFs, we plot the source computation separately for each worker configuration, since source computation and ST2 might affect each other in the online setting. However, this concern is ungrounded: the source computation's behavior is nearly identical, regardless of worker configuration. This is consistent with our results from \cref{online-sc}, where we found that varying worker configurations does not have an impact on source computation performance.

Due to the blocking TCP behavior discussed in \cref{online-sc}, we cannot confirm whether ST2 is able to keep up with the source computation from these CDFs alone. Considering our results from \cref{cdfs}, however, we argue that it should be able to keep up, and does not block due to low performance on ST2's end. Using the results from \cref{fig:online_latency_cdf}, we are also able to confirm our hypotheses from the previous section: Across all epoch sizes, both worker configurations track the source computation latency distribution closely for most epochs --- they are bottlenecked by its performance.

Compared to the offline experiment, latency is more unstable across epochs. We argue that this is to be expected when introducing the network stack as an additional variable.

\cref{fig:online_throughput_cdf} visualizes the online throughput CDFs. It also plots separate source computation CDFs for each worker configuration. Again, the source computation behaves nearly identical for both worker configurations. While most epochs are able to match or exceed the source computation's throughput, fat tails with relatively small throughputs exist, especially at high epoch sizes. We assume that the major cause for this is the blocking TCP behavior as well.

\subsection{Summary}

In this section, we examined the source computation's and ST2's behavior when running it online. ST2's favorable offline results translate to the online setting, if a sufficiently powerful (or parallelized) TCP connection and network bandwidth is provided. In our benchmarks, we observed the negative impact on scaling and especially throughput that a bottlenecked connection can have, but also found evidence supporting our claim that ST2 can easily keep up with a source computation in an online setting.

\section{Comparison with SnailTrail~1}
\label{st1comp}

As its name suggests, ST2 is the successor to SnailTrail~1. One of ST2's goals is to be able to match, if not surpass its predecessor's performance. This allows us to justify our implementation choices --- in particular the richer PAG semantics ---, and also lets us piggyback on SnailTrail~1's benchmark results: SnailTrail~1 reported ``a throughput two orders of magnitude larger than the event rate observed in all log files'' it ran on \cite{hoffmann2018}, making it suitable for profiling even high-performance streaming jobs. We evaluate whether ST2 has been able to reach that goal by comparing both systems' benchmark results.

\subsection{Setup}

SnailTrail~1's benchmarks ran on comparable hardware to ours (cf.~\cref{hardware-and-software-specifications}), making a comparison based on the numbers reported in the original paper \cite{hoffmann2018} straightforward. Note that in its benchmarks, SnailTrail~1 already considers downstream analyses --- the critical participation metric --- which require an additional traversal of the constructed PAG. This should be kept in mind when directly comparing SnailTrail~1's with ST2's performance results. SnailTrail~1 uses fixed window semantics: Log events are collected over a fixed time frame (\enquote{snapshot}) and then combined into a PAG. Increasing the window size leads to more events that have to be considered for each PAG construction. For our comparison, we equate such a PAG window with ST2's concept of an epoch.
 
\subsection{Results}
 
\citeauthor{hoffmann2018} \cite{hoffmann2018} report almost linear latency scaling with increasing window size. We extrapolate from its most favorable latency result, which uses a snapshot size of 1 second to create a PAG from \numprint{30000} events in \numprint{0.06} seconds. This would translate to a \numprint{1000000}-event epoch latency of 2 seconds. In comparison, ST2 can create a \numprint{1000000}-event epoch in less than \numprint{0.1} seconds. \citeauthor{hoffmann2018} \cite{hoffmann2018} further report SnailTrail~1's maximum achieved throughput at \numprint{1200000} events per second. In comparison, ST2 reaches throughputs in excess of \numprint{10000000} events.

There are multiple reasons why ST2 is able to exceed SnailTrail~1's performance. In general, SnailTrail~1 sacrifices latency for throughput. While computing the critical participation metric, it exchanges each window's data to a single worker in a round-robin fashion. Thus, individual window latency increases. In comparison, ST2 already exchanges data within an epoch. In ST2, we introduced richer time semantics than a simple one-second fixed window (cf.~\cref{subsec:impl_windows}). Still, the PAG is constructed without the use of blocking operators. The profiling library (cf.~\cref{subsec:impl_online_offline}) implementing these semantics has been built for speed: It discards any events that are non-essential to the PAG construction before they are written to file or transmitted via TCP. The custom replay operator is able to throttle the number of concurrent in-flight epochs, which reduces progress tracking overhead. With more control over how events are replayed to the PAG construction, we can also take advantage of ordering. As a result, the local edge construction is implemented as non-blocking, quasi-stateless \texttt{map}-like operator instead of a more expensive join (cf. \cref{sec:impl_pag}). Lastly, ST2 builds on a newer version of Timely, which might provide a performance boost in its own. Taken together, these changes lead to the significantly higher performance of ST2.

\subsection{Summary}

In this section, we compared ST2's performance to the results published for SnailTrail~1 \cite{hoffmann2018}. Primarily by exploiting knowledge about the relationship between profiled and profiling application, ST2 is able to increase its performance by an order of magnitude. This should make it suitable for profiling many common streaming computations in an online setting. 

\section{PAG Evaluation Summary}

The benchmarks and comparisons discussed in this chapter confirm that ST2 is able to construct and maintain a source computation's PAG in an efficient and scalable manner.  

An implementation using Timely provides the highest possible performance and even exceeds SnailTrail~1's results by an order of magnitude, especially due to optimizations in the hand-off between source computation and PAG construction. It is well-suited for profiling nearly any source computation, and can even be scaled beyond the source computation if necessary.

While a differential PAG implementation is possible and provides high-level abstractions, it cannot leverage most of Differential's unique features. Incurring a 10x performance penalty is therefore not worth the trade-off for ST2.

In the online setting, ST2's throughput and epoch latency are limited by the source computation's performance. The source computation itself behaves identical in both settings, under one condition: networking, and especially the TCP communication, must provide the necessary bandwidth and buffer size to not become a bottleneck.

If this condition is met, the PAG construction is scalable enough to accommodate nearly any source computation, while still leaving enough room for further analytics.

\chapter{Functional Evaluation}\label{func-evaluation}

In this chapter, we evaluate ST2's functionality. Compared to \cref{pag-evaluation}, this chapter focuses on the downstream analytics that are enabled by the PAG construction instead of the PAG construction itself. We also do not discuss the technical implementation of ST2's data analytics. For that, refer to \cref{sec:impl_analysis}.

First, we present ST2's command-line interface (\cref{cli}). It enables users to conduct analyses on top of a source computation's PAG. We then discuss ST2's real-time dashboard (\cref{dashboard}), which leverages the commands to provide users with a graphical way of interacting with the analyses. We also present an exemplary case study to evaluate ST2's ability to spot issues when using the dashboard in an online setting.

\section{Command-Line Interface}\label{cli}

In this section, we introduce ST2's command-line interface (CLI). We first provide a general overview of its functionality (\cref{cli-overview}), before discussing the aggregate metrics command in \cref{metrics}. Then, we present its invariant checker (\cref{invariants}), which can be used to track the source computation's adherence to custom temporal rules, and the graph patterns command (\cref{patterns}), which can be used to spot non-trivial inter-worker-dependency related issues. Lastly, we discuss the performance impact these analytics have on top of the PAG construction in \cref{cli-performance}.

\subsection{Overview}\label{cli-overview}

\begin{figure}[htb]
\centering
\includegraphics[width=0.9\textwidth]{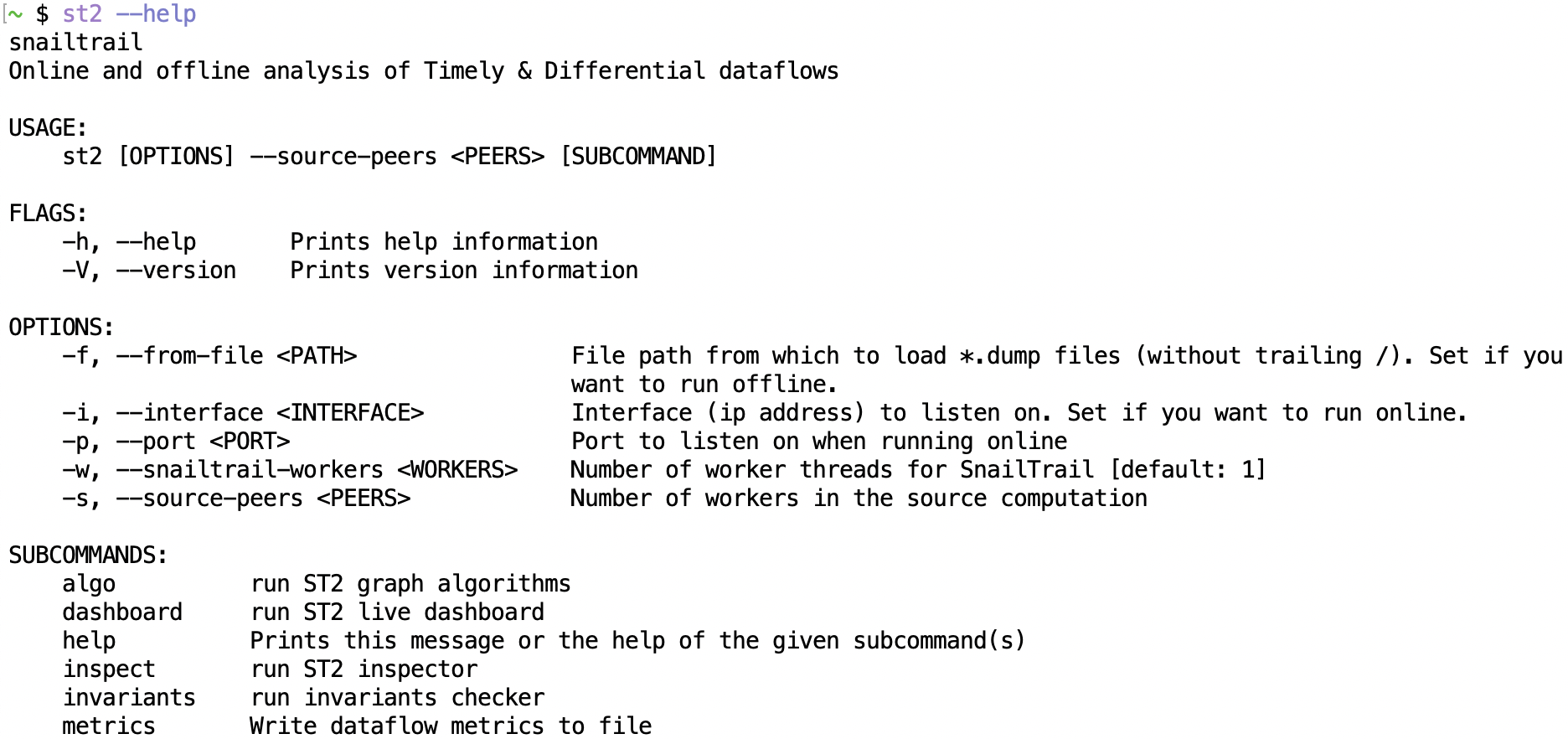}
\caption{Command-line interface documentation}
\label{fig:cli}
\end{figure}

ST2's CLI is self-documenting. If run with the \texttt{-{}-help} flag, it will print the documentation depicted in \cref{fig:cli}.

The CLI provides multiple options. In an offline setting, the CLI should be run with the \texttt{-{}-from-file} flag. This allows specifying the path from which the offline traces used by ST2 should be loaded. In an online setting, \texttt{-{}-interface} and \texttt{-{}-port} allow specifying on which IP address and port the source computation will publish log events. The source computation then sets these using the \texttt{SNAILTRAIL\_ADDR} environment variable. ST2 must know how many connections it has to expect. For this, the \texttt{-{}-source-peers} flag is used. If the source computation is running on a load balance factor greater than 1, then $source\ peers = \#source\ computation\ workers * LBF$. Similarly, the \texttt{-{}-snailtrail-workers} flag specifies on how many worker threads ST2 should run. By default, it is set to 1.

The CLI also allows to specify which subcommand to run. \texttt{metrics}, \texttt{invariants}, and \texttt{algo} are discussed in the following sections. \texttt{dashboard} is separately discussed in \cref{dashboard}. \texttt{inspect} is internally used to debug and benchmark ST2.

\subsection{Metrics}\label{metrics}

\begin{figure}[htb]
\centering
\includegraphics[width=0.9\textwidth]{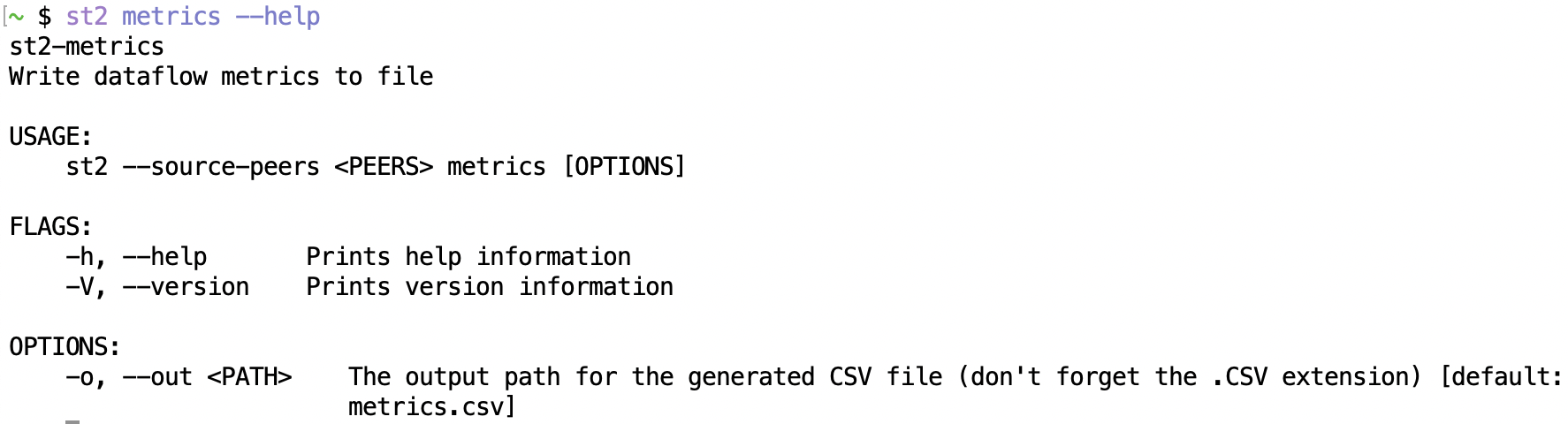}
\caption{Metrics subcommand documentation}
\label{fig:metrics}
\end{figure}

Using the \texttt{metrics} subcommand, aggregate metrics for ST2 can be produced and written to file. The subcommand's documentation is depicted in \cref{fig:metrics}. In addition to the command-line arguments discussed in \cref{cli-overview}, \texttt{-{}-out} should be passed with a path to which the aggregate metrics are written. Running the subcommand creates a file that contains the following information:

\begin{description}[leftmargin=!,labelwidth=\widthof{\bfseries activity\_type}]
	\item [epoch] Describes for which PAG epoch the metrics are exported.
	\item [from\_worker] Describes from which worker the activity originates.
	\item [to\_worker] Describes at which worker the activity is targeted.
	\item [activity\_type] Describes the activity's type (e.g.\ data or control messages, processing, waiting).
	\item [\#(activities)] Describes the number of activities that occurred for the specified epoch, workers, and type.
	\item [t(activities)] Describes the total time in nanoseconds that these activities took.
	\item [\#(records)] Describes how many records these activities carried, if applicable (e.g.\ data messages or processing).
\end{description}

Data is grouped by the first four attributes. For example, if a line in the created file reads the following: $4,0,1,DataMessage,240,350000,15000$, this means that in epoch 4, 240 data messages were sent from worker 0 to worker 1. This took \numprint{0.35} milliseconds, and the data messages carried \numprint{15000} records.

\begin{figure}[htb]
\centering
\includegraphics[width=0.9\textwidth]{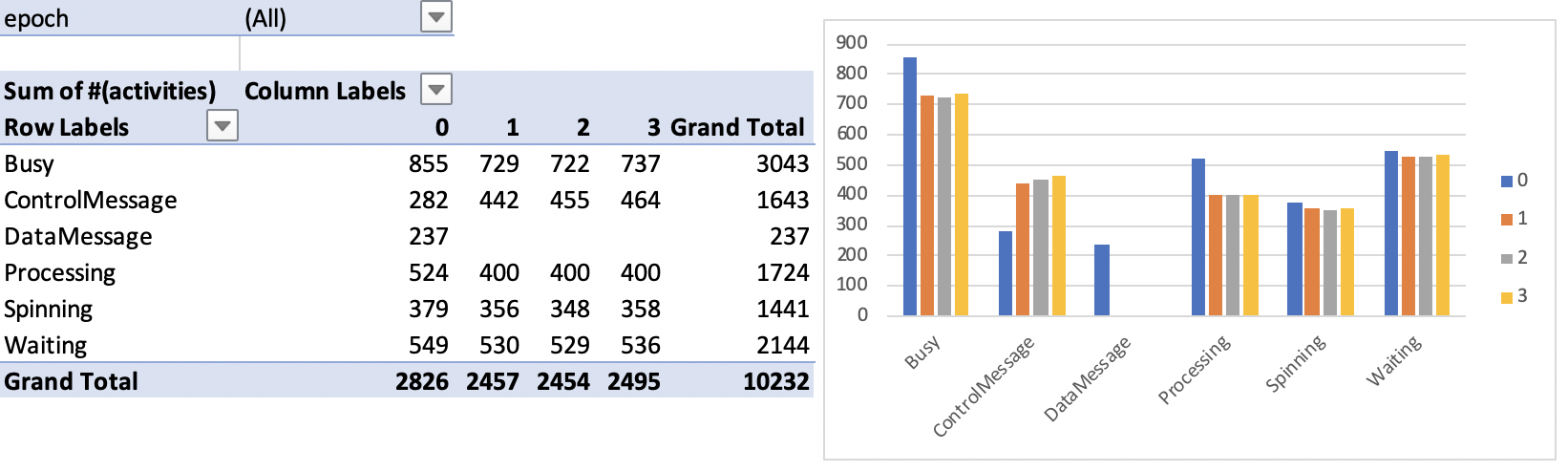}
\caption{Metrics analysis using pivot tables}
\label{fig:metrics_pivot}
\end{figure}

The exported data can then be used for OLAP-style data analysis. For example, \cref{fig:metrics_pivot} shows the number of activities across all epochs, split by worker origin. It uses pivot tables and pivot charts to slice and dice the data into the desired format. This way, the metrics subcommand allows the simple creation of e.g.\ activity, worker, and operator summaries.

\subsection{Invariants}\label{invariants}

\begin{figure}[htb]
\centering
\includegraphics[width=0.9\textwidth]{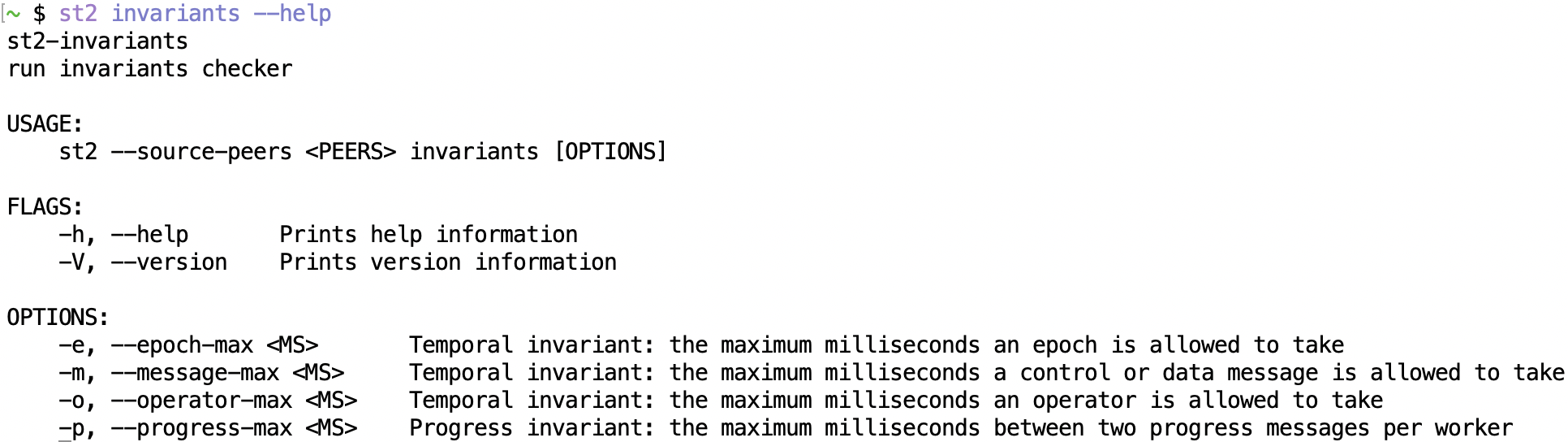}
\caption{Invariants subcommand documentation}
\label{fig:invariants}
\end{figure}

\begin{figure}[htb]
\centering
\includegraphics[width=0.9\textwidth]{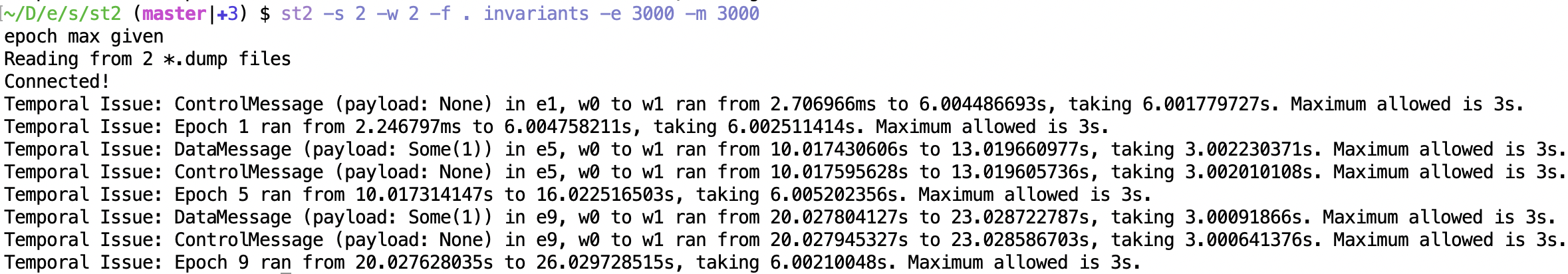}
\caption{Invariants subcommand demo}
\label{fig:invariants_demo}
\end{figure}

With the \texttt{invariants} subcommand, users can register rules that should be validated during the source computation's execution. Its documentation is shown in \cref{fig:invariants}. The listed arguments are optional, such that a subset of invariants can be (de)activated at will. \texttt{-{}-epoch-max} enforces a ceiling on the time that a PAG epoch is allowed to take. Similarly, \texttt{-{}-message-max} checks whether data and control messages stay within certain time bounds, and \texttt{-{}-operator-max} detects operator slowdowns. Lastly, \texttt{-{}-progress-max} checks whether the computation continues to make progress within a given frame of time. In addition to the user-configurable invariants, a non-temporal progress invariant is also checked, which ensures that there is at least one progress message seen per epoch. That way, grave bugs such as stalling dataflows become discoverable. \cref{fig:invariants_demo} shows the invariant checker in action; it runs with \texttt{-{}-epoch-max} and \texttt{-{}-message-max} set to \numprint{3000}ms, and profiles a source computation that experiences major slowdowns on multiple epochs and messages.

While progress invariants are helpful to straight-out debug source computations --- if a progress invariant is violated, there likely exists a bug in the stream processor itself~---, temporal invariants are also useful for more nuanced analyses. In a business setting, service-level-agreements might dictate that messages in the system shall not exceed a certain latency threshold. This can be monitored using invariants. They also provide a straightforward way to detect straggling operators, or to audit source computation performance during long sessions: By letting them run side-by-side, effects over time such as accumulating state or shifting data skew can be detected early.

\subsection{Graph Patterns}\label{patterns}

\begin{figure}[htb]
\centering
\includegraphics[width=0.4\textwidth]{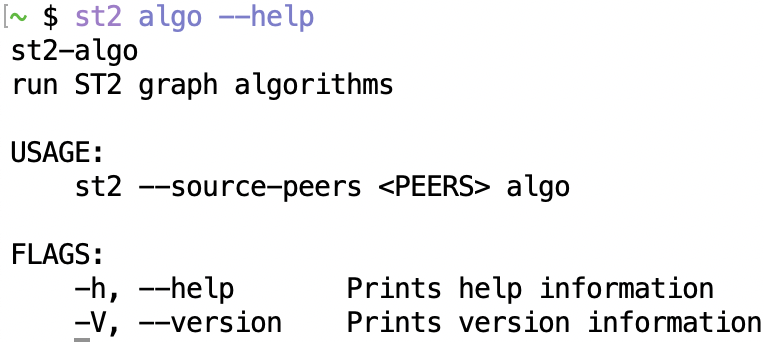}
\caption{Algo subcommand documentation}
\label{fig:algo}
\end{figure}

The \texttt{algo} subcommand can be used to run graph pattern algorithms on a source computation's PAG. For the current version of ST2, the k-hops pattern and weighted k-hops pattern are supported. The subcommand's documentation is depicted in \cref{fig:algo}. Compared to previous subcommands, no custom parameters have to be passed, as the subcommand is predominantly used in conjunction with the dashboard (cf.~\cref{dashboard}). However, adding support for parameterized hop counts would be straightforward, as discussed in \cref{impl-khops}.

The k-hop pattern's goal is to find recurring patterns  that are potential bottleneck causes for the source computation. This is similar to SnailTrail~1's critical participation metric (cf.~\cite{hoffmann2018}) --- however, it does not require complete PAG traversals, but only partial exploration, given a set of starting PAG nodes. For example, if a waiting activity always concludes with the reception of a specific data message, this data message is likely causing the bottleneck. Still, the root cause of the bottleneck might lie even further upstream, such that it is worthwhile to not only explore the direct neighborhood of a waiting activity, but also transitive predecessors up to a certain depth. This is what ST2's k-hop pattern does; starting from the end of waiting activity edges, it traverses the PAG backwards up to a certain depth, and produces aggregated information about the activity types and other information that the visited edges provide. The weighted k-hop pattern is slightly more opinionated --- it \enquote{interprets} the results by weighting them, e.g.\ with activity durations of the respective edges. That way, edges that are major contributors to a waiting activity are surfaced more quickly than if using uniformly distributed edge weights.

A graph pattern use case is presented in \cref{data-skew}; in our experiments, the patterns were particularly useful in guiding issue identification in combination with other ST2 tools, especially when highlighted in the dashboard's PAG visualization.

\subsection{Performance Impact}\label{cli-performance}

While most of ST2's analyses require only lightweight implementations (cf.~\cref{sec:impl_analysis}), by definition, any additional computation will have an impact on performance. We therefore also benchmark the CLI commands on top of the PAG to ascertain that its promising benchmark results hold. We benchmark each command individually, as they can be used in isolation, and, due to the dataflow model, do not interfere with each other. This way, we also do not have to repeat the experiments for the dashboard backend, which composes a subset of the CLI commands. We use the same experimental setting as in \cref{experimental-setup}: We run the offline experiments on \numprint{500000} log events epochs generated by a 32-worker triangles source computation and profile these log traces with a 32-worker ST2 instance that first creates the PAG and then runs a command on top of it.

\begin{figure}[p]
\centering
\includegraphics[width=1\textwidth]{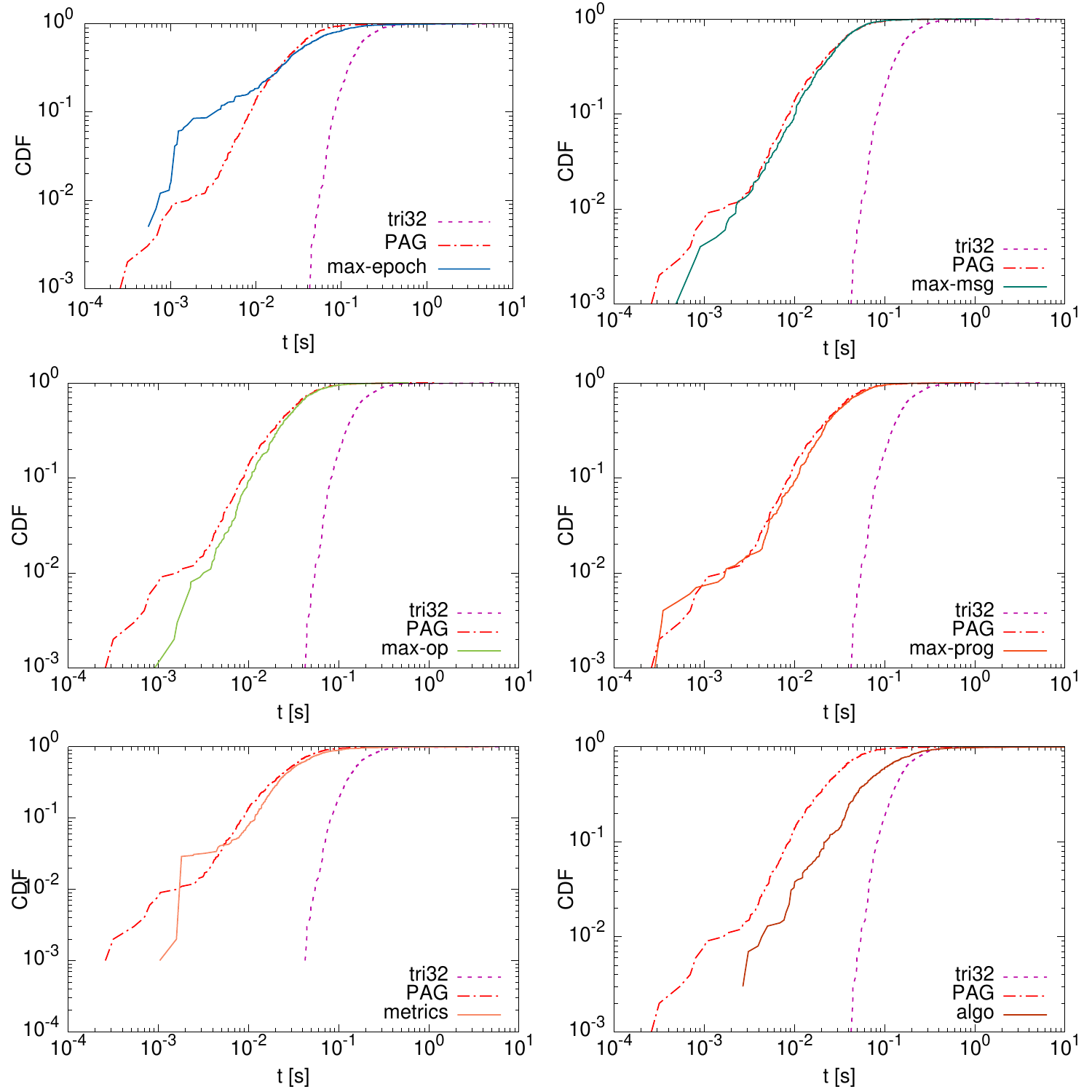}
\caption[Command latency CDFs]{Command latency CDFs for an epoch size of \numprint{500000} and 32 ST2 workers. \texttt{tri32} denotes the triangles source computation. \texttt{PAG} denotes the Timely PAG construction. For a given point, its y coordinate denotes the percentage of epochs that take less time than the corresponding x coordinate in seconds.}
\label{fig:cli-lat}
\end{figure}

The latency CDFs are depicted in \cref{fig:cli-lat}. As expected, there are latency differences between the PAG construction and commands. For all invariants (\texttt{max-epoch}, \texttt{max-message}, \texttt{max-operator}, and \texttt{max-progress}) these are relatively small. \texttt{max-epoch} is slightly more unstable, as it requires a more complex \texttt{aggregate} operator that acts on event batches that have been previously delayed to the epoch boundary using \texttt{delay\_batched}. For all invariants, the overall latency is slightly worse, and there is some larger variance for around 1\% of events. However, as the commands are relatively cheap to run --- they mostly consist of \texttt{map} and \texttt{filter} operators --- none affect performance in a major way. This is important, since it is particularly useful to keep invariants running side-by-side with the profiled source computation (cf.~\cref{invariants}). The aggregate metrics command (\texttt{metrics}) performs acceptable, with the caveat that it also has to make use of \texttt{aggregate} and \texttt{delay\_batched} operators. The k-hops graph pattern CDF (\texttt{algo} in \cref{fig:cli-lat}) states the worst latency: Around 10\% of events are not significantly faster than the source computation, and there is a significant overall increase in latency compared to the PAG construction. This does not come as a surprise; the k-hops pattern is the most computationally intensive command tested --- for example, it involves multiple customized \texttt{join} operators during its first hop to traverse the PAG (cf.~\cref{impl-khops}). However, following the same argumentation as in \cref{pag-evaluation}, we believe that ST2 can still comfortably hold its own for most streaming workloads, as the triangles computation is very efficient, particularly taxing due to its implementation in Timely, and we are not making use of the load balance factor to further speed up ST2. We therefore conclude that from a latency standpoint, even with the added overhead, all commands are still able to keep up with the 32-worker triangles computation, and also leave some room to be composed together for complex analysis combinations.

\begin{figure}[p]
\centering
\includegraphics[width=1\textwidth]{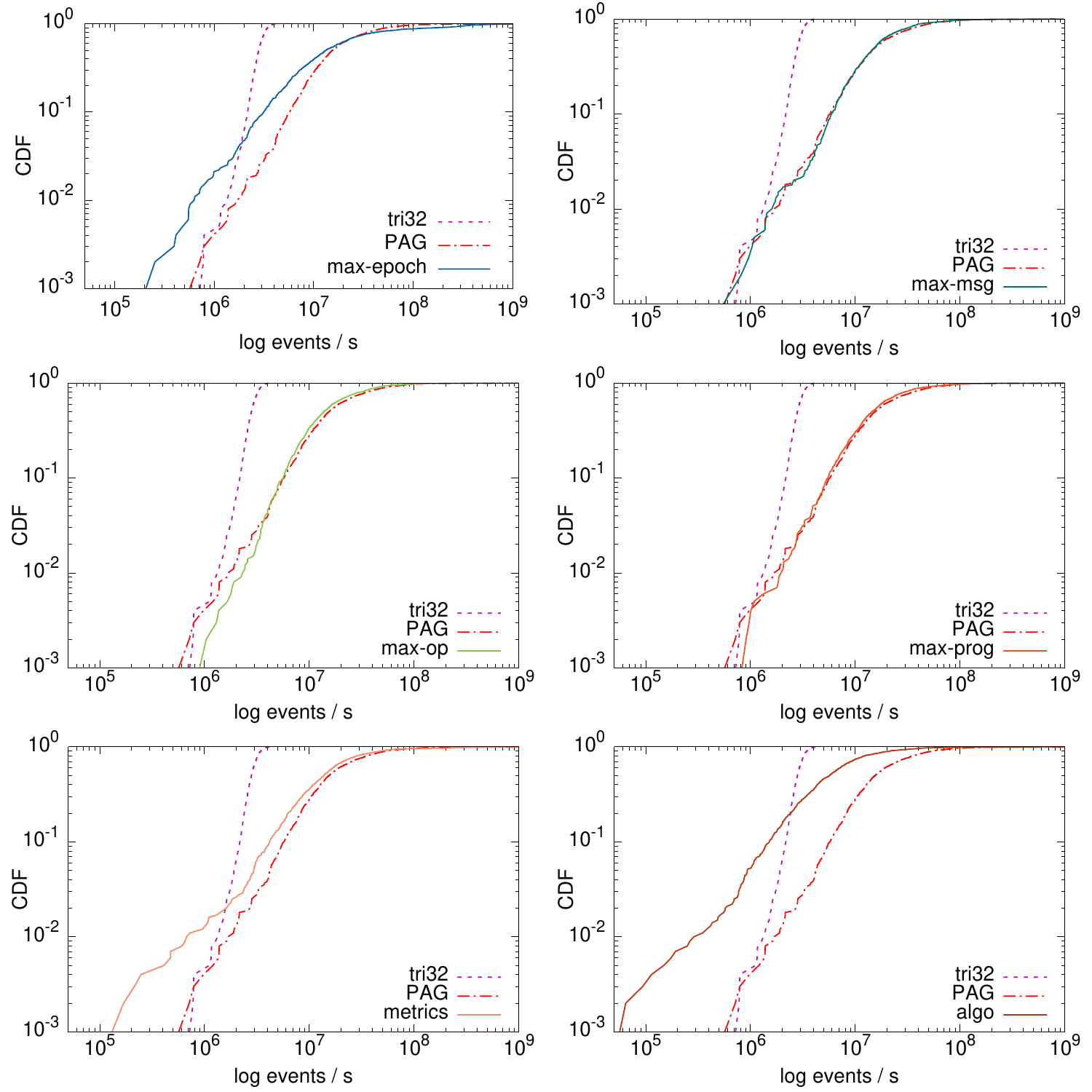}
\caption[Command throughput CDFs]{Command throughput CDFs for an epoch size of \numprint{500000} and 32 ST2 workers. \texttt{tri32} denotes the triangles source computation. \texttt{PAG} denotes the Timely PAG construction. For a given point, its y coordinate denotes the percentage of epochs that process less events per second than the corresponding x coordinate.}
\label{fig:cli-throughput}
\end{figure}

The throughput CDFs results are shown in \cref{fig:cli-throughput}. Again, performance differences between the PAG construction and running additional commands become apparent upon closer inspection. For most invariants (\texttt{max-message}, \texttt{max-operator}, and \texttt{max-progress}) the differences in throughput are negligible. \texttt{max-epoch} achieves a similar throughput overall, but it is also more unstable; some epochs' throughput even drops below the source computation. Again, we attribute this behavior to the more complex operators at play. This also explains why the aggregate metrics throughput is slightly lower than the invariants. The k-hop algorithm's costliness is also reflected in its CDF: Around 20\% of epochs achieve lower throughput than the source computation, and nearly all events' throughput is significantly lower than the Timely PAG construction.

We conclude that while all commands have a negative impact on performance, for most commands, this impact is negligible. For particularly taxing algorithms such as graph pattern matching, it might be worthwhile to look into making use of the load balance factor to boost ST2's speed when facing very demanding computations, or when running many commands simultaneously. However, in nearly all common scenarios, ST2 can comfortably keep up. 

\subsection{Summary}

In the previous sections, we have introduced ST2's CLI. The CLI allows a user to control ST2 without having to occupy herself with its technical details. For all commands, general options allow configuring ST2 to run in online or offline mode. In a second step, invariants, aggregate metrics, and graph patterns can be configured and individually run. We also evaluated the commands' performance. More complex operator implementations led to higher latencies and lower throughputs. Apart from slight decreases for the \texttt{max-operator} invariant and the \texttt{metrics} subcommand, these performance drops were barely noticeable. Only the graph pattern algorithms, which make use of multiple joins, were significantly slower than a mere PAG construction. However, they were still able to keep up with the demanding source computation, and could also be sped up further using a higher load balance factor. We therefore concluded that for all common use cases, ST2 is able to run with one or multiple commands enabled in an online setting. 

\section{Dashboard}\label{dashboard}

We now discuss ST2's real-time frontend: a dashboard that combines a number of analytics introduced in \cref{cli} and presents them to the user in a visual manner. We do not dive into the technical details behind the dashboard and its connection to the ST2 backend --- for that, refer to \cref{frontend}. Instead, we first give an overview of the dashboard in \cref{dashboard-overview} and then evaluate its functionality in a case study (\cref{data-skew}). In the case study, we manipulate a source computation in order to introduce synthetical performance issues and bottlenecks. We then run the ST2 dashboard and analyze its behavior.

\subsection{Overview}\label{dashboard-overview}

\begin{figure}[htb]
\centering
\includegraphics[width=0.9\textwidth]{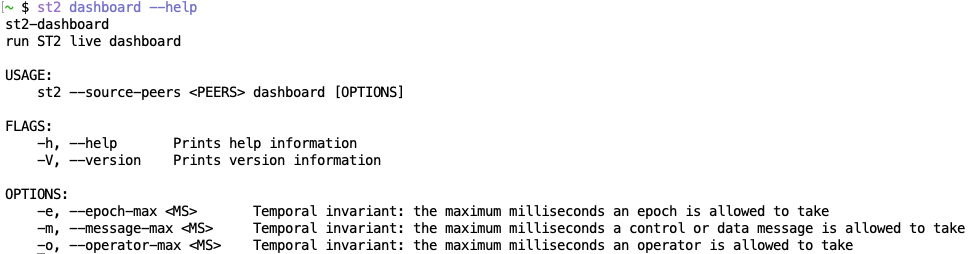}
\caption{Dashboard subcommand documentation}
\label{fig:dashboard-cli}
\end{figure}

Before running the dashboard frontend, its backend has to be started. For this, the CLI's \texttt{dashboard} command is used. Its documentation is shown in \cref{fig:dashboard-cli}. Similarly to previous commands (cf.~\cref{cli}), it can also be configured using optional arguments: \texttt{-{}-epoch-max}, \texttt{-{}-message-max}, and \texttt{-{}-operator-max}. These provide the same functionality as the invariant flags discussed in \cref{invariants}: they control at what point invariant violation alerts are triggered and displayed in the frontend. Once the backend has been started and connected to a source computation --- again, this is either done by providing a path to the \texttt{-{}-from-file} flag or by running the source computation with a \texttt{SNAILTRAIL\_ADDR} environment variable that matches the \texttt{-{}-interface} and \texttt{-{}-port} flags ---, the dashboard can be opened and will automatically connect to the backend.

\begin{figure}[p]
\centering
\includegraphics[width=1\textwidth]{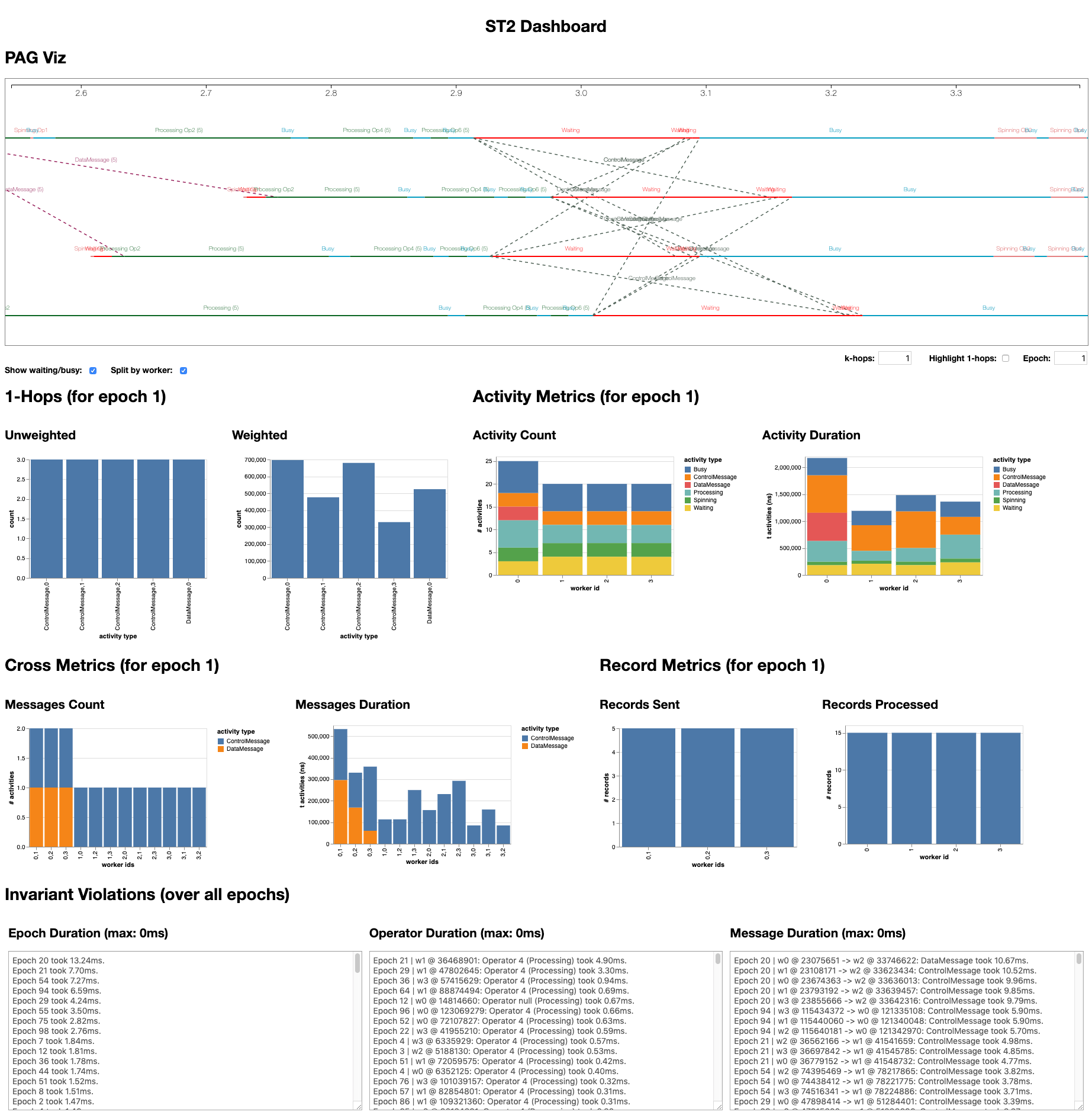}
\caption{Dashboard overview}
\label{fig:dashboard-overview}
\end{figure}

\begin{figure}[htb]
\centering
\includegraphics[width=1\textwidth]{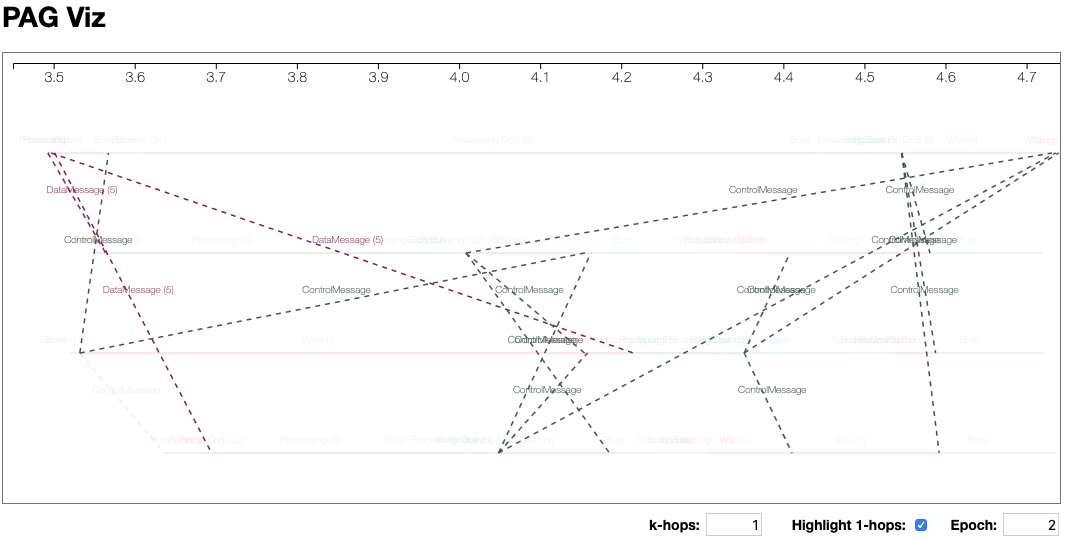}
\caption{Dashboard k-hops PAG visualization}
\label{fig:dashboard-khops}
\end{figure}

The started dashboard is shown in \cref{fig:dashboard-overview}. At the top (\enquote{PAG Viz}), the PAG is visualized. Edges are labeled --- if applicable, also with operator identifiers and record counts ---, activity types are color-coded, and assigned to the correct worker (y-axis) and timeline (x-axis). Additional information can be obtained as tooltip by hovering over an edge. The PAG visualization is also dynamically zoomable and scrollable, to enable analyzing even brief activities. The PAG's epoch can be selected using the configuration inputs to the bottom right of the PAG visualization. If run online, this will only have an effect if the source computation is already processing the selected epoch. However, as long as this is the case, the dashboard will update right away with the new epoch information. Furthermore, the configuration settings allow to highlight the edges reached by the k-hops algorithm (cf.~\cref{patterns}) in the PAG visualization. Such a highlighted version of the PAG for the first hop is depicted in \cref{fig:dashboard-khops}.

Below the PAG visualization, charts visualize graph patterns and metrics. At the top left (\enquote{K-Hops}), the activity type distribution for the currently selected k-hop depth is shown unweighted and weighted with the activities' durations. At the top right (\enquote{Activity Metrics}), activity counts and durations for the current PAG epoch, split by activity type, are depicted. The lower left (\enquote{Cross Metrics}) visualizes the counts and durations for remote control and data messages. The lower right (\enquote{Record Metrics}) informs the user about the number of records that were carried by data messages and processed by operators during the selected epoch. For all of these diagrams, \enquote{waiting} and \enquote{busy} activities can be hidden, as they often are not as interesting for performance root cause analysis. Furthermore, all diagrams can either be reported on a per-worker basis (for cross metrics, this displays each permutation of source and destination worker) or aggregated over all workers.

Lastly, at the bottom of the dashboard, invariant violations are recorded. They contain information about the epoch the violation occurred in, its duration, source worker, unique identifier (which can be used to track down the violation in the PAG visualization), and, if applicable, the affected operator and activity type. While the rest of the dashboard visualizes information depending on the selected epoch, invariant alerts are tracked across epochs to make sure that a user is able to quickly discover issues in the computation. Overall, the dashboard should be able to provide insightful analyses in an interactive manner, without flooding a user with non-critical information. We evaluate whether this is the case in the next section.

\subsection{Case Study: Data Skew}\label{data-skew}

In this case study, we use the dashboard to analyze a source computation that suffers from data skew. Data skew can emerge when data is fed to workers in a nonuniform way, for example due to a misconfigured data source. It also can appear if exchange pacts within the dataflow exchange data unevenly, for example due to skewed join keys. These issues are often hard to track down --- it is not trivial to inspect large volumes of data, high-level operators that exchange under the hood are often treated as black boxes, and adverse exchanges might have unforeseen downstream consequences that conceal the true root cause.

To simulate data skew, we manipulate an exchange operator to exchange all records onto a single worker. We then run this manipulated source computation with four workers over ten epochs. In each epoch, we pass \numprint{2000} records to each worker. The resulting source computation's code is listed in \cref{lst:data-skew}.  

\begin{lstlisting}[float=t,language=Rust,caption={Data-skewed computation},label={lst:data-skew}]
fn main() {
  timely::execute_from_args(std::env::args(), |worker| {
    let adapter = Adapter::attach(worker);

    // dataflow
    let mut input = InputHandle::new();
    let probe = worker.dataflow(|scope| {
      scope.input_from(&mut input)
        .exchange(|x| 0)          // malicious exchange
        .map(|x| x + 1 as u64)
        .probe()
    });

    // input
    for round in 0..10 {
      for x in 0 .. 2000 {
        input.send(x);
      }

      input.advance_to(round + 1);
      while probe.less_than(input.time()) {
        worker.step_or_park(None);
      }
      adapter.tick_epoch();
    }
  }).unwrap();
}
\end{lstlisting}

\begin{figure}[htb]
\centering
\subfloat[\texttt{message-max} invariant]{\includegraphics[width=0.5\textwidth]{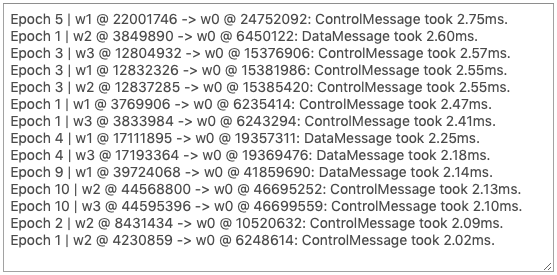}}
\subfloat[Activity duration]{\includegraphics[width=0.5\textwidth]{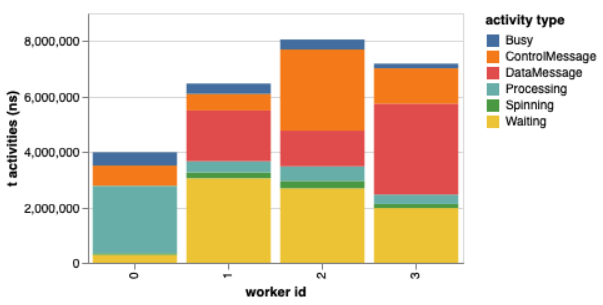}} 
\caption{Data skew invariant and activity duration}
\label{fig:data-skew2}
\end{figure}

\begin{figure}[htb]
\centering
\includegraphics[width=1\textwidth]{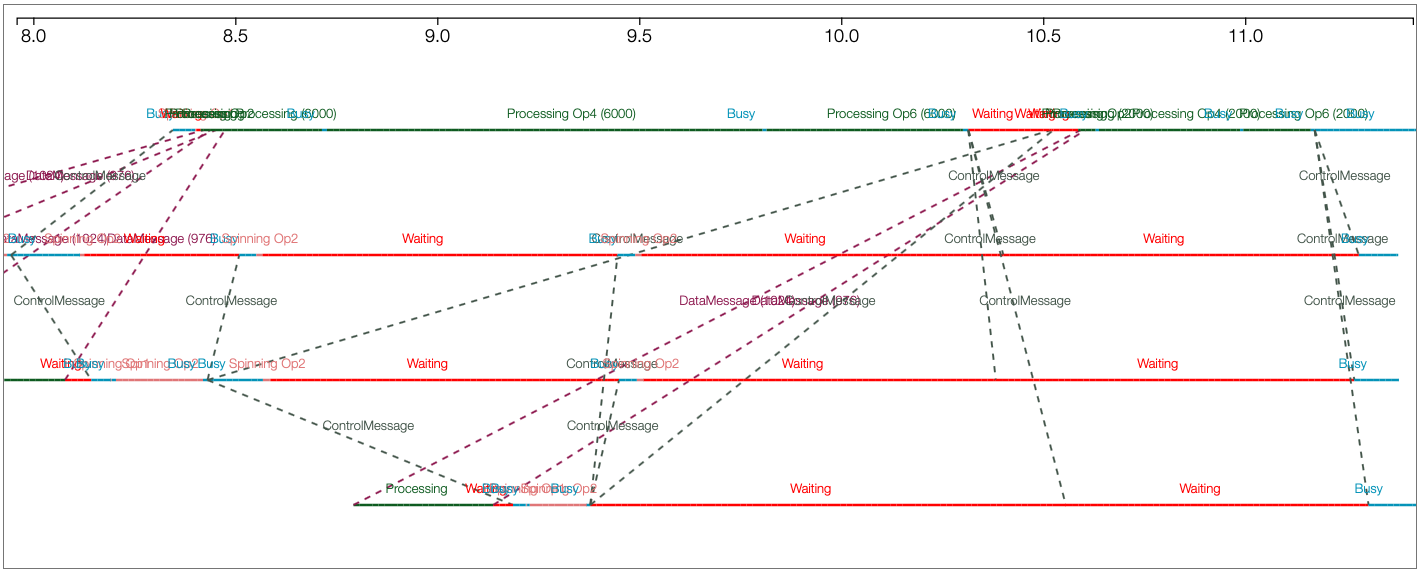}
\caption{Data skew PAG visualization}
\label{fig:data-skew1}
\end{figure}

Upon starting the dashboard, the first thing we note is that the \texttt{message-max} invariant is triggered multiple times, and in each case, the straggling control and data messages are directed at worker 0 (cf.~\cref{fig:data-skew2} \emph{(a)}). This already hints at worker 0 being busy with other tasks (as we know, it has to process thousands of records), and not being able to take the messages off of its internal queue, thereby increasing message latencies. To support this hypothesis, we inspect the culprits triggering the invariant alerts in the PAG visualization. There, it becomes apparent that three of four workers spend most of their time waiting, and only the first worker actively processes records across the epoch's lifetime (cf.~\cref{fig:data-skew1}). The activity duration plot depicted in \cref{fig:data-skew2} \emph{(b)} confirms this suspicion: worker 1--3 spend most of their time waiting (in yellow) or sending messages (in red and orange), while worker 0 is busy with processing (in turquoise).

\begin{figure}[htb]
\centering
\subfloat[Unweighted]{\includegraphics[width=0.5\textwidth]{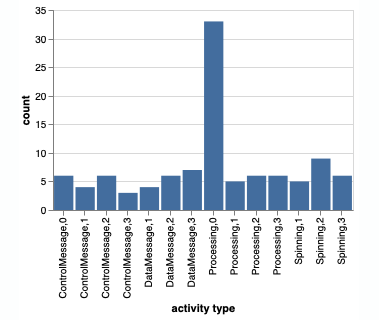}} 
\subfloat[Weighted with activity duration]{\includegraphics[width=0.5\textwidth]{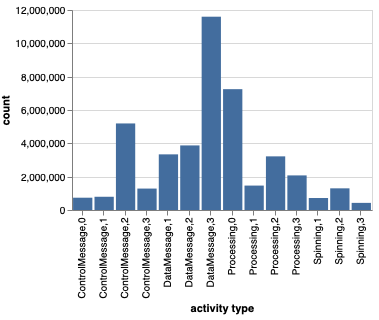}}
\caption{Data skew 10-hop pattern}
\label{fig:data-skew3}
\end{figure}

We can now use the k-hop graph pattern to search for the (potentially cross-worker) cause of this behavior. The resulting 10-hop charts depicted in \cref{fig:data-skew3} reveal our manipulation: Data messages from workers 1, 2, and 3 dominate the weighted graph --- their duration is disproportionally high, as the message receipt is blocked by worker 0. Worker 0's processing activities also take a lot of time and happen frequently: they overshadow the unweighted graph. This makes sense, as processing of previous records in itself becomes a bottleneck for later processing and activities of other workers; the k-hop algorithm correctly identifies them as recurring critical path participants.

\begin{figure}[htb]
\centering
\subfloat[Records sent]{\includegraphics[width=0.5\textwidth]{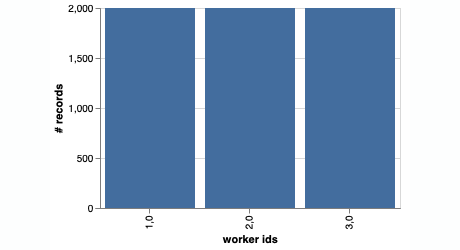}} 
\subfloat[Messages count]{\includegraphics[width=0.5\textwidth]{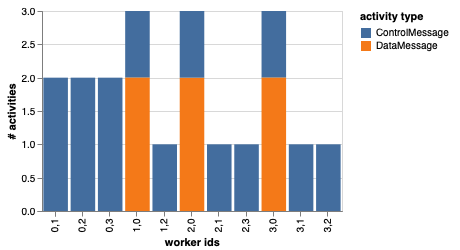}}
\caption{Data skew cross and record metrics}
\label{fig:data-skew4}
\end{figure}

Lastly, the cross and record metrics unveil the unfavorable exchange pact beyond doubt, as shown in \cref{fig:data-skew4}: Records carried by data messages originate only at worker 1--3, and are delivered only to worker 0. Consequently, worker 0 also has to frequently update all other workers with its progress using control messages, but does not send any data messages itself. It also is the only worker that processes any records at all. Hence, we are sure that a defective exchange pact is causing the long waiting activities and slowing down the overall computation, and have successfully used the dashboard to solve the challenge posed by the case study.

Using the analytics provided by the dashboard, we were able to identify the source computation's performance issues. While the presented use case is rather simple and might not have required us to make use of all analyses we consulted, it has highlighted important points. In many cases, invariant violation alerts will be the first to attract user attention. Subsequently, the PAG visualization provides an intuitive way to explore the highlighted hotspots. The remaining analyses can then be used to identify problematic patterns such as unexpected activity distributions, unhealthy relationships between worker threads, or faulty remote messaging and message contents. For identifying cross-worker causes, the k-hops pattern is particularly useful, and it can also be combined with the PAG visualization to arrive at further insights. Overall, a combination of metrics, invariants, patterns, and visual aids is essential to track down the root cause of performance issues or dataflow bugs. 


\subsection{Summary}

In the previous sections, we have reviewed ST2's real-time frontend, which provides a user interface to interactively explore a source computation's health and behavior. The dashboard can be started similarly to the CLI discussed in \cref{cli}, and also offers some configuration options for its integrated invariant checker. It consists of a configurable epoch-wise PAG visualization that is able to selectively highlight the results of the k-hops graph patterns, multiple interactive visualizations to guide the analysis of patterns and metrics, and the invariant checker that alerts users of issues during dataflow execution. We also evaluated the dashboard in a real-world use case, where data skew caused by an erroneous exchange pact impedes an otherwise healthy source computation. Using a combination of the dashboard's tools, we were able to quickly track down the faulty exchange operator. While the presented use case was a simple one, in our experiments, a combination of the visualizations and analyses provided by the dashboard were always sufficient to guide dataflow debugging and performance optimization. During our evaluation, we also remained unaffected by performance issues of the dashboard itself, which matches our expectations from \cref{cli-performance}. We therefore conclude that the dashboard is functionally apt to analyze non-trivial source computations in an online fashion.

\section{Functional Evaluation Summary}

In this chapter, we evaluated ST2's functionality. In its first part, we introduced the CLI, which enables users to interact with ST2 without diving into technical details. We presented multiple commands that can be used for source computation analysis, and also benchmarked them with positive results: For common streaming applications, ST2 is able to keep up in an online setting, even when running multiple invariants and aggregate metrics simultaneously, and, if used responsibly, also when making use of more complex graph patterns. We then introduced ST2's dashboard, which combines the functionality of multiple commands and presents them to the user in a visual manner. After giving an overview of the dashboard, we evaluated it in a real-world use case, in which we created synthetic bottlenecks in a source computation for ST2 to spot. In the presented case study and all other tests, we were able to identify the bottleneck causes with the dashboard successfully and without performance limitations. We therefore conclude that ST2 is not only usable from a performance perspective, but also functionally: It is capable of providing insights into real-world stream processing jobs and can be effectively used to debug and optimize them.

\chapter{Conclusion}\label{conclusion}

In this thesis, we have presented the design, implementation and functionality of ST2. ST2, just like its predecessor SnailTrail~1, is used to tackle the challenging task of analyzing distributed dataflows in an online setting.

During this work, we introduced methods to analyze distributed dataflows and associated challenges, and provided insight on ST2's foundations: the data-centric dataflow programming model, Timely Dataflow and its unique progress tracking mechanism, possible extensions provided by Differential, and its predecessor SnailTrail~1. 

We discussed ST2's implementation: starting with the computational considerations that influenced its design, we presented its decoupled adapter architecture making use of a profiling contract, and the PAG construction based on epochal window semantics: a graph-based representation of log event traces on top of which analyses are implemented and made accessible through a CLI and real-time dashboard.

We evaluated the PAG construction in a multitude of settings, contrasted a Timely and a Differential PAG implementation, and compared our results to SnailTrail~1. We found that the Timely implementation is preferable in our use case, since the PAG construction cannot leverage most of Differential's unique features. We were also able to demonstrate that the PAG construction, if configured correctly, is well-suited to profile nearly any source computation offline and online --- if necessary, it can even be scaled beyond the source computation's configuration ---, and that it exceeds even SnailTrail~1's performance by a significant margin.

Lastly, we implemented multiple data analytics --- aggregate metrics, temporal and progress invariant checking, and graph pattern matching --- on top of the PAG. We were able to demonstrate that the analyses did not have a significant impact on ST2's performance. In a case study, we used the analytics and auxiliary visualizations from the dashboard, and were able to showcase that they are useful tools for debugging, monitoring, and optimizing a source computation in practical settings.

We therefore believe that ST2 is a valuable system for analyzing source computations in the Timely ecosystem. In the future, thanks to its decoupled architecture it could also easily be adapted to support other stream processors, and distributed systems in general. Similarly, the PAG opens the door to the implementation of new analysis algorithms, which can be created without having to consider underlying implementation details. With some care, it might even be possible to leverage e.g.\ similarity metrics between PAG epochs to create a truly differential PAG. This would open up ST2 to performance benefits provided by differential and incremental computation, and also to completely new analyses targeting changes in the source computation over time. All in all, we believe that ST2 --- powered by Timely's correctness guarantees, expressivity, and performance characteristics --- paves the way for users and developers alike to better understand and refine their distributed dataflows.



\appendix{}

\microtypesetup{protrusion=false}

\setlength\cftbeforeloftitleskip{0pt}%
\setlength\cftafterloftitleskip{1em}%
\setlength\cftbeforelottitleskip{2.5em}%
\setlength\cftafterlottitleskip{1em}%

\cleardoublepage

\pagenumbering{roman}
\setcounter{page}{6}

\ifthenelse{\equal{\mylanguage}{de}}{
\pdfbookmark[0]{Abbildungsverzeichnis}{listoffigures}
}{
\pdfbookmark[0]{List of Figures}{listoffigures}
}
\listoffigures{}

\begingroup
\let\cleardoublepage\relax
\let\clearpage\relax
\vspace{1em}
\addtocontents{toc}{%
  \edef\protect\SavedTocDepth{\protect\the\protect\value{tocdepth}}%
}%
\addtocontents{toc}{\protect\setcounter{tocdepth}{-10}}
\ifthenelse{\equal{\mylanguage}{de}}{%
\renewcommand{\lstlistlistingname}{Programmverzeichnis\texorpdfstring{\vspace{-0.5em}}{}}
}{%
\renewcommand{\lstlistlistingname}{List of Listings\texorpdfstring{\vspace{-0.5em}}{}}
}
\lstlistoflistings
\addtocontents{toc}{\protect\setcounter{tocdepth}{\protect\SavedTocDepth}}
\endgroup

\ifthenelse{\equal{\mylanguage}{de}}{
\pdfbookmark[0]{Tabellenverzeichnis}{listoftables}
}{
\pdfbookmark[0]{List of Tables}{listoftables}
}

\listoftables{}
\microtypesetup{protrusion=true}
\cleardoublepage
\sloppy\printbibliography[heading=bibintoc]\fussy

\cleardoublepage
\appendix

\chapter{Appendix}\label{appendix}
  
\section{Timely PAG Benchmark}

\begin{figure}[H]
\centering
\includegraphics[width=.8\textwidth]{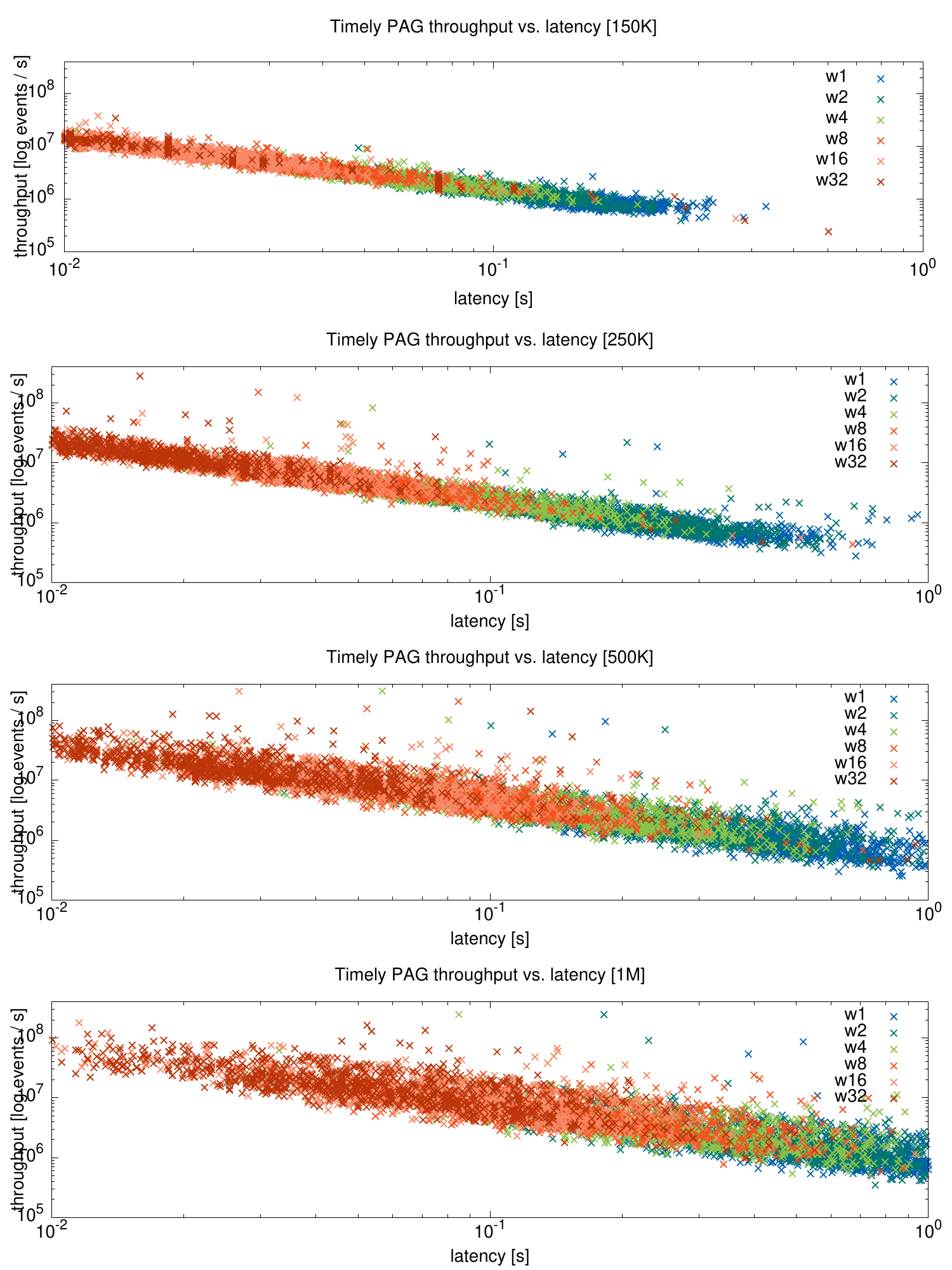}
\caption{Timely PAG throughput vs.\ latency (per epoch)}
\label{fig:st_latvsthroughcloud}
\end{figure}

\begin{figure}[H]
\centering
\includegraphics[width=.95\textwidth]{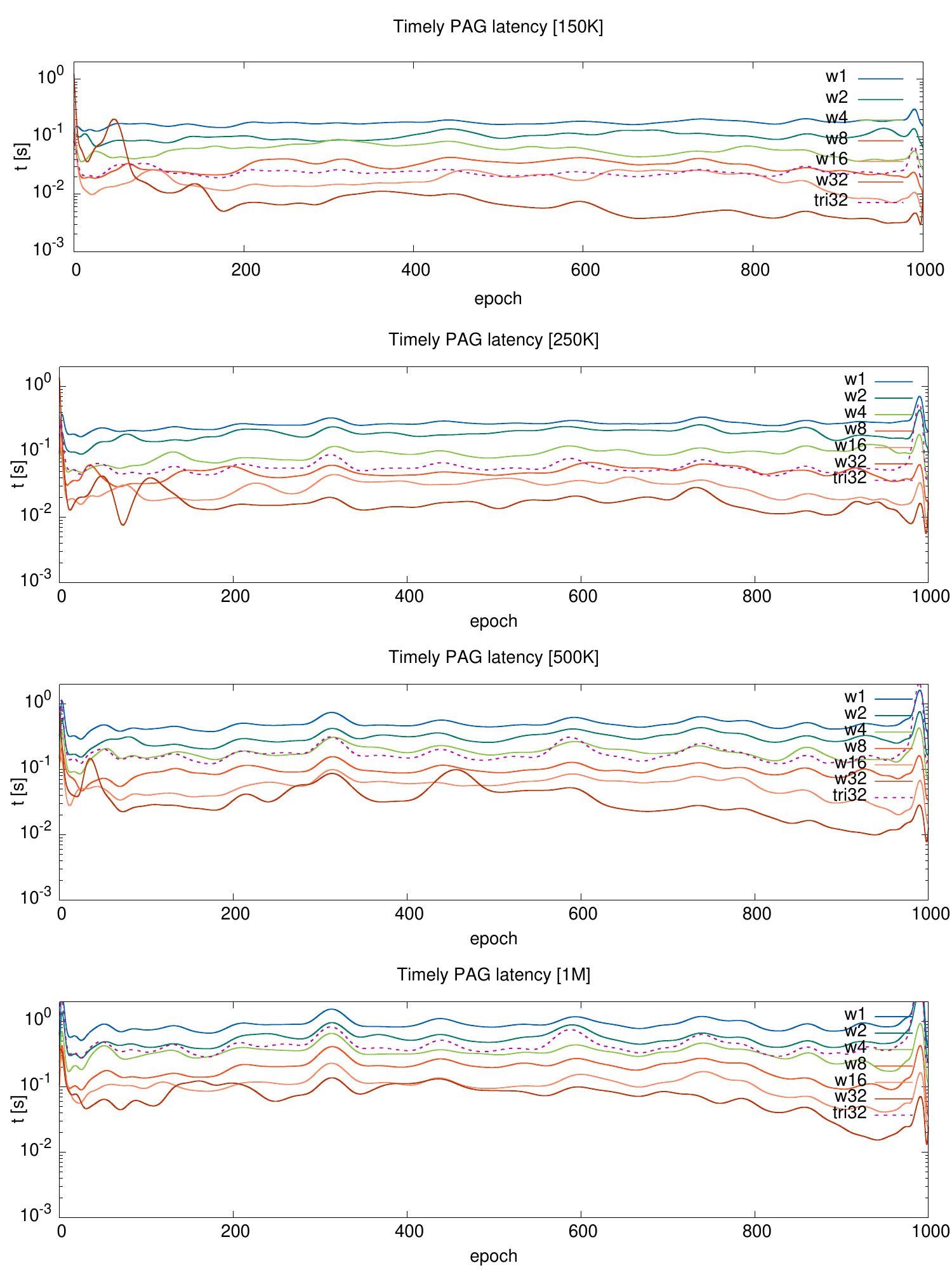}
\caption[Timely PAG latency]{Timely PAG latency for multiple epoch sizes and worker configurations. Epoch size provided in brackets. Dashed line denotes the triangles source computation.}
\label{fig:latency}
\end{figure}

\begin{figure}[H]
\centering
\includegraphics[width=.95\textwidth]{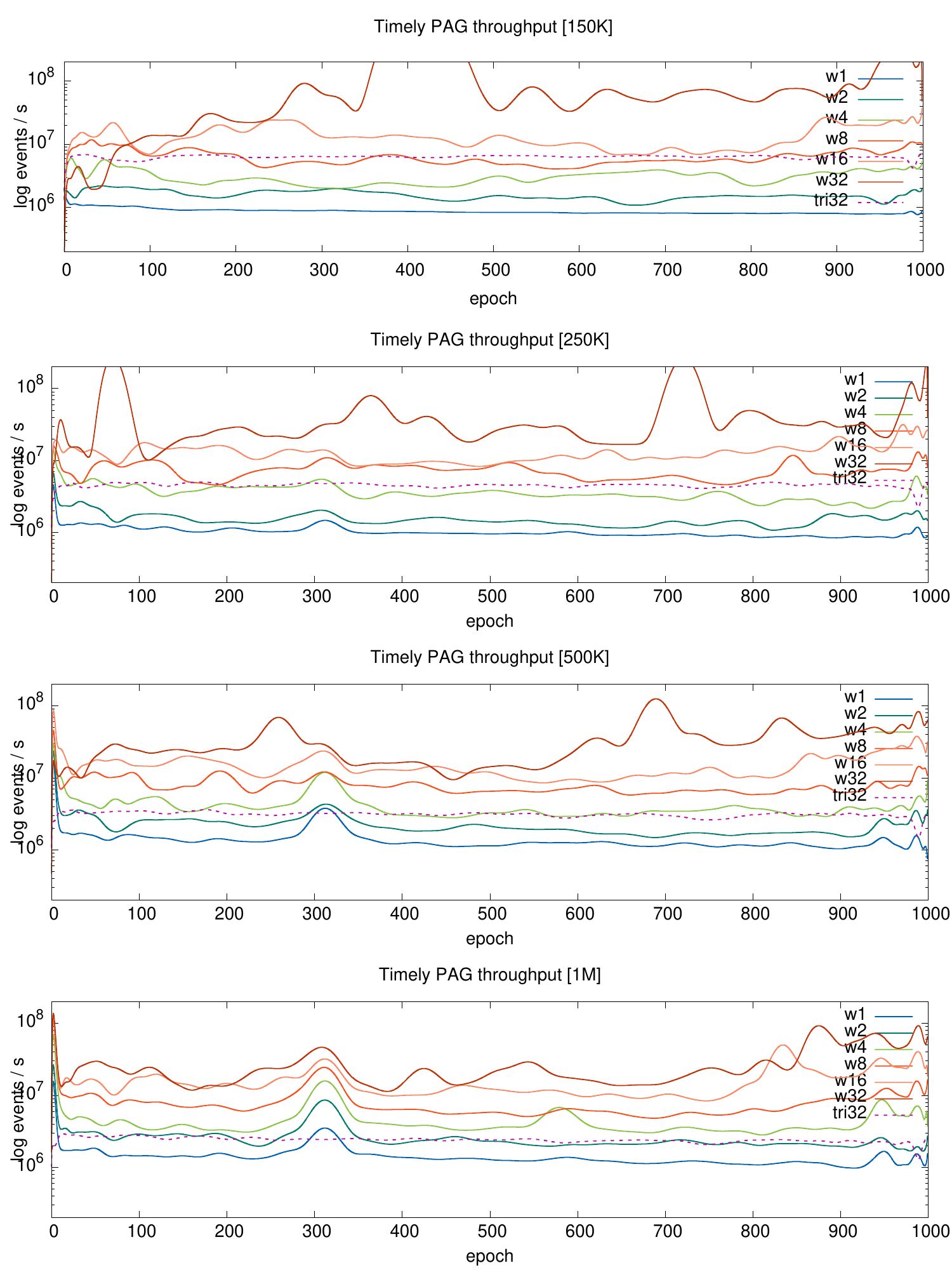}
\caption[Timely PAG throughput]{Timely PAG throughput for multiple epoch sizes and worker configurations. Epoch size provided in brackets. Dashed line denotes the triangles source computation.}
\label{fig:throughput}
\end{figure}

\section{Differential PAG Benchmark}

\begin{figure}[H]
\centering
\includegraphics[width=.9\textwidth]{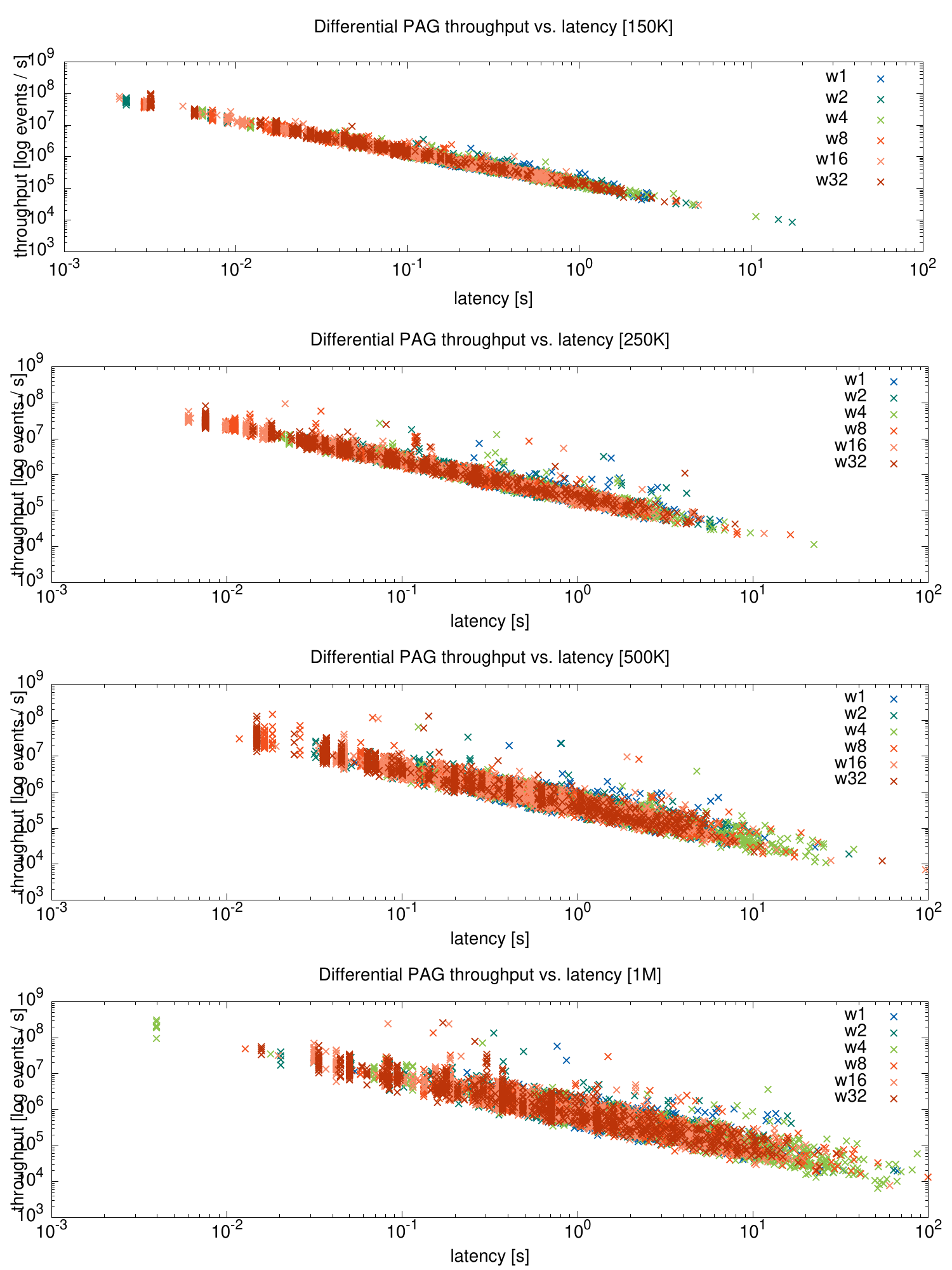}
\caption{Differential PAG throughput vs.\ latency (per epoch)}
\label{fig:diff_st_latvsthroughcloud}
\end{figure}

\begin{figure}[H]
\centering
\includegraphics[width=.95\textwidth]{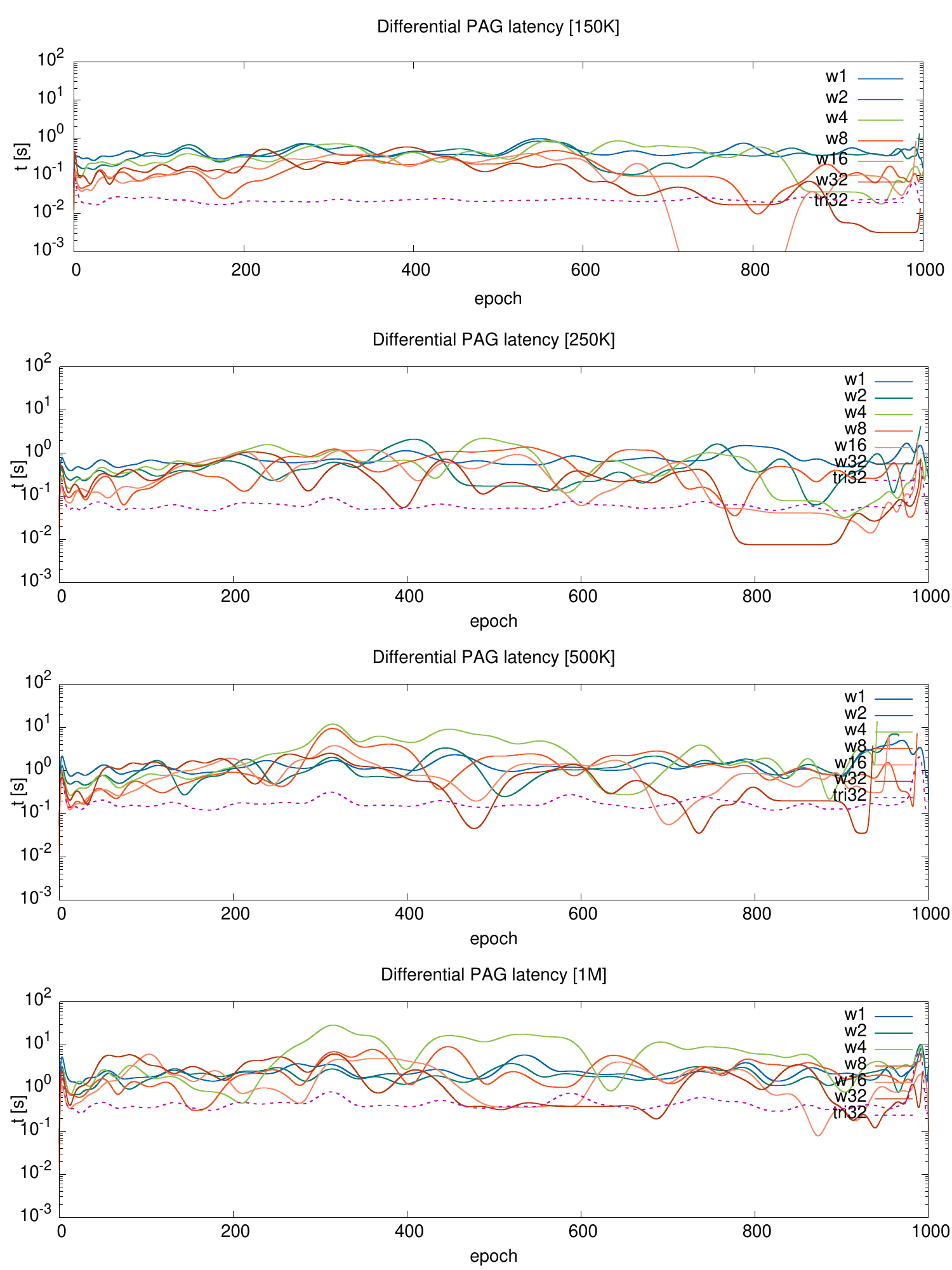}
\caption[Differential PAG latency]{Differential PAG latency for multiple epoch sizes and worker configurations. Epoch size provided in brackets. Dashed line denotes the triangles source computation.}
\label{fig:diff_latency}
\end{figure}

\begin{figure}[H]
\centering
\includegraphics[width=.95\textwidth]{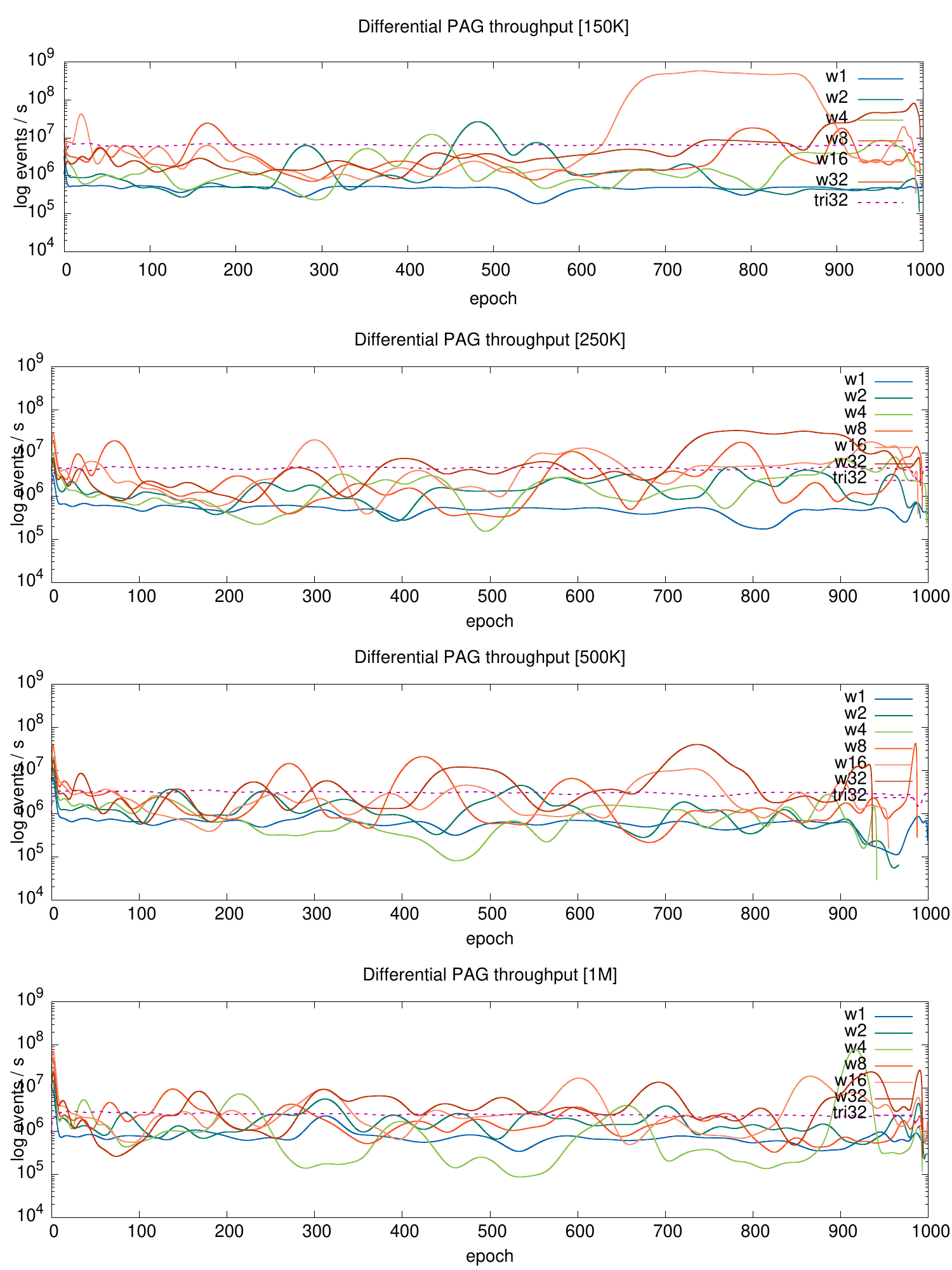}
\caption[Differential PAG throughput]{Differential PAG throughput for multiple epoch sizes and worker configurations. Epoch size provided in brackets. Dashed line denotes the triangles source computation.}
\label{fig:diff_throughput}
\end{figure}


\addtocontents{toc}{\protect\setcounter{tocdepth}{2}}

\end{document}